\documentclass[10pt,journal,compsoc,twoside]{IEEEtran}

\usepackage{graphicx} \graphicspath{{images/}}
\usepackage[nocompress]{cite} 
\usepackage{amsmath}
\usepackage{booktabs} 
\usepackage{bm}
\usepackage{enumitem}
\usepackage{algorithm}
\usepackage[noend]{algpseudocode}
\usepackage{subcaption}

\renewcommand{\vec}[1]{\boldsymbol{#1}}
\newcommand{\norm}[1]{\left\lVert#1\right\rVert}

\begin{document}

\title{Fast and Scalable\\ Position-Based Layout Synthesis}

\author{Tomer~Weiss,
	Alan~Litteneker,
	Noah~Duncan,
        Masaki~Nakada,
	Chenfanfu~Jiang,
	Lap-Fai~Yu,~\IEEEmembership{Member,~IEEE,} and
        Demetri~Terzopoulos,~\IEEEmembership{Fellow,~IEEE}%
\IEEEcompsocitemizethanks{\IEEEcompsocthanksitem T.~Weiss, A.~Litteneker, M.~Nakada, and D.~Terzopoulos are with the University of California, Los Angeles.
\IEEEcompsocthanksitem N.~Duncan is with WorkPatterns, Inc.
\IEEEcompsocthanksitem C.~Jiang is with the University of Pennsylvania.
\IEEEcompsocthanksitem L.-F.~Yu is with the University of Massachusetts, Boston.}
\thanks{Manuscript received ?? ??, 2018; revised ?? ??, 2018.}
}

\markboth
{Transactions on Visualization and Computer Graphics, VOL.~??, NO.~??}%
{Weiss \MakeLowercase{\textit{et al.}}: Fast and Scalable Position-Based Layout Synthesis}

\IEEEtitleabstractindextext{%
\begin{abstract}
The arrangement of objects into a layout can be challenging for
non-experts, as is affirmed by the existence of interior design
professionals. Recent research into the automation of this task has
yielded methods that can synthesize layouts of objects respecting
aesthetic and functional constraints that are non-linear and
competing. These methods usually adopt a stochastic optimization
scheme, which samples from different layout configurations, a process
that is slow and inefficient. We introduce an physics-motivated,
continuous layout synthesis technique, which results in a significant
gain in speed and is readily scalable. We demonstrate our method on a
variety of examples and show that it achieves results similar to
conventional layout synthesis based on Markov chain Monte Carlo (McMC)
state-search, but is faster by at least an order of magnitude and can
handle layouts of unprecedented size as well as tightly-packed layouts
that can overwhelm McMC.
\end{abstract}
}

\maketitle

\IEEEdisplaynontitleabstractindextext

\IEEEraisesectionheading{\section{Introduction}
\label{sec:introduction}}

\IEEEPARstart{T}{he} arrangement of objects into a desirable layout is
an everyday problem that is nonetheless surprisingly complex. For
example, to find a desirable furniture arrangement for a living-room,
one must consider the visibility of the television, a suitable
separation of sofas, and access to adjacent rooms, among other factors
that differ according to taste and style. It is often difficult for
people to solve layout problems, as is affirmed by the existence of
professional interior layout designers and self-help resources.

\begin{figure}[t]
\begin{subfigure}{0.99\columnwidth}
\includegraphics[width=\columnwidth]{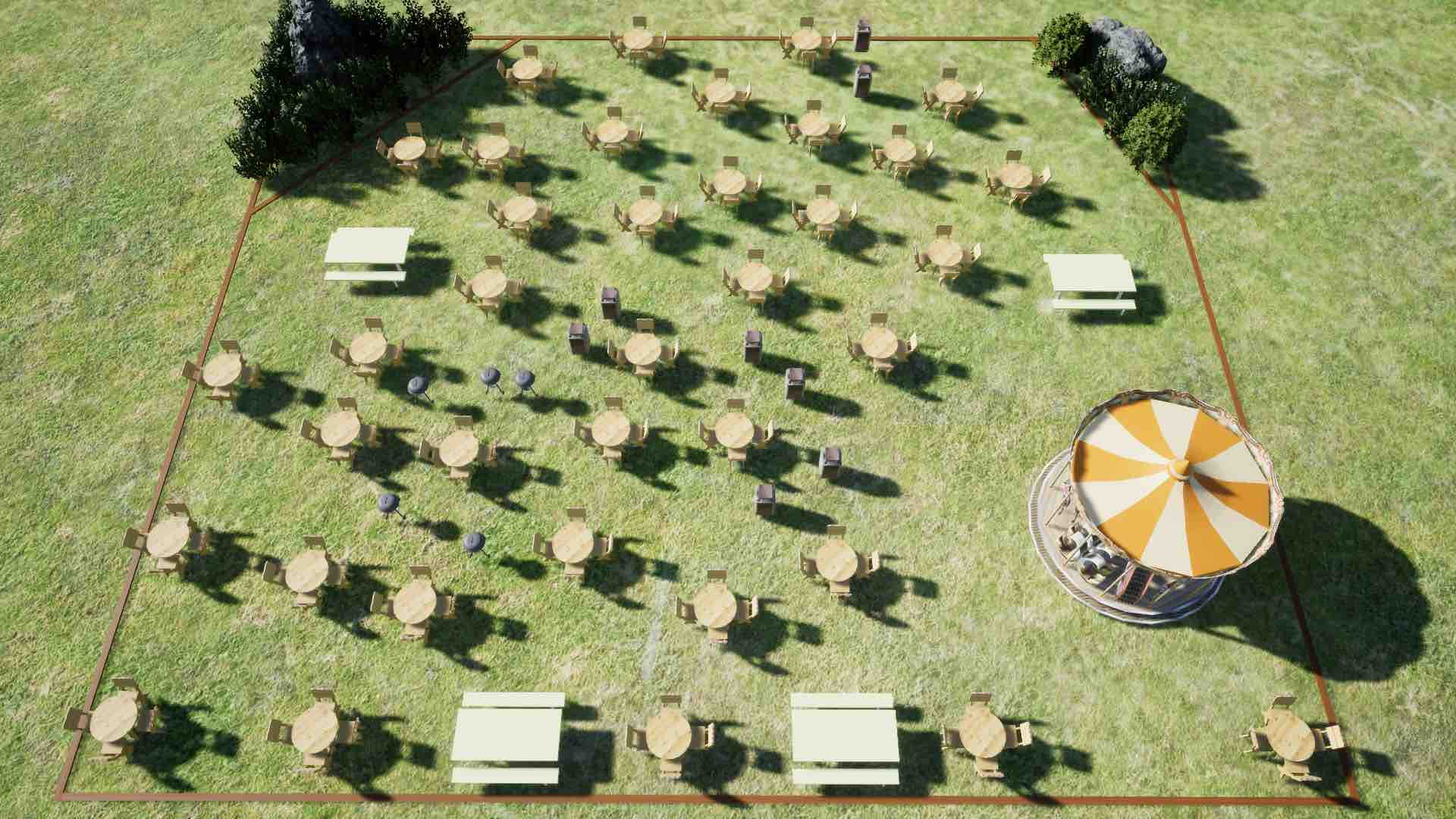}
\end{subfigure}
\\[2pt]
\begin{subfigure}{0.99\columnwidth}
\includegraphics[width=\columnwidth]{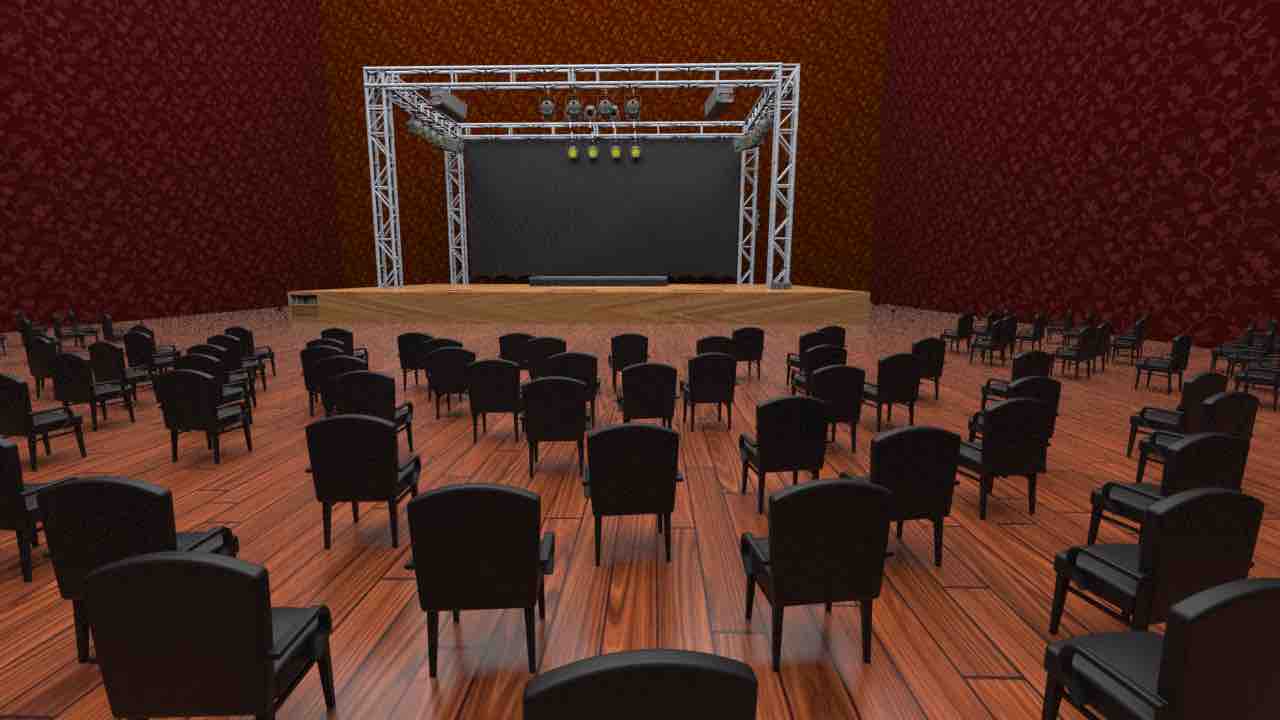}
\end{subfigure}
\caption{A tightly-packed picnic layout (top), and a theater layout
with a large number of chairs (bottom), automatically placed by our
method given user-specified constraints that include distance, viewing
angle, and spaciousness criteria.
\label{fig:teaser}
}
\end{figure}

\begin{figure*}
\centering
\begin{subfigure}{0.33\textwidth}
\includegraphics[width=\textwidth]{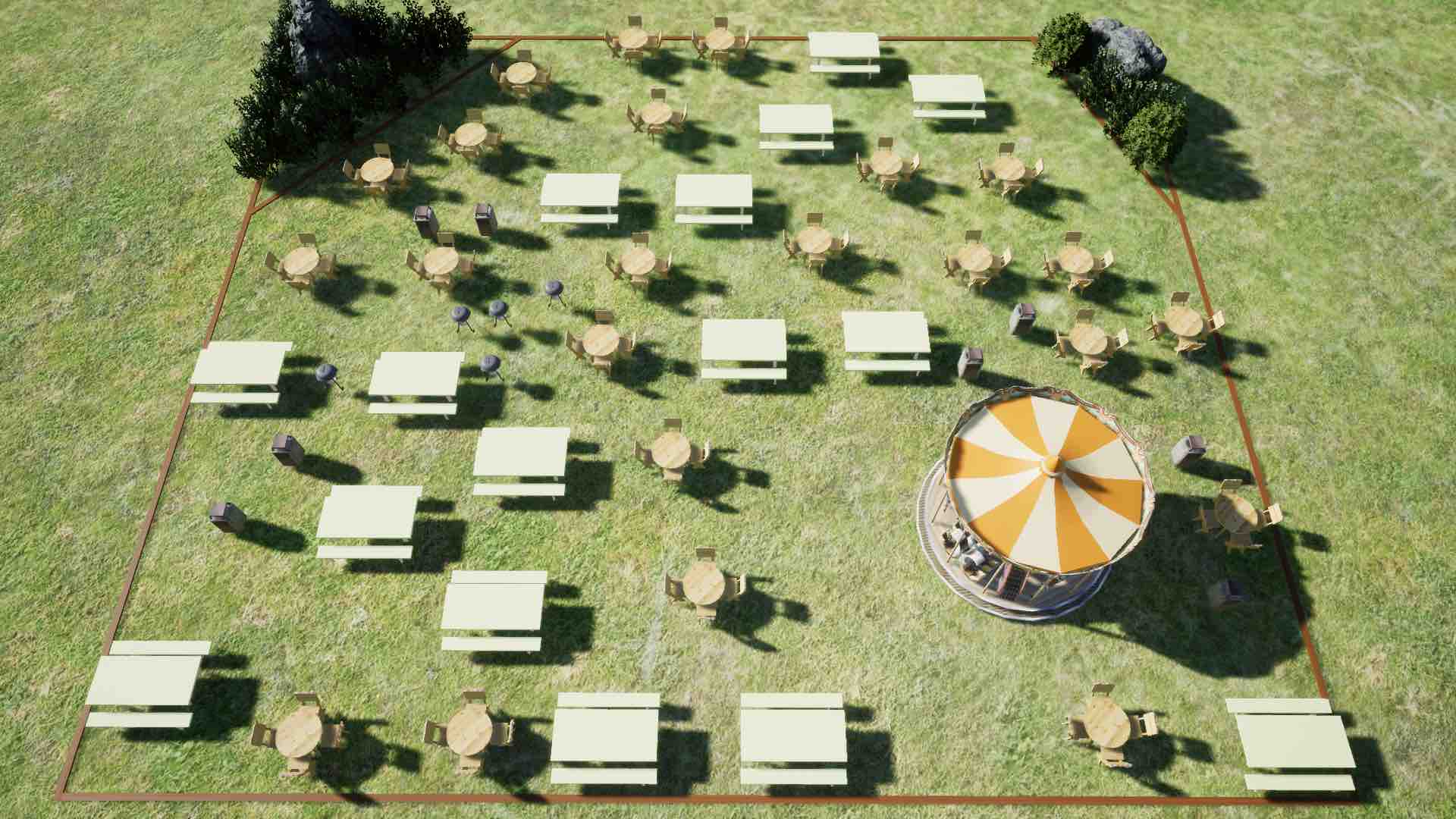}
\label{fig:tp_pic1}
\end{subfigure}\hfill
\begin{subfigure}{0.33\textwidth}
\includegraphics[width=\textwidth]{tp_picnic2_small}
\label{fig:tp_pic2}
\end{subfigure}\hfill
\begin{subfigure}{0.33\textwidth}
\includegraphics[width=\textwidth]{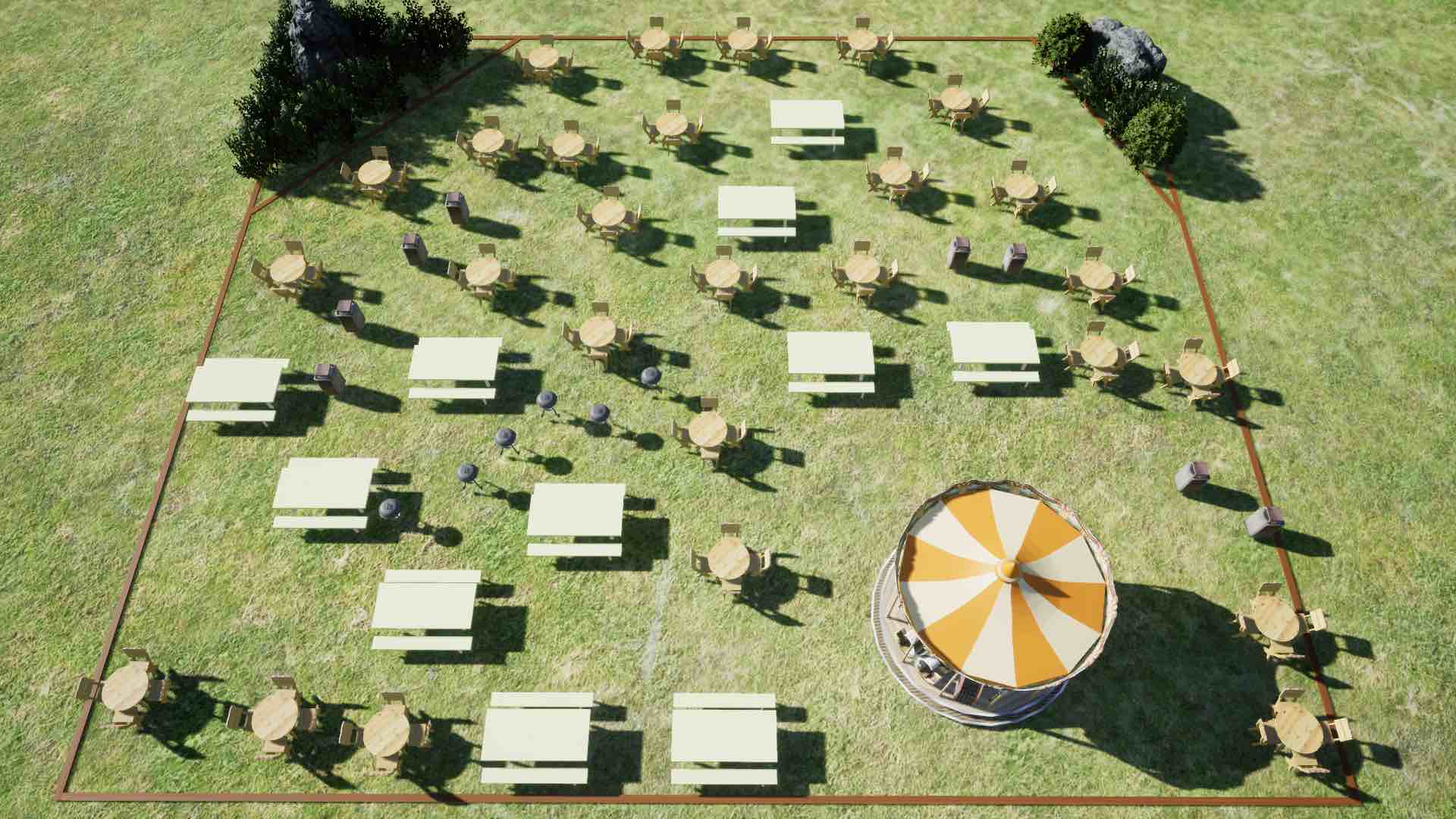}
\label{fig:tp_pic3}
\end{subfigure}
\caption{Tightly-packed picnic layouts (comprised of various numbers
of different object types) synthesized by our method.
\label{fig:tp_picnic}}
\end{figure*}

\begin{figure*}
\centering
\includegraphics[width=0.33\textwidth]{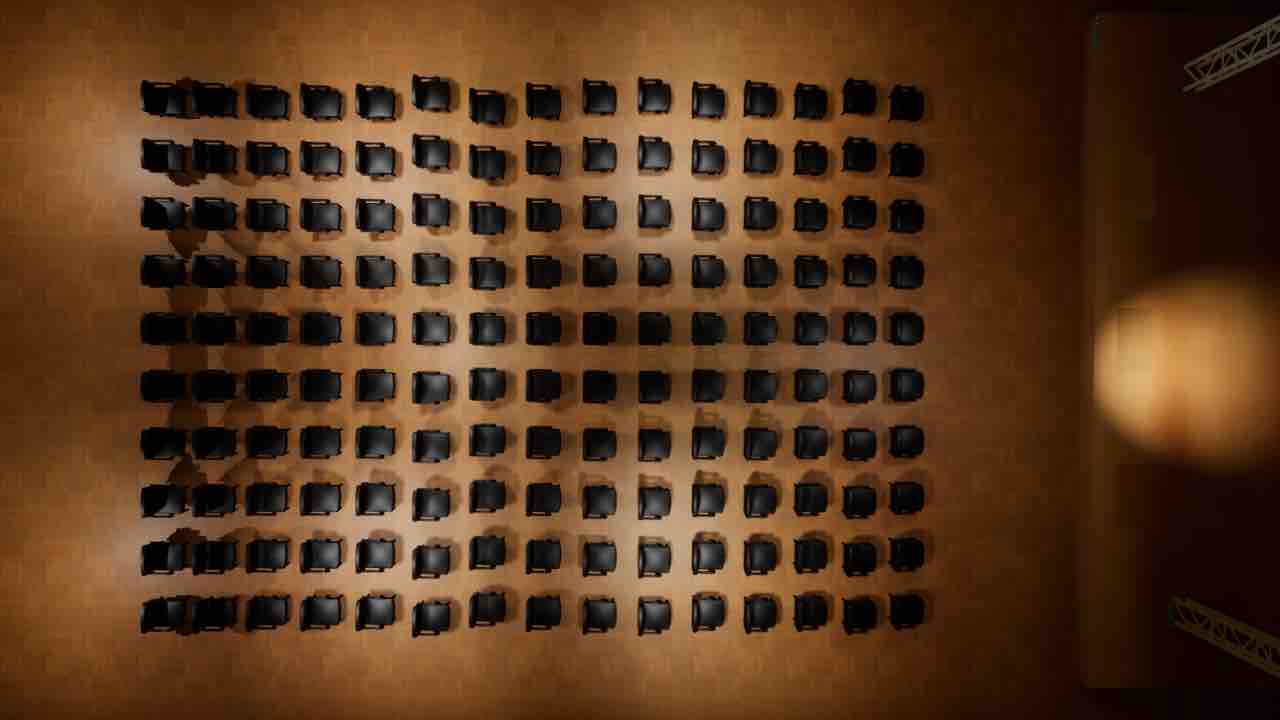}\hfill
\includegraphics[width=0.33\textwidth]{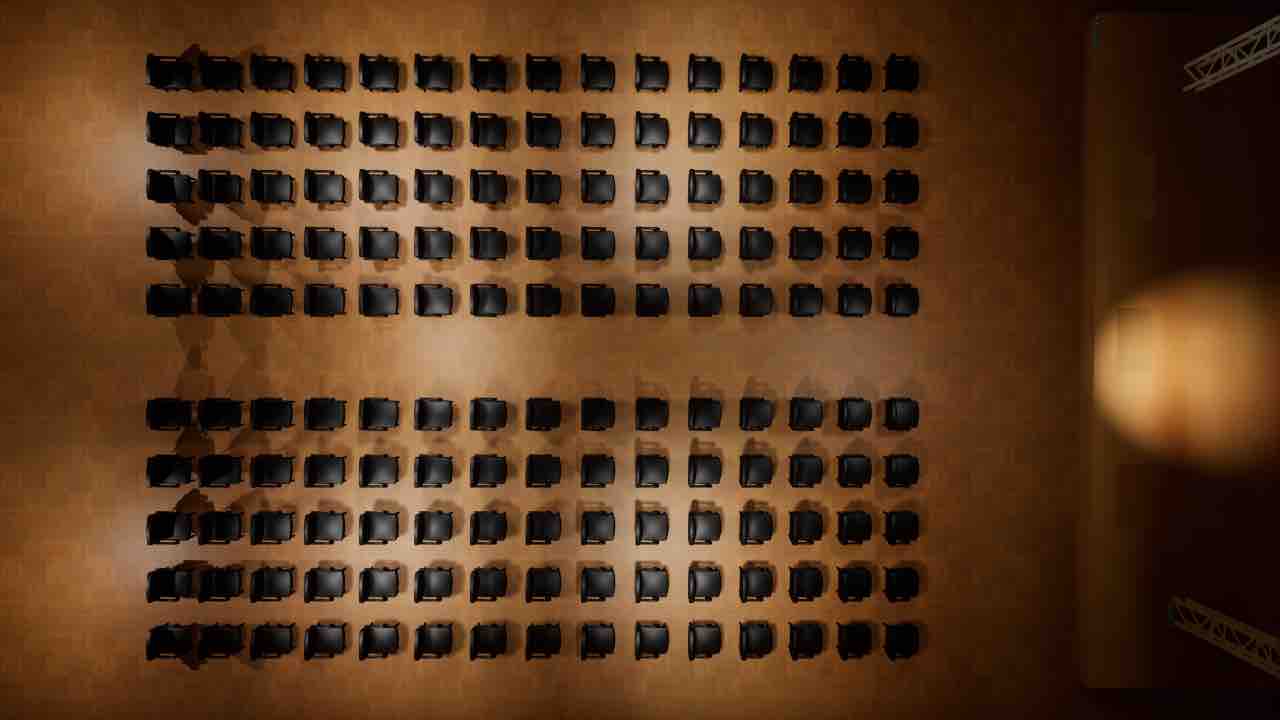}\hfill
\includegraphics[width=0.33\textwidth]{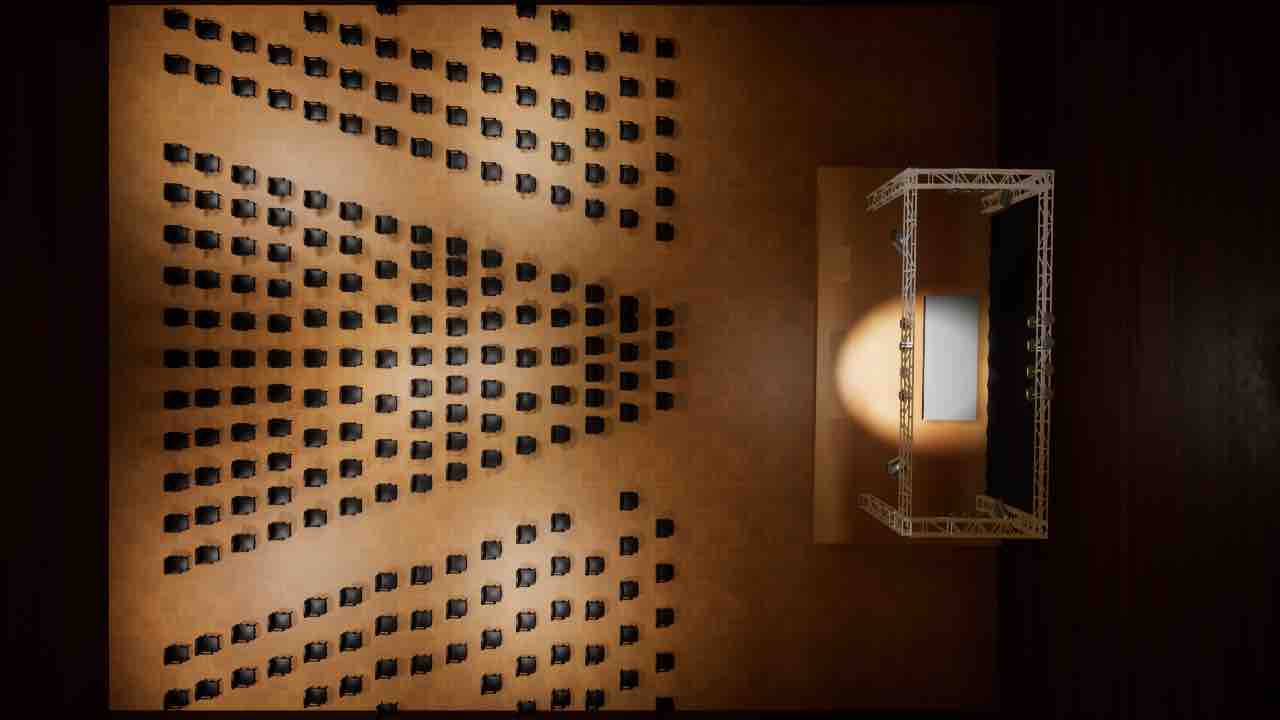}\\[2pt]
\includegraphics[width=0.33\textwidth]{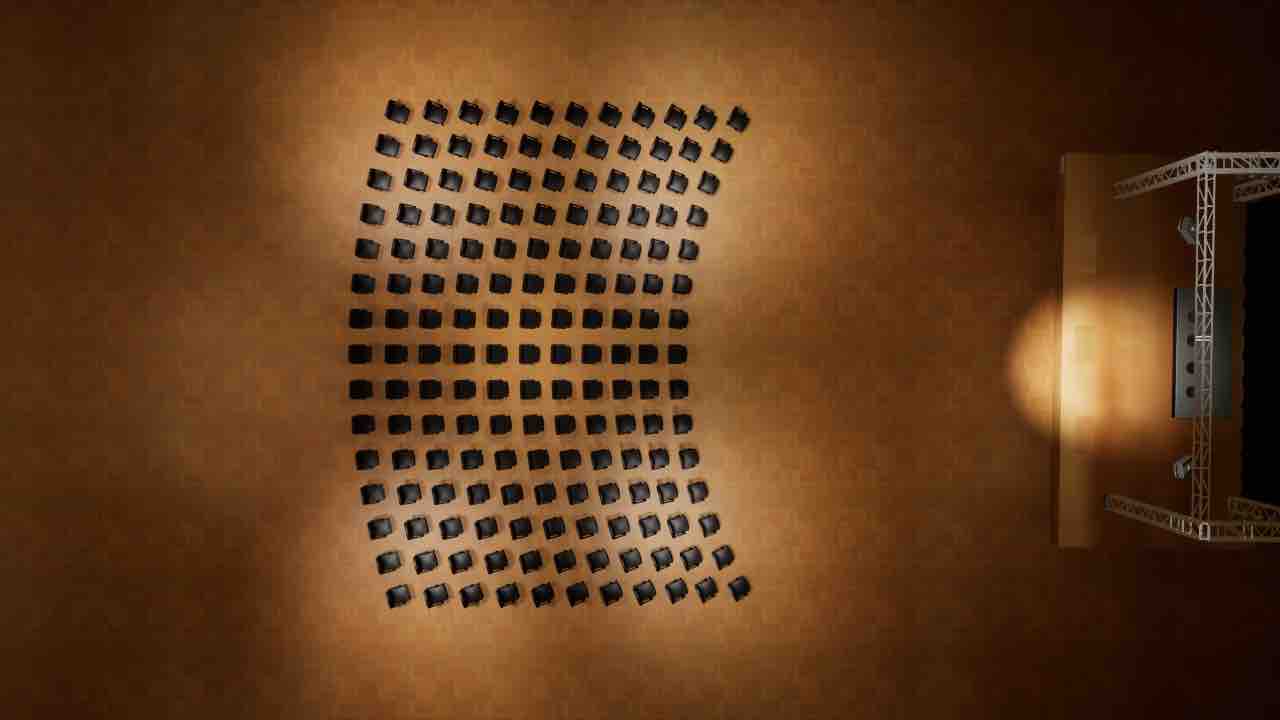}\hfill
\includegraphics[width=0.33\textwidth]{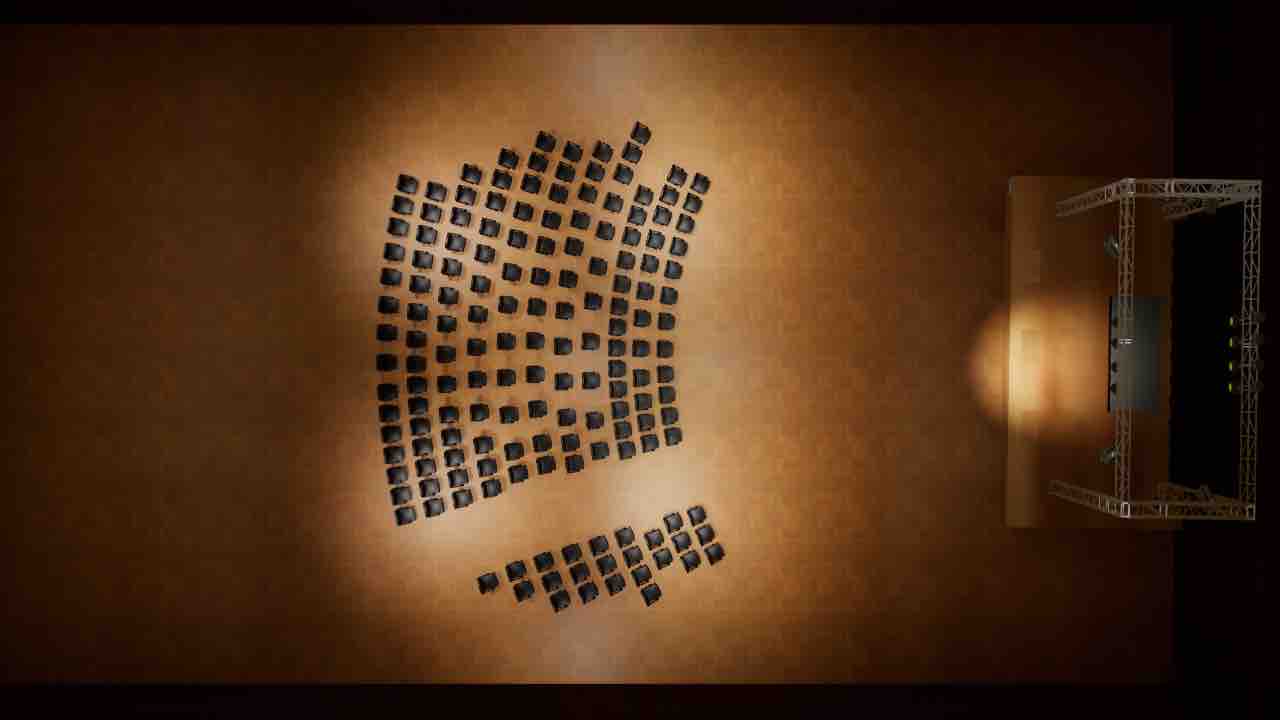}\hfill
\includegraphics[width=0.33\textwidth]{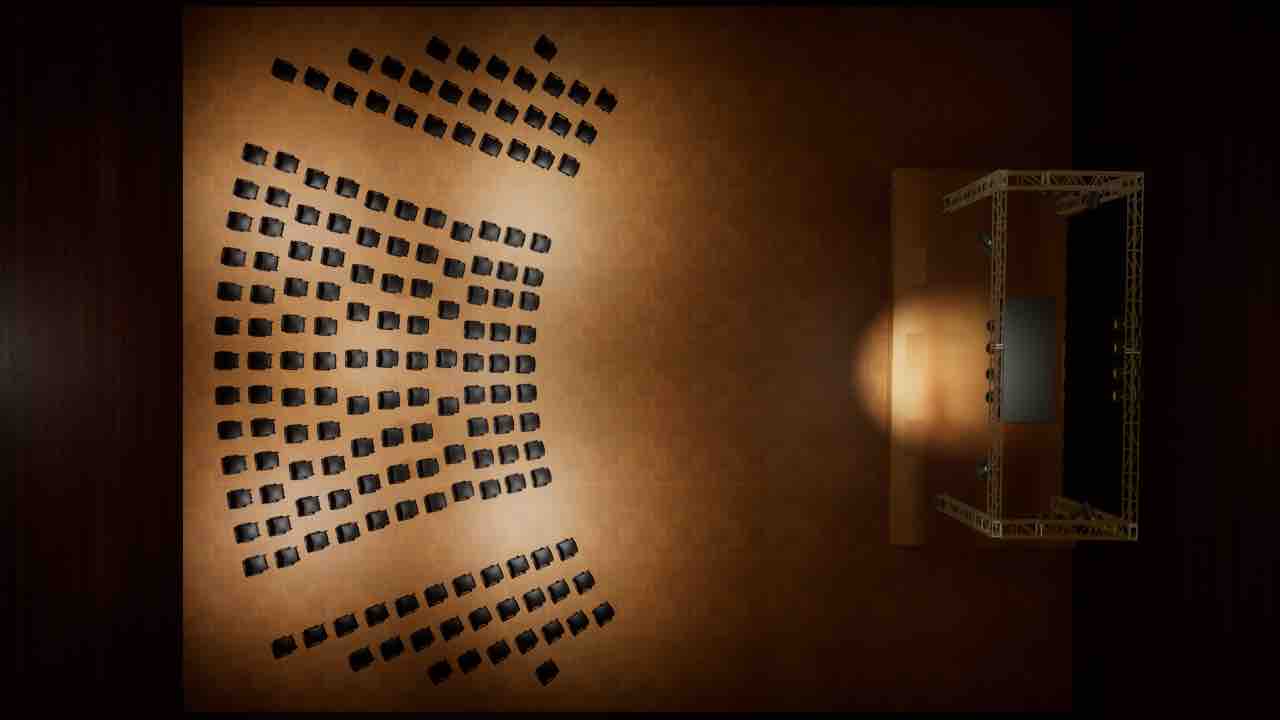}
\caption{Variations of the theater scenes synthesized by our method.}
\label{fig:theater_variations}
\end{figure*}

The principal motivation in computer graphics for automatic or
semi-automatic layout synthesis is the need to model realistic virtual
worlds \cite{smelik2014survey}. Methods that can automatically
synthesize realistic, large-scale virtual environments
(Fig.~\ref{fig:teaser}) are useful for gaming, educational, and
training purposes. Such methods are more useful in practice if they
can generate larger, more complex environments and execute faster.

In recent years, researchers have proposed several methods for
synthesizing layouts that pose layout synthesis as a highly non-convex
optimization problem subject to numerous constraints. Due to the
challenging nature of these problems, previous work applied stochastic
optimization to sample viable layout candidates. Markov chain Monte
Carlo (McMC) methods \cite{chib1995understanding} are the preferred
technique because the constraints are often difficult to express as
differentiable functions. Unfortunately, these techniques become
inefficient when dealing with large numbers of objects.

To overcome this problem, we introduce a continuous framework for
layout synthesis. Our main observation is that there are commonalities
between layout synthesis and the elastic simulation of deformable
objects. Elasticity penalizes the deformation of an object---the
energy increases proportionally to the magnitude of the
deformation---and layout synthesis penalizes the magnitude of
constraint violation. Both can be formulated as optimization problems,
and both can be tackled using continuous optimization procedures. Our
new, continuous approach enables the fast generation of large-scale,
tightly-packed layouts that are intractable using previous approaches.
However, like stochastic methods, our deterministic approach can
synthesize multiple viable layouts for a given environment
(Fig.~\ref{fig:tp_picnic}, Fig.~\ref{fig:theater_variations}). To our
knowledge, ours is the first physics-motivated approach to layout
synthesis.

Our method takes as input an environment, a set of objects, and
prescribed aesthetic and/or functional layout constraints that can be
easily modified. Initially the positions and orientations of the
objects are randomized, which is analogous to choosing a random
initial guess in an iterative solver. Object positions and
orientations are iteratively modified to achieve a viable layout. At
each iteration, the objects are moved so as to satisfy competing
constraints. Hard constraints such as observing layout boundaries and
preventing collisions between layout objects are enforced by default.
The procedure converges when all the prescribed constraints are
adequately satisfied. We show that a diverse set of constraints, which
have been applied in prior layout synthesis schemes, can be formulated
within our framework. Layout objects can also be grouped, and each
group assigned aesthetic or functional layout constraints. Groups are
reusable in defining other layout groups, and they are also easily
modifiable both in terms of the participating objects and the group
layout constraints.

Our main contributions in this paper are as follows:
\begin{enumerate}
\item We propose a novel, physics-motivated approach to layout
synthesis.
\item We formulate a set of common layout constraints within our
framework.
\item We develop a novel, continuous and deterministic layout
synthesis algorithm with significantly reduced computational cost,
making large-scale layout synthesis problems tractable.
\end{enumerate}

The remainder of the paper is organized as follows:
Section~\ref{sec:related-work} reviews relevant prior work.
Section~\ref{sec:algorithm} presents the technical details of our
approach. Section~\ref{sec:expr} describes our experiments and
presents our results. In Section~\ref{sec:disc-concl}, we further
discuss our framework, including its strengths and limitations, and
suggest avenues for future work.

\section{Related Work}
\label{sec:related-work}

\subsection{Layout Synthesis}

We focus on layout synthesis problems in which a set of objects is to
be arranged in an open space. The objects are assumed to be rigid
bodies. The goal of the layout problem is to position and orient the
objects such that they satisfy several functional and aesthetic
criteria. These criteria are encoded as the terms of a non-convex
objective function. The main challenge stems from finding an
arrangement that respects conflicting terms, resulting in a multitude
of possible layout outcomes, some of which may be unsatisfactory.
Relevant publications in this category include the following:

Yu et al.~\cite{yu2011make} and Merrell et
al.~\cite{merrell2011interactive} introduced an McMC-based approach to
furniture arrangement. Yeh et al.~\cite{yeh2012synthesizing}
formulated layout constraints with factor graphs, allowing a variable
number of elements in the synthesized layout. These McMC-based,
stochastic sampling methods can synthesize scenes that respect a
complex and conflicting set of constraints, but they have been shown
to work only on layouts with a limited number of objects and relaxed
spacing. The underlying inefficiency of these approaches stems from
the fact that they do not employ local gradient information---they
merely sample a new position of a furniture item by applying a shift
move to the current position.

Fu et al.~\cite{fu_siga17} synthesize layouts from object relation
graphs learned from a database of floor plans. Feng et
al.~\cite{feng2016crowd} optimized mid-scale layouts by stochastically
optimizing an objective function derived from agent-based simulation.
Fisher et al.~\cite{fisher2012example} generated small, local layouts
of objects in a scene, guided by exemplars---e.g., the layout of items
on a desk. Layouts are generated by sampling probabilistically from an
occurrence model distribution. Additionally, the authors report that
since the layout generation is probabilistic, it cannot handle hard
constraints such as rigid grid layouts or exact alignment
relationships.

Peng et al.~\cite{peng2014computing} introduced a method that creates
layouts from deformable templates, which differs from prior work that
assumed objects are rigid. They incorporated a continuous method in
their approach, but use it only to deform objects, whereas we use ours
to position objects. Recently, Wu et al.~\cite{wu2018} proposed a
mixed integer-linear programming formulation for floor plan synthesis.

Layout research has also addressed contexts other than interior
design. Majerowicz et al.~\cite{majerowicz2014filling} focused on
adding objects to shelves in a 2D setting. Bao et
al.~\cite{Bao:2013:GEG:2461912.2461977} uses a combined stochastic and
numerical optimization approach to explore and refine building
layouts. Zhu et al.~\cite{zhu2012motion} synthesized layouts of
mechanical components to control the motion of a toy. Cao et
al.~\cite{cao2012automatic,cao2014look} synthesized manga layouts.
Reinert et al.~\cite{reinert2013interactive} introduced an interactive
layout generation method of arranging shapes according to aesthetic
attributes, such as color and size. We focus on interior and exterior
design layouts, but our method generalizes to other contexts.

\subsection{Physics-Based vs Position-Based Methods}

Physics-based modeling techniques have been used in various contexts,
from animation \cite{terzopoulos1987elastically,
manteaux2016adaptive}, to geometric design \cite{qin1996d,
attar2009physics}, to architectural floor plan design
\cite{harada1995interactive, arvin2002modeling}. Position-Based
Dynamics (PBD), independently introduced by Muller et
al.~\cite{muller2007position} and by Stam \cite{stam2009nucleus}, was
originally proposed as a means of simulating physical models in
situations, such as games, where speed and robustness takes priority
over physical realism. Researchers have since applied the approach to
a variety of simulations, from soft and rigid bodies
\cite{deul2016position}, fluids \cite{macklin2013position}, to crowd
simulation \cite{weiss_mig17}. PBD is part of a larger family of
simulation methods called position-based methods. Bender et
al.~\cite{bender2014survey} present a survey. The common
characteristic of these methods is that they work directly with
positions rather than with forces as does true Newtonian dynamics (so
PBD should more properly be called Position-Based Kinematics). The
method works by iteratively adjusting particle positions to satisfy a
set of constraints. To our knowledge, we are the first to pursue this
approach in the context of layout synthesis.

\section{Algorithm}
\label{sec:algorithm}

Starting from random initial layouts, our method explores sequences of
possible object arrangements by iteratively solving user-prescribed
layout constraints.

\begin{algorithm}[t]
\caption{ }
\label{alg:pbd_opt}
\begin{algorithmic}[1]
\For{Object $i$}
\State Initialize $\vec p_i =\vec p_{i}^0$; ~ $\theta_i =\theta_i^0$ \label{loop:2}
\EndFor
\State Set $l=1$
\While{SolverIteration $ l < \mathrm{max}$} \label{loop:5}
\State UpdateStiffnesses($C_1,\ldots,C_m$) \label{loop:3}
\State ProjectConstraints($C_1,\ldots,C_m$) \label{loop:6}
\For{Object i}
\State GenerateCollisionConstraints() \label{loop:7}
\EndFor
\State ProjectCollisionConstraints() \label{loop:8}
\EndWhile
\end{algorithmic}
\end{algorithm}

A layout is represented by a set of $n$ oriented particles, each of
which denotes the position and orientation of an associated 3D object
mesh. Each particle $i$ has three attributes:
\begin{enumerate}
\item a position $\vec p_i$,
\item an orientation $\theta_i$, and
\item a mass $m_i$, and corresponding inverse mass $w_i=1/m_i$,
\end{enumerate}
determined by the volume of the bounding box of the associated object.

Our algorithm modifies the position and orientation of each particle
in order to satisfy a set of $m$ layout constraints, which restrict
the positions and orientations of several layout items. A constraint
comprises
\begin{itemize}
\item a scalar constraint function $C$,
\item a stiffness parameter $k\in [0, 1]$, and
\item a constraint type,
\end{itemize}
either an \emph{equality} constraint $C(\vec p)=0$ or an
\emph{inequality} constraint $C(\vec p) \geq 0$, where $\vec p =[\vec
p_1^T, \theta_1,\ldots, \vec p_n^T, \theta_n]^T$.

Starting from a uniformly distributed random initial position for each
object in the layout, our algorithm solves each constraint
independently. The constraint's spring-like stiffness determines the
magnitude of the positional correction toward satisfying the
constraint. The positional corrections are either processed
sequentially, or averaged in a batch.

To measure the quality of a layout, we employ the energy function
\begin{equation}
E = \biggl( \sum_{j=1}^m \gamma_j C_j^2 \biggr)^{1/2},
\label{eq:energy}
\end{equation}
where $C_j$ denotes constraint $j$ with respective weight $\gamma_j$.

Algorithm~\ref{alg:pbd_opt} overviews our method. Line~\ref{loop:2}
initializes object $i$ to a random location $\vec p_i^0$ and
orientation $\theta_i^0$. In each iteration of the main loop, starting
at Line~\ref{loop:5}, each constraint $C_i$ is calculated and
immediately projected, so that the next constraint uses the updated
result (details in Section~\ref{constraint:layout}). Collision
constraints require special treatment, since they may change in each
iteration, and are generated in Line~\ref{loop:7} using a spatial hash
(details in Section~\ref{constraint:col}).

Each constraint $C_i$ has an associated stiffness parameter $k_i$
which is updated in each iteration (Line~\ref{loop:3}). Depending on
the constraint type, the stiffness either decreases, increases or
remains constant. For example, the pairwise distance constraint
decreases over time, the collision constraint increases, and for hard
constraints, like the layout boundary, the stiffness is constant. This
is a type of numerical continuation method \cite{Algower2003}.
Increasing the number of iterations results in more
physically-plausible solutions. In addition, since some constraints
are conflicting, we evaluate them in different orders, by interleaving
them, similar to Stam's proposal~\cite{stam2009nucleus}.


\subsection{Constraint Projection}
\label{constraint:proj}

PBD satisfies a system of constraints by iteratively solving each
constraint independently. According to Bender et
al.~\cite{bender2014survey}, the correction $\Delta\vec p$ is derived
using the first-order approximation $0 = C(\vec p + \Delta \vec p)
\approx C(\vec p) + \nabla_{\vec p} C(\vec p) \cdot \Delta \vec p$ and
restricting the correction to be in the direction of the constraint
gradient: i.e., $\Delta\vec p = \lambda\nabla_{\vec p} C(\vec p)$.
This leads to the following formula for the positional correction to
particle $i$:
\begin{equation}
\Delta \vec p_i = - s \nabla_{\vec p_i} C (\vec p),
\label{eq:p-correction}
\end{equation}
with scale factor
\begin{equation}
s = \frac{k w_i C(\vec p)}{\sum_{i=1}^n w_i \norm{\nabla_{\vec p_i} C (\vec
p)}^2 },
\end{equation}
where the stiffness parameter $k$ determines the influence of the
constraint. The stiffness $k^l$ at each iteration $l$ varies according
to $k^l = 1-(1-k^0)^{M/l}$, where $k^0$ is the initial stiffness and
$M \geq 1$ determines the rate at which $k^l$ approaches zero.

We employ two different schemes for satisfying constraints. Each
constraint is either solved independently and projected, the updated
particle position $\vec p_i$ immediately becoming visible to other
constraints, or a subset of constraints is solved as a batch. In the
batch case, we average the positional corrections of all the
constraints affecting $\vec p_i$, with averaging coefficient
1.2, as suggested by Macklin et al.~\cite{macklin:2014:unified}.

A formula similar to (\ref{eq:p-correction}) can be written for
constraints involving particle orientations, but we treat them
differently. We simply determine the smallest rotational correction
that satisfies the constraint, and apply it to rotate the
corresponding layout object (see
Sections~\ref{constraint_orient_pairwise} and
\ref{constraint_orient_wall}).

\subsection{Parenting and Grouping}
\label{constraint:group}

Layout objects can be grouped; for example, tiers of seats in a
theater, or a table and chairs. In our framework, the group is also
represented by an oriented particle. The dimensions and size of the
group is approximated by a bounding box. Furthermore, we can
hierarchically define layout constraints within the group.

Constraints internal to a group can be rigid or nonrigid. In the rigid
case, we simply apply positional corrections only to the particle
representing the group, such that the grouped objects remain fixed
relative to each other. In the nonrigid case, the particle
representing the group can move, but so can the grouped objects with
respect to one another subject to the layout constraints internal to
the group.

\begin{figure}
\centering
\begin{subfigure}{0.49\columnwidth}
\includegraphics[width=\columnwidth]{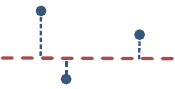}
\caption{ }
\end{subfigure}
\begin{subfigure}{0.49\columnwidth}
\includegraphics[width=\columnwidth]{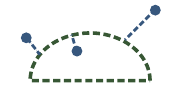}
\caption{ }
\end{subfigure}
\caption{Layout objects can be constrained relative to curves, such as
line segments (a) or circular arcs (b), enabling the imposition of
nonrigid group relationships.}
\label{fig:segments_and_arcs}
\end{figure}

For example, Fig.~\ref{fig:segments_and_arcs} illustrates objects
grouped along line segments and circular arcs, which enables us to
design seating tiers and add pathways in the theater scenes of
Fig.~\ref{fig:theater_variations} while maintaining pairwise distances
between the chairs. Segments are defined by the medial axis of the
group's bounding box. An arc is defined by its endpoints along with
the center of the circle. When the particle representing the group
moves, for each object in the group, a pairwise distance constraint
(Section~\ref{constraint:distance}) is applied between a member object
and the nearest point on the curve, which is represented by a particle
with infinite mass. Since the curve is represented parametrically, the
parametric ordering of the associated layout objects can be used to
apply the pairwise distance and other group constraints, and the
ordered application of constraints can improve convergence. We
interleave the application of constraints, as proposed by Umetani et
al.~\cite{umetani2014position}.

\subsection{Constraint Types}
\label{constraint:layout}

The following sections discuss the constraints that we employ to
produce layouts that are consistent with design standards
\cite{dechiara2001time}.

\subsubsection{Pairwise Distance Constraint}
\label{constraint:distance}

\begin{figure}
\centering
\begin{subfigure}{0.49\columnwidth}
\includegraphics[width=\columnwidth]{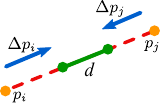}
\caption{Pairwise distance}
\label{fig:c-1}
\end{subfigure}
\begin{subfigure}{0.49\columnwidth}
\centering
\includegraphics[width=0.6\columnwidth]{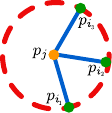}
\caption{Focal point distance}
\label{fig:c-3}
\end{subfigure}
\\[10pt]
\begin{subfigure}{0.49\columnwidth}
\includegraphics[width=\columnwidth]{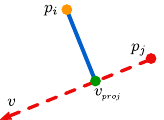}
\caption{Traffic lane}
\label{fig:c-2}
\end{subfigure}
\begin{subfigure}{0.49\columnwidth}
\includegraphics[width=\columnwidth]{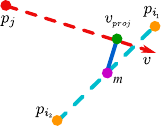}
\caption{Focal point symmetry}
\label{fig:c-4}
\end{subfigure}
\caption{Different positional layout constraints. Note that in (d),
$C$ denotes the center of mass of particles $j_1$ and $j_2$. $\vec
v_\mathrm{proj}$ denotes the projection of $C$ onto the vector
starting at focal point $i$.}
\end{figure}

In interior design, two furniture objects $i$ and $j$ (e.g., a chair
and a table) are often required to be at a certain distance from each
other in order for the layout to be deemed comfortable. We impose a
desired distance $d$ between the particles $i$ and $j$ representing
the objects (Fig.~\ref{fig:c-1}) with the constraint function
\begin{equation}
C(\vec p)=\norm{\vec p_{ij}} - d,
\label{pairwise-distance-c}
\end{equation}
where $\vec p_{ij} =\vec p_i-\vec p_j$, with gradients
\begin{equation}
\nabla_{\vec p_i} C(\vec p) = \hat{\vec p}_{ij}; \quad \nabla_{\vec
p_j} C(\vec p) = - \hat{\vec p}_{ij},
\label{pairwise-distance-c-d}
\end{equation}
where $\hat{\vec p}_{ij} = \vec p_{ij} / \norm{\vec p_{ij}}$. Imposed
as an equality constraint $C(\vec p) = 0$, the particle positional
corrections are \cite{muller2007position}
\begin{equation}
\Delta \vec p_i = - \frac{w_i C(\vec p)}{w_i+w_j} \hat{\vec p}_{ij}; \quad
\Delta \vec p_j = \frac{w_j C(\vec p)}{w_i+w_j} \hat{\vec p}_{ij}.
\label{eq:pairwise-distance-corrections}
\end{equation}

\subsubsection{Focal Point Distance Constraint}

An object, such as the stage in a theater or a TV in a living-room,
can be deemed a \emph{focal point} for a group of objects
\cite{jones2014beginnings,deasy1990designing}, and the objects may be
constrained to be at a distance $d$ from the focal point. We enforce
such constraints simply by adding a pairwise distance constraint
(\ref{pairwise-distance-c}) between the particle $\vec p_j$ that
represents the focal point object and each of the surrounding objects
represented by particles $\vec p_i$ (Fig.~\ref{fig:c-3}). The focal
point object can be prevented from correcting its position by setting
its inverse mass $w_j$ to zero.

\subsubsection{Traffic Lane Constraint}
\label{constraint:lane}

Objects should often be arranged to accommodate traffic lanes that
introduce space between objects or groups of objects to allow easy
access \cite{jones2014beginnings, deasy1990designing}; e.g., walkways
in a theater (Fig.~\ref{fig:theater_variations}). To this end, we
enforce a clearance around a vector extending from a particle. This is
implemented analogously to a pairwise distance constraint
(\ref{pairwise-distance-c}) between a particle $\vec p_i$ and the
closest point $\vec p_{\vec v_\mathrm{proj}}$ on a vector $\vec v$
from another particle $\vec p_j$ (Fig.~\ref{fig:c-2}):
\begin{equation}
C(\vec p) = \norm{\vec p_i - \vec p_{\vec v_\mathrm{proj}}} - d ,
\label{traffic-lane-c}
\end{equation}
where
\begin{equation}
\vec p_{\vec v_\mathrm{proj}} = \vec p_j + \frac{\vec p_{ij} \cdot
\vec v}{\vec v \cdot \vec v} \vec v ,
\end{equation}
is the point along $\vec v$ nearest to $\vec p_i$, and $d$ is the
desired minimal distance of $\vec p_i$ from $\vec v$; i.e.,
(\ref{traffic-lane-c}) is enforced as an inequality constraint,
$C(\vec p)\geq 0$. We treat $\vec p_{\vec v_\mathrm{proj}}$ as a
\emph{ghost} particle that is rigidly attached to $\vec p_j$. Hence,
any positional correction of $\vec p_{\vec v_\mathrm{proj}}$ is
applied to $\vec p_j$. The ghost particle has inverse mass $w_j$,
which may be set to 0 if need be to prevent the constraint from
affecting the position of $\vec p_j$.

\subsubsection{Heat Point Constraint}
\label{constraint:heat}

For an object or group of $n$ objects, the \emph{heat point}
$\tilde{\vec p}$ is the position where the center of mass, $\vec m = w
\sum_{i=1}^n m_i \vec p_i$ with $w = 1/m$ and $m = \sum_{i=1}^n m_i$,
of the particles $\vec p_i$ representing the objects is required to
be. For example, a computer display, keyboard, and mouse should be
located near the middle of the front of a tabletop for easy access. To
this end, we define the constraint function
\begin{equation}
C(\vec p)= \frac{1}{2} \norm{\vec m - \tilde{\vec p}}^2 ,
\label{heat-point-c}
\end{equation}
whose gradient for particle $\vec p_j$ is
\begin{equation}
\nabla_{\vec p_j} C(\vec p) = m_j w (\vec m - \tilde{\vec p}).
\label{heat-point-c-d}
\end{equation}

\subsubsection{Focal Point Symmetry Constraint}

We can constrain a group of objects to be positioned symmetrically
around a vector $\vec v$ directed away from the group's focal point
$\vec p_j$---e.g., two chairs positioned symmetrically in front of the
television---by projecting the center of mass $\vec m$ of their
representative particles onto $\vec v$ (Fig.~\ref{fig:c-4}). We
constrain $\vec m$ to coincide with the projection $\vec p_{\vec
v_\mathrm{proj}}$ using a constraint function analogous to
(\ref{heat-point-c}):
\begin{equation}
C(\vec p)= \frac{1}{2}\norm{\vec m - \vec p_{\vec v_\mathrm{proj}}}^2.
\label{symmetry-c}
\end{equation}

\subsubsection{Visual Balance Constraint}

\begin{figure}
  \includegraphics[width=\columnwidth]{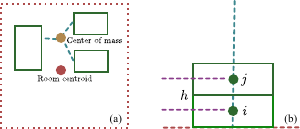}
\caption{(a) Visual balance. (b) Stacking constraint: $h$ denotes the
vertical distance between the centers of objects $i$ and $j$.
\label{fig:vis_and_stack}
}
\end{figure}

Placing all the furnishings at one end of a room would create an
imbalanced, inharmonious ambiance, which prompted Merrell et
al.~\cite{merrell2011interactive} to propose a visual balance
constraint. To create a sense of equilibrium, we want to arrange the
furnishings such that the mean of the visual weights is close to the
center of the room \cite{jones2014beginnings,lok2004evaluation}. Since
larger objects have more visual weight, we define the visual weight of
an object in accordance to its dimensions projected onto the ground
plane. We implement a visual balance constraint analogously to
(\ref{heat-point-c}) between the room's centroid $\vec c$ and the
center of mass $\vec m$ of particles representing the visual weights
of furnishings (Fig.~\ref{fig:vis_and_stack}a):
\begin{equation}
C(\vec p)= \frac{1}{2} \norm{\vec m - \vec c}^2 ,
\label{balance-c}
\end{equation}
where denotes the center of the room. Since the room's centroid is
static, particle $\vec c$ is assigned zero inverse mass.

\subsubsection{Layout Boundaries and Distance to Wall Constraint}
\label{wall_constraint}

Large furnishings usually work best when placed near a wall
\cite{jones2014beginnings}. For example, we usually avoid placing
bookshelves in the center of a room. Also, in realistic use cases,
furnishings should not collide with walls. Finally, all layout objects
are constrained by layout boundaries.

\begin{figure}
\centering
\begin{subfigure}{0.44\columnwidth}
\includegraphics[width=\columnwidth]{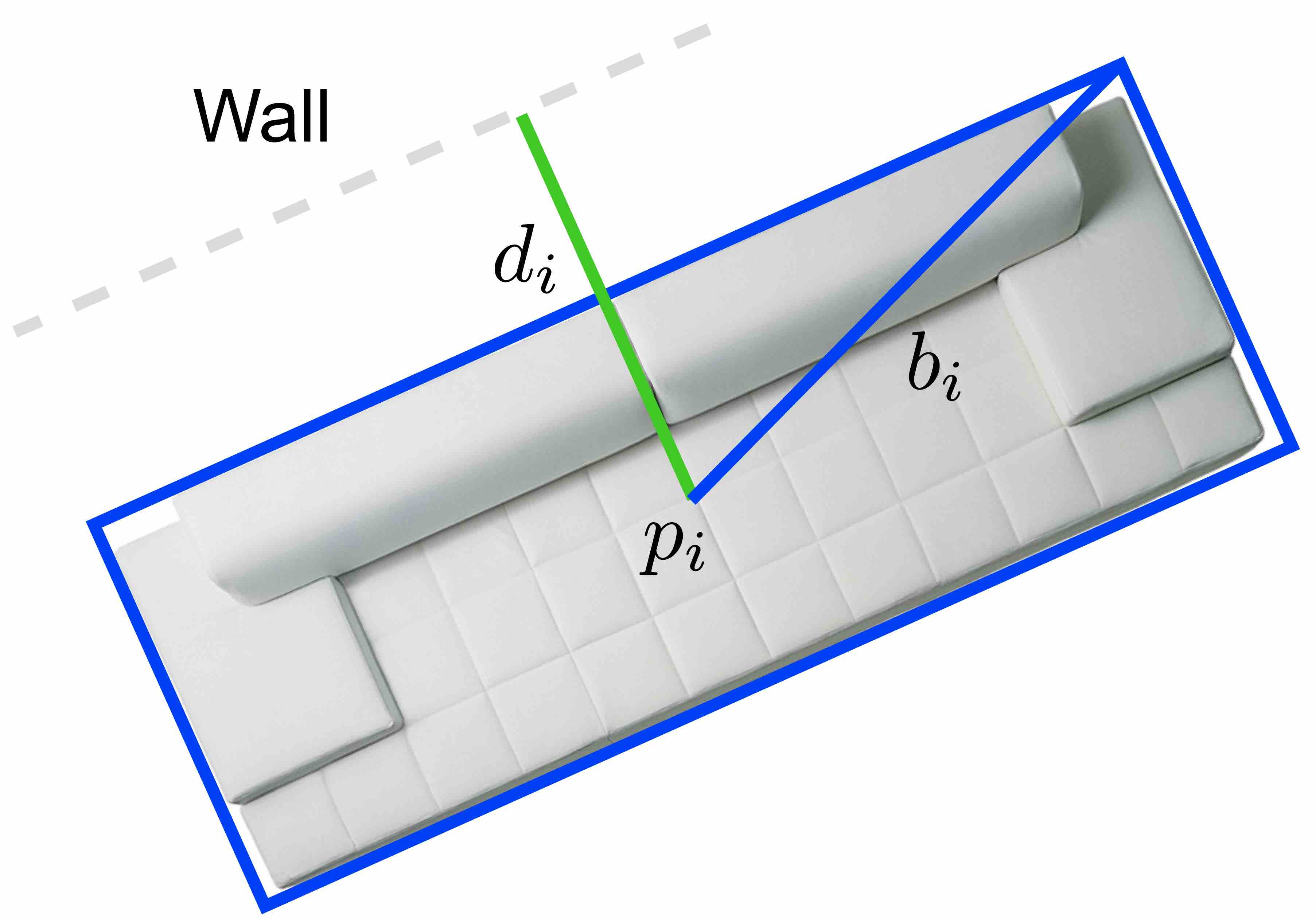}
\caption{ }
\label{fig:distance-to-wall}
\end{subfigure}
\begin{subfigure}{0.52\columnwidth}
\includegraphics[width=\columnwidth]{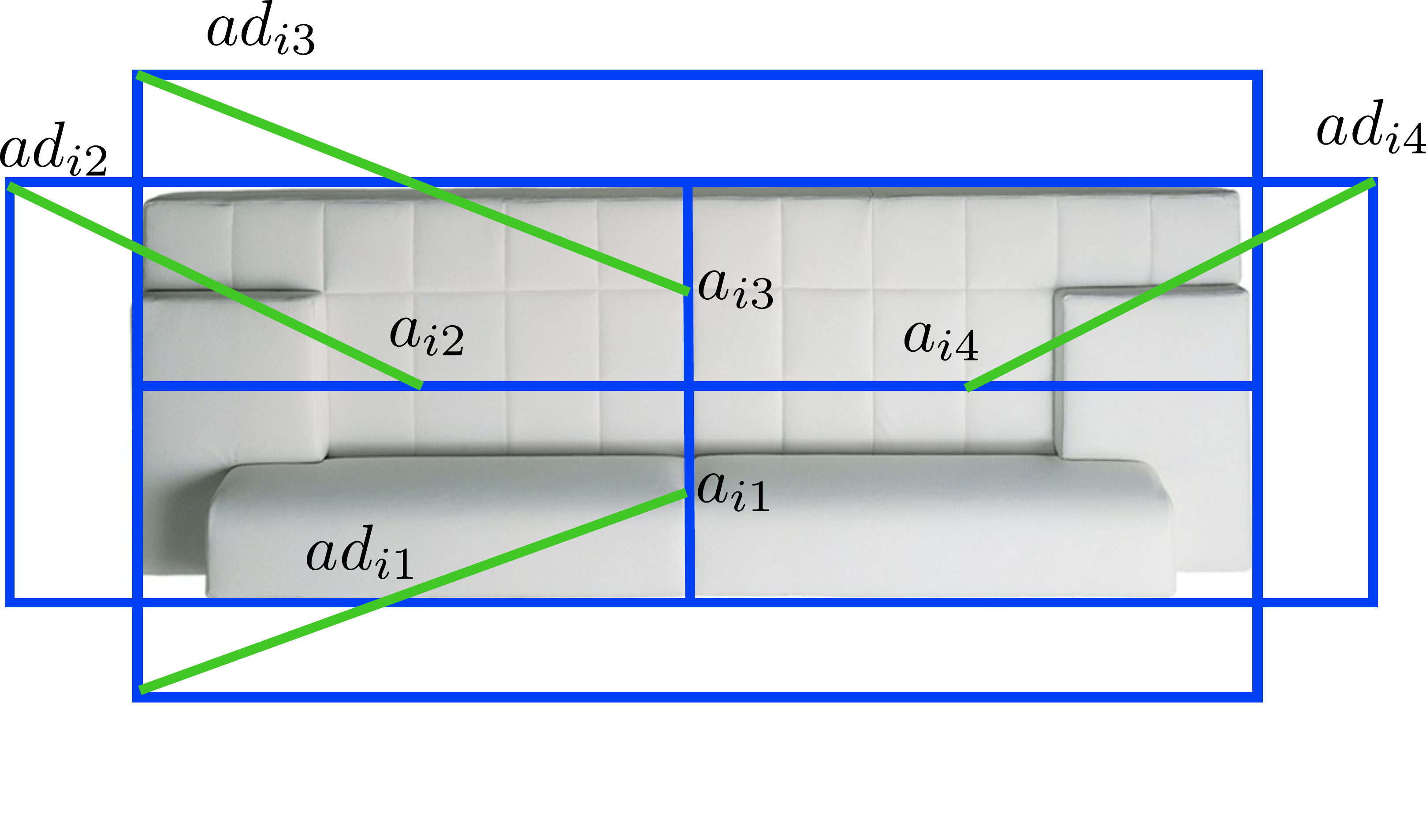}
\caption{ }
\label{fig:accessible-space}
\end{subfigure}
\caption{(a) Wall distance and orientation constraint ($\vec p_i$
denotes the center of object $i$ and $b_i$ is the size of its bounding
box). (b) $ad_{ik}$ denotes the accessibility distance of the
respective accessibility center $\vec a_{ik}$.}
\end{figure}

We define both inequality and equality constraints for object $i$
constrained to be near a wall with distance $d$
(Fig.~\ref{fig:distance-to-wall}) as follows:
\begin{equation}
C(\vec p)= \norm{\vec p_i -\vec p_\mathrm{wall}} - d ,
\label{wall-c}
\end{equation}
where $\vec p_\mathrm{wall}$ is the point on the wall nearest to the
position $\vec p_i$ of the particle representing object $i$ (in case
of multiple nearest points, we simply choose the first point found).
The positional corrections are analogous to
(\ref{eq:pairwise-distance-corrections}), but note that since $\vec
p_\mathrm{wall}$ is fixed, corrections are applied only to $\vec
p_i$.

\subsubsection{Accessibility Constraint}
\label{constraint:access}

Clearance between furniture items is essential for human comfort
\cite{jones2014beginnings}. For instance, a coffee table should be
close, but not too close, to a sofa in a living-room. We employ a
modified version of an accessibility constraint proposed by Yu et
al.~\cite{yu2011make}. Every object $j$ is associated with a bounding
box, where the faces of the bounding box that are orthogonal to the
ground plane are identified with accessibility centers $\vec a_{jk}$,
where $k \in \{1,2,3,4\}$ correspond to the back, left, front, and
right faces, respectively.

Our accessibility constraint is defined using a pairwise distance
constraint function between particle $\vec p_i$ representing object
$i$ and $\vec a_{jk}$, accessibility center $k$ of object $j$
(Fig.~\ref{fig:accessible-space}):
\begin{equation}
C(\vec p)= \norm{\vec p_i - \vec a_{jk}} - d.
\label{accessibility-c}
\end{equation}
Distance $d = b_i + ad_{jk}$, where $b_i$ is the diagonal of the
bounding box of the object, and $ad_{jk}$ is the diagonal of the
accessibility center. The accessibility constraint is an inequality
constraint $C(\vec p) \geq 0$, which is activated only if the
respective cuboids defined by the accessibility distances intersect.
The positional correction to $\vec p_j$ is applied as if accessibility
point $\vec a_{jk}$ is rigidly attached.

\subsubsection{Collision Constraint}
\label{constraint:col}

\begin{figure}
\begin{subfigure}{0.49\columnwidth}
\includegraphics[width=\columnwidth]{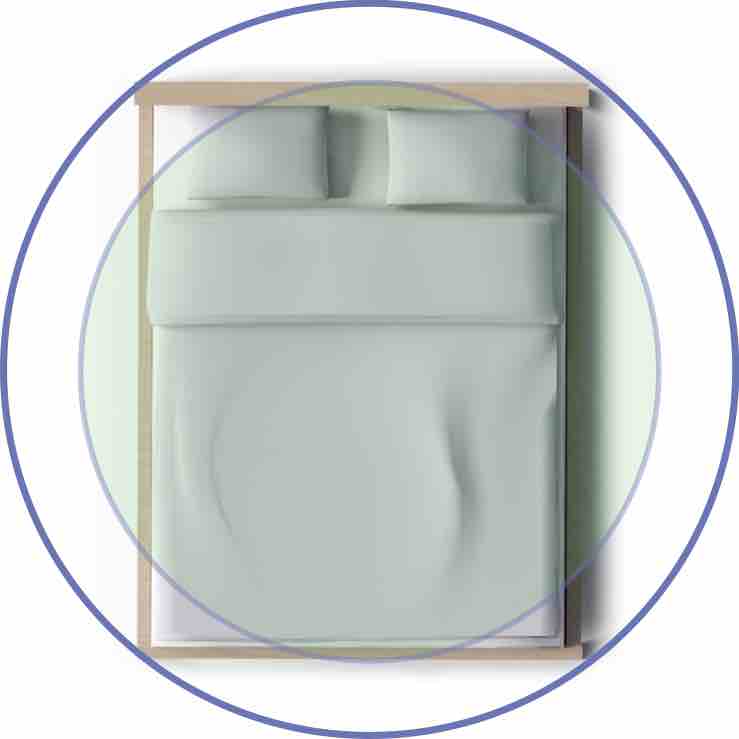}
\caption{ }
\label{fig:bounding_sphere}
\end{subfigure}
\begin{subfigure}{0.49\columnwidth}
\includegraphics[width=\columnwidth]{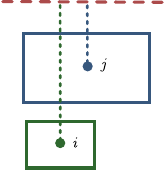}
\caption{ }
\label{fig:collision_and_wall}
\end{subfigure}
\caption{(a) Our method resolves collisions between layout item's by
resolving collisions between the respective bounding spheres. (b)
Additional collision constraint between two layout items $i$ and $j$
that are also constrained to be next to the wall.}
\end{figure}

The collision constraint is complementary to the accessibility
constraint. Simply put, to ensure a realistic layout, the bounding
cuboids of the layout objects should not collide. To that end,
collisions between objects are resolved by a pairwise distance
inequality constraint between the bounding spheres
(Fig.~\ref{fig:bounding_sphere}). Let $\vec p_i$ and $\vec p_j$
represent objects $i$ and $j$ whose bounding sphere radii are $r_i$
and $r_j$. Our inequality collision constraint function is
\begin{equation}
C(\vec p)= \norm{\vec p_i - \vec p_j} - (r_i+r_j).
\label{collision-c}
\end{equation}
Since checking for all pairwise object collisions is computationally
expensive, we employ a spatial hash in order to reduce the number of
collision checks.

If during the constraint projection objects $i$ and $j$ are colliding,
and both are constrained to be next to a wall
(Section~\ref{wall_constraint}), we perform an additional collision
constraint projection between their respective closest wall points
(Fig.~\ref{fig:collision_and_wall}). These wall points are considered
rigidly attached ghost particles. This allows us to satisfy both the
pairwise distance constraint and the distance to wall constraint. If
there are multiple candidate wall points, we simply choose the first
one found.

\begin{figure*}
\parbox[b]{2mm}{\rotatebox[origin=b]{90}{\emph{Theater (overhead)}}}
\begin{subfigure}{0.32\textwidth}
\includegraphics[width=\textwidth]{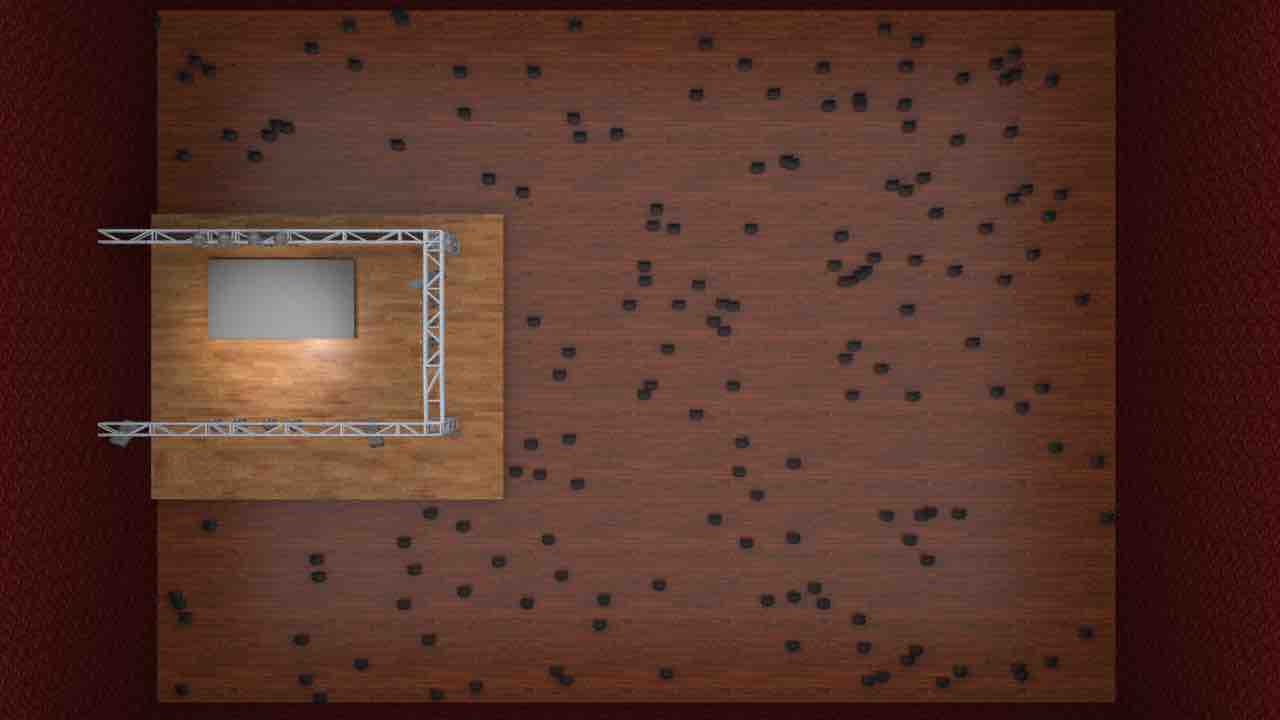}
\label{fig:stage1}
\end{subfigure}\hfill
\begin{subfigure}{0.32\textwidth}
\includegraphics[width=\textwidth]{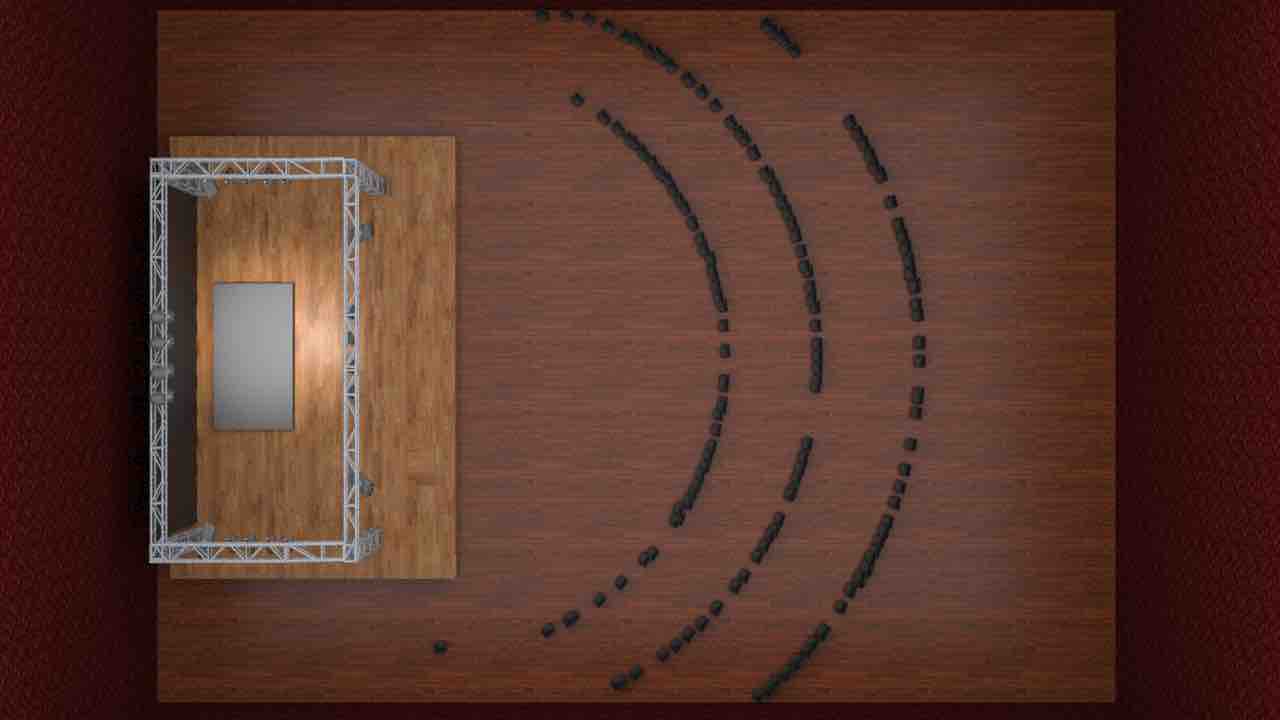}
\label{fig:stage2}
\end{subfigure}\hfill
\begin{subfigure}{0.32\textwidth}
\includegraphics[width=\textwidth]{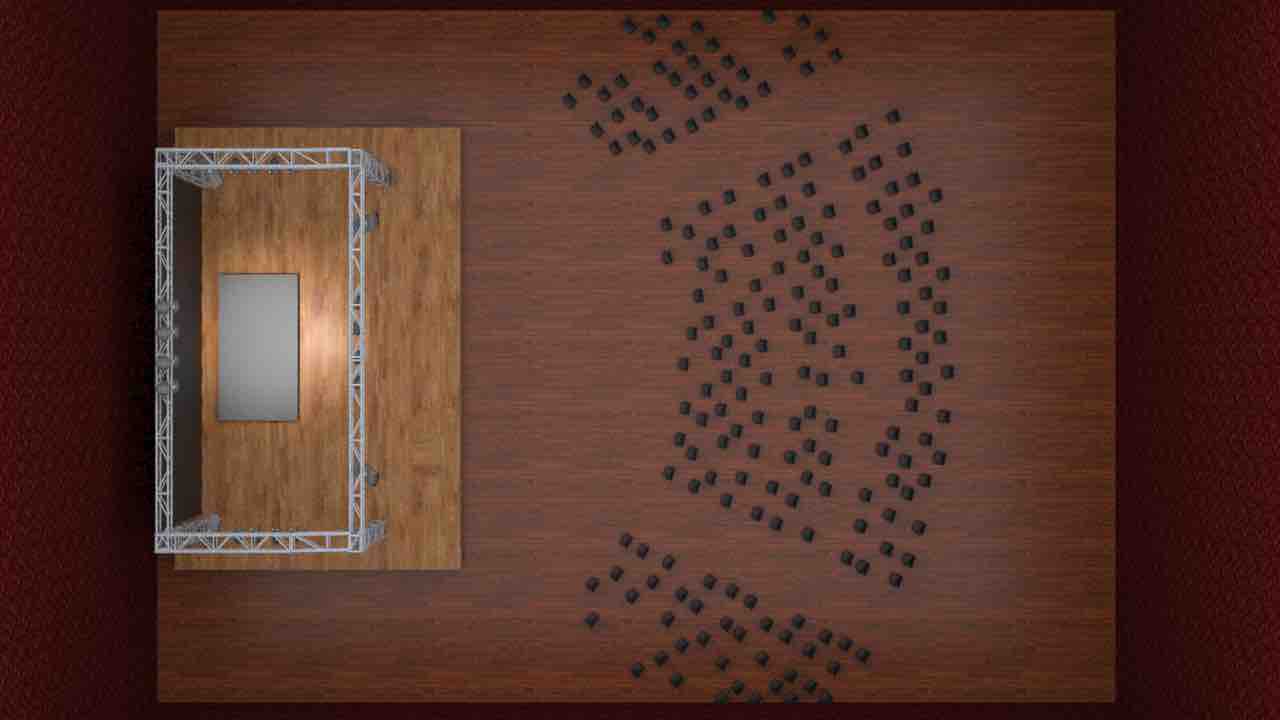}
\label{fig:stage3}
\end{subfigure}\\[-1pt]
\parbox[b]{2mm}{\rotatebox[origin=b]{90}{\emph{Theater (frontal)}}}
\begin{subfigure}{0.32\textwidth}
\includegraphics[width=\textwidth]{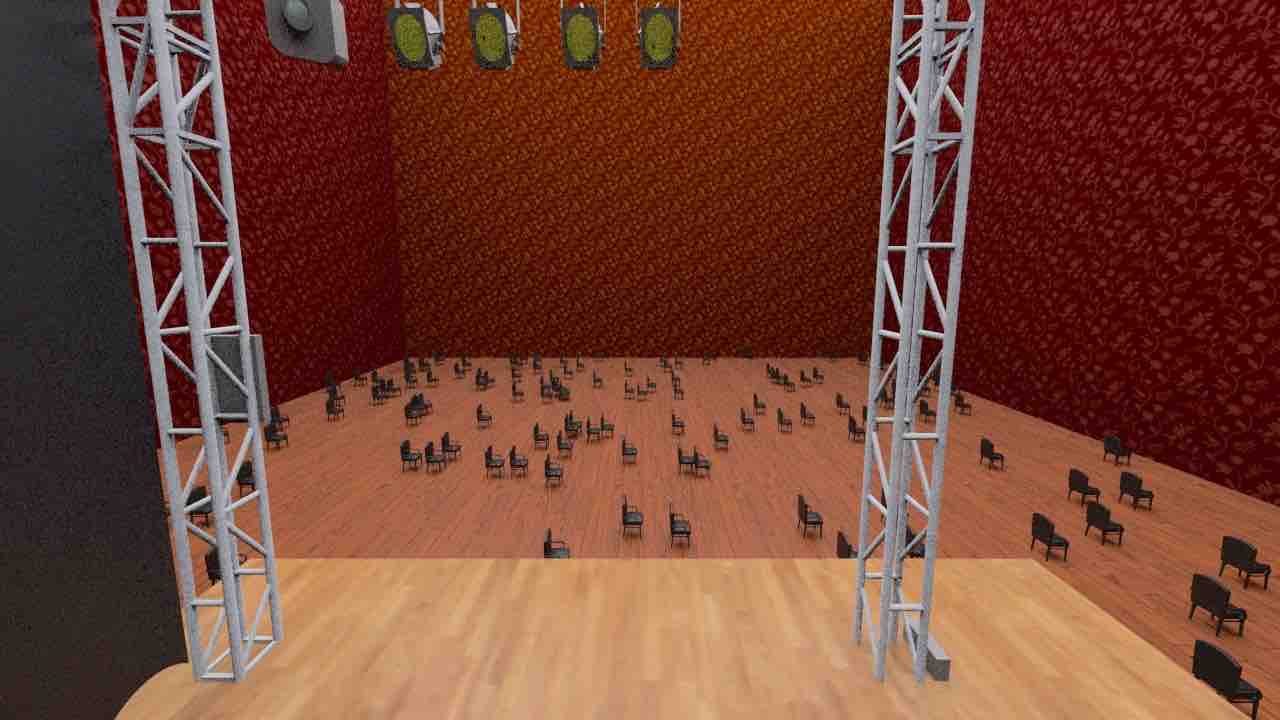}
\label{fig:stage1_cam2}
\end{subfigure}\hfill
\begin{subfigure}{0.32\textwidth}
\includegraphics[width=\textwidth]{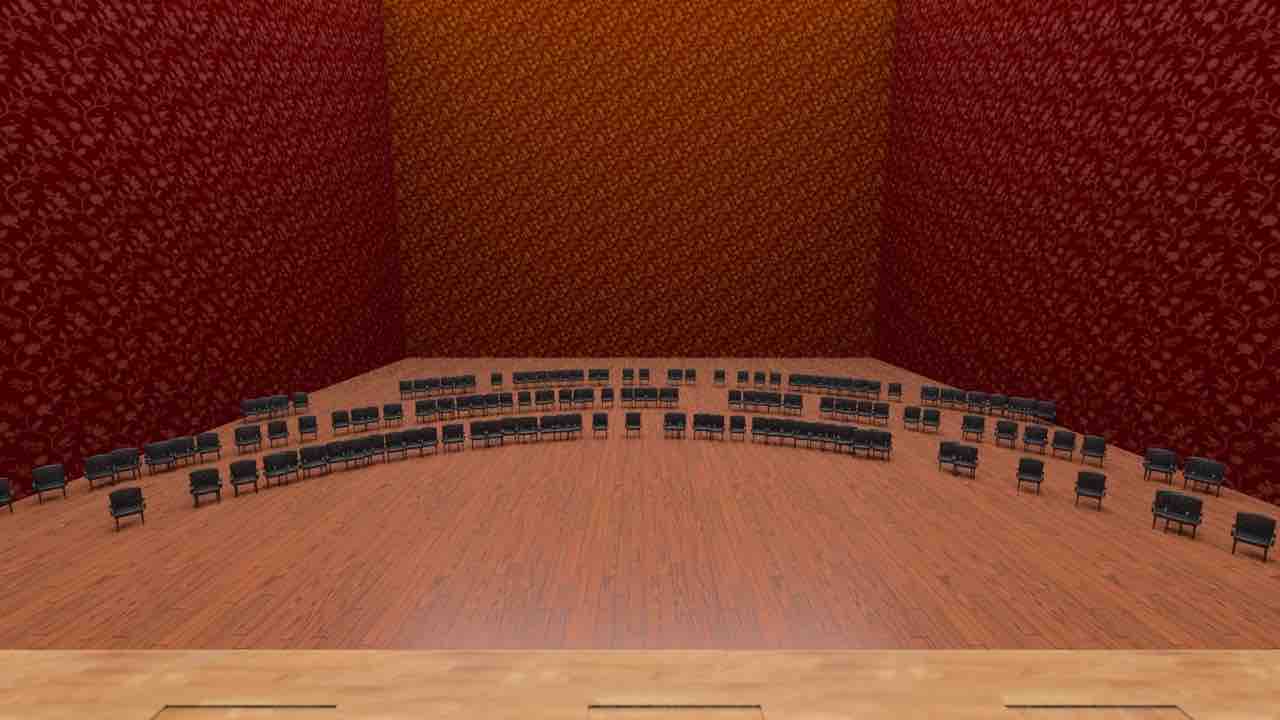}
\label{fig:stage2_cam2}
\end{subfigure}\hfill
\begin{subfigure}{0.32\textwidth}
\includegraphics[width=\textwidth]{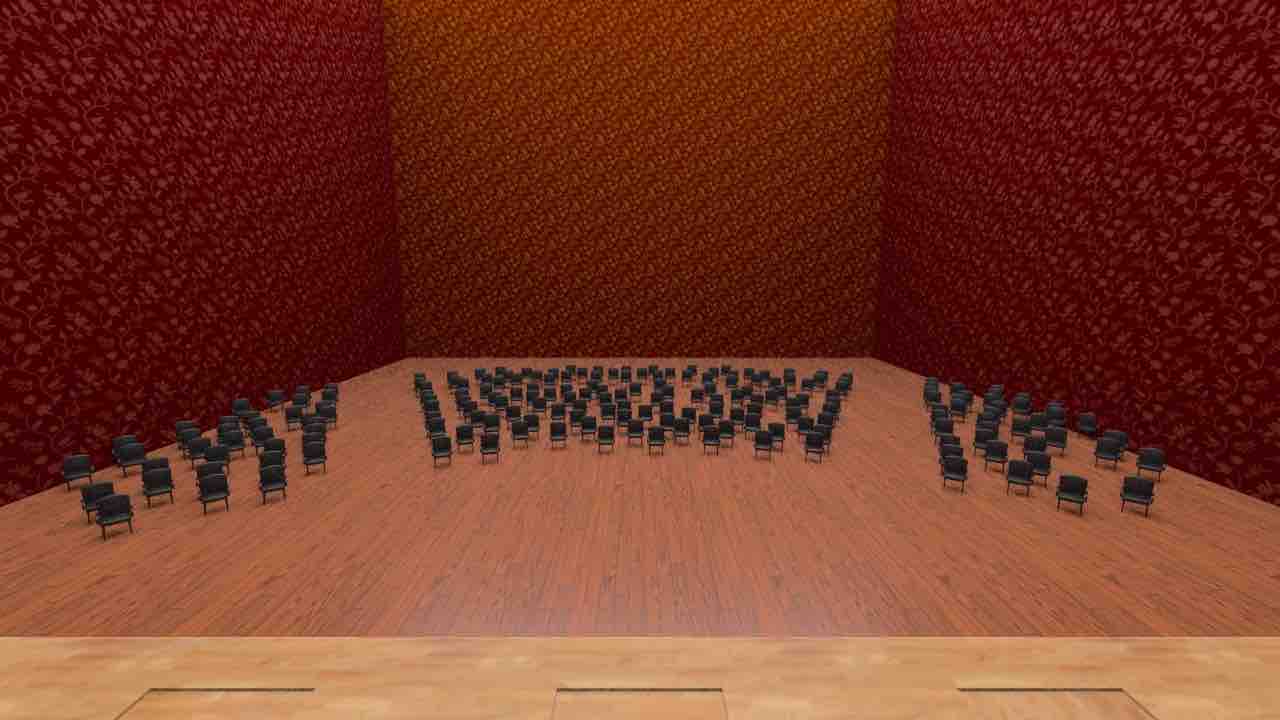}
\label{fig:stage3_cam2}
\end{subfigure}\\[-1pt]
\parbox[b]{2mm}{\rotatebox[origin=b]{90}{\emph{Picnic}}}
\begin{subfigure}{0.32\textwidth}
\includegraphics[width=\textwidth]{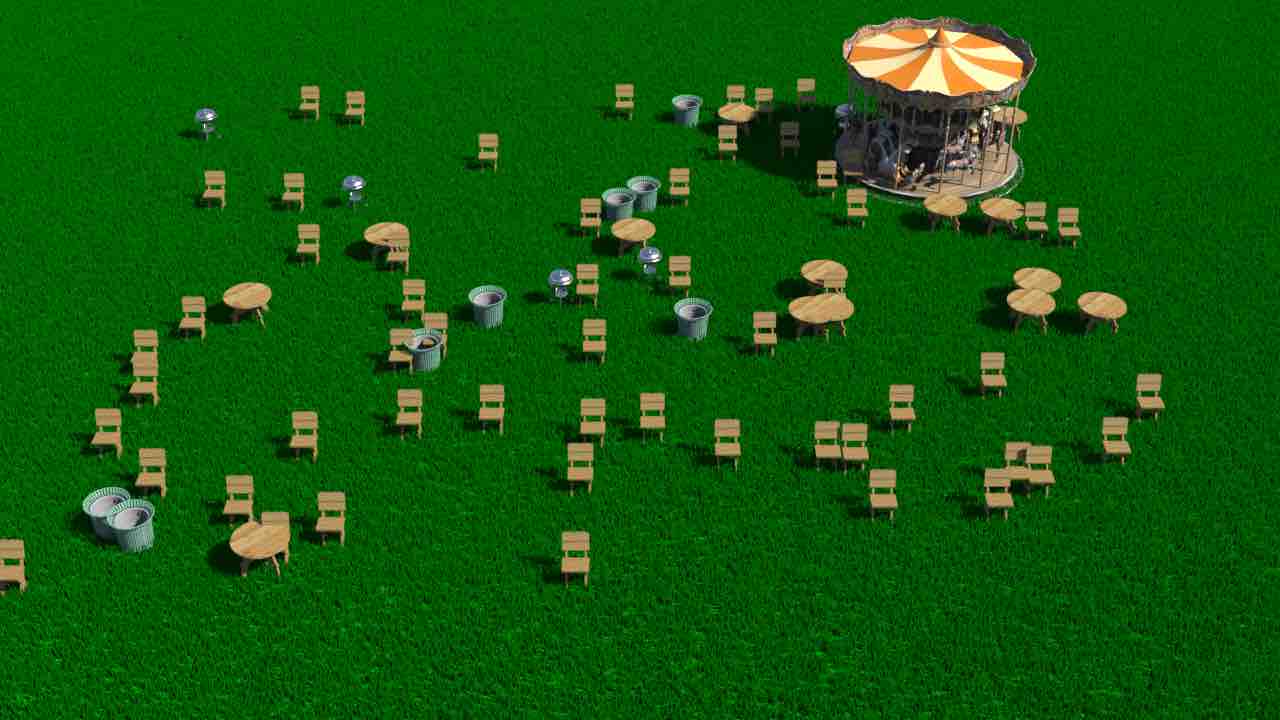}
\label{fig:picinc1}
\end{subfigure}\hfill
\begin{subfigure}{0.32\textwidth}
\includegraphics[width=\textwidth]{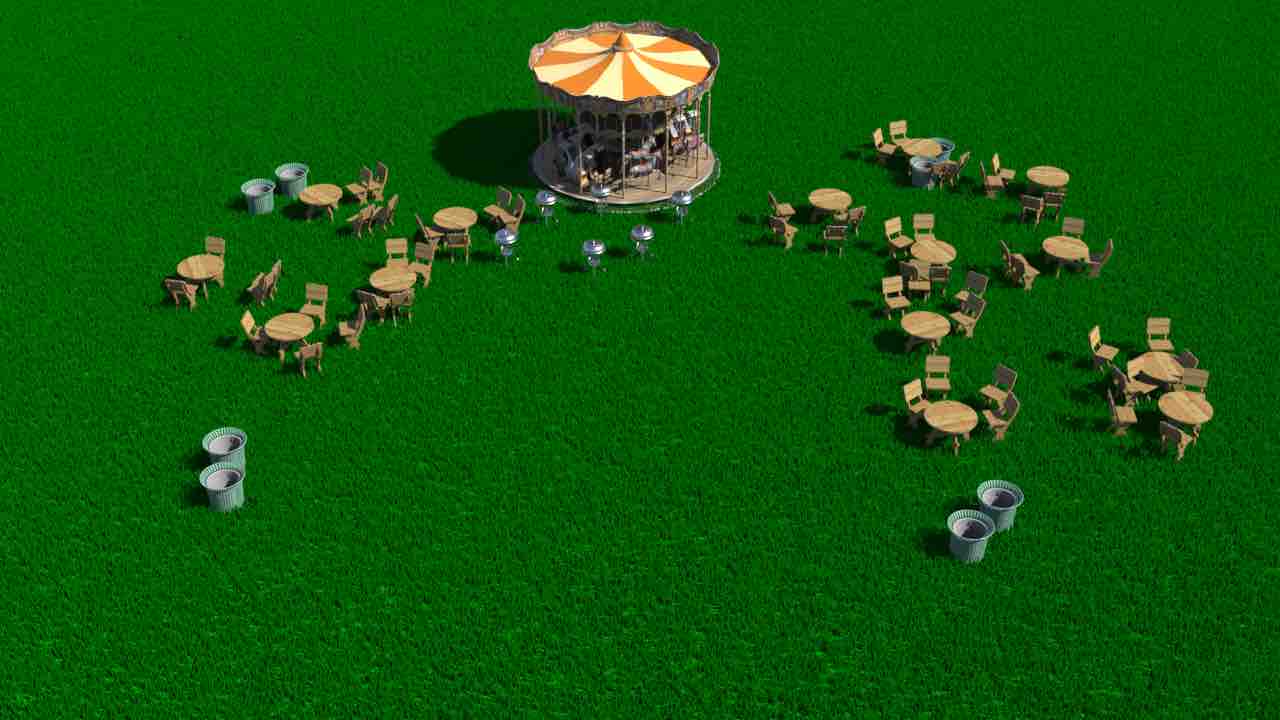}
\label{fig:picinc2}
\end{subfigure}\hfill
\begin{subfigure}{0.32\textwidth}
\includegraphics[width=\textwidth]{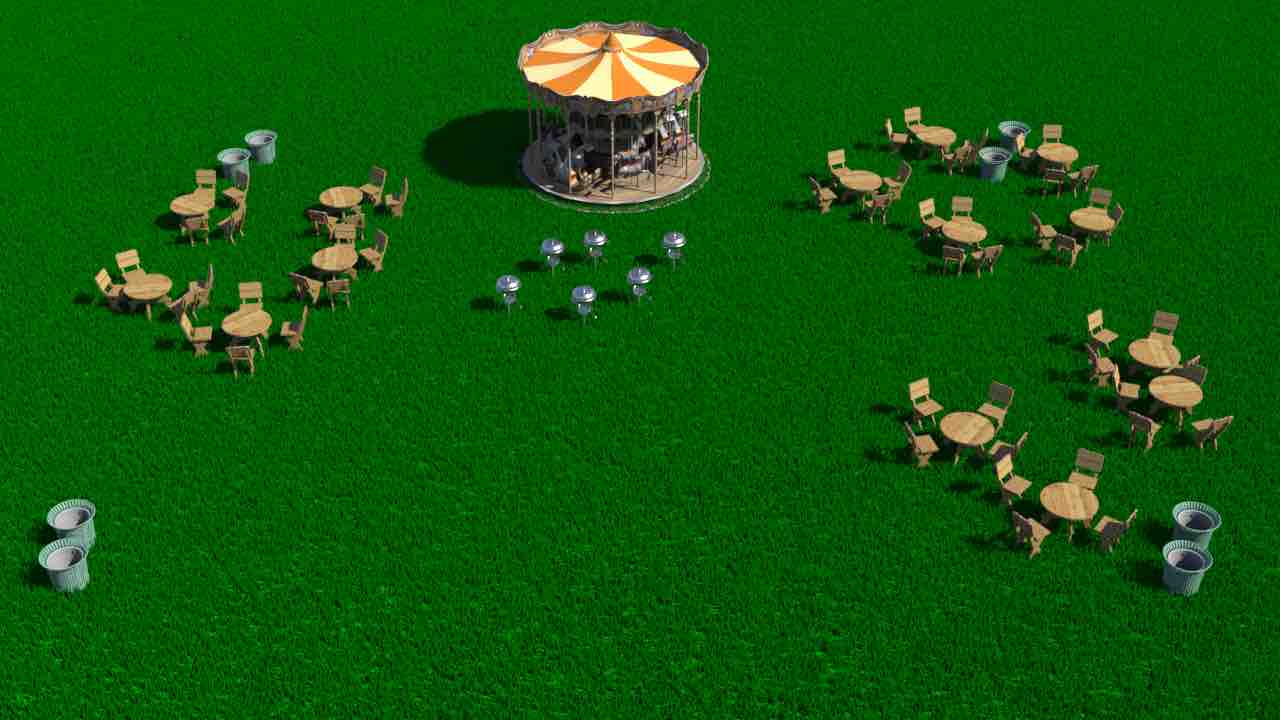}
\label{fig:picnic3}
\end{subfigure}\\[-1pt]
\parbox[b]{2mm}{\rotatebox[origin=b]{90}{\emph{Living-Room}}}
\begin{subfigure}{0.32\textwidth}
\includegraphics[width=\textwidth]{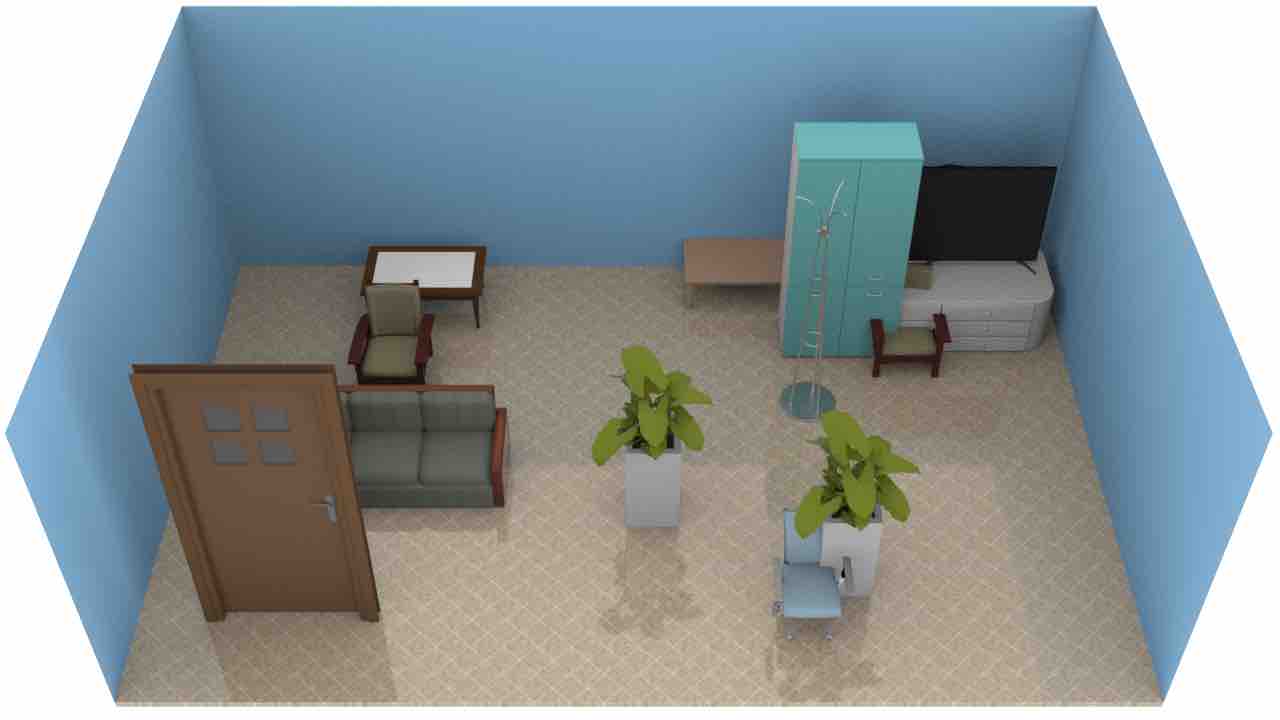}
\label{fig:living1}
\end{subfigure}\hfill
\begin{subfigure}{0.32\textwidth}
\includegraphics[width=\textwidth]{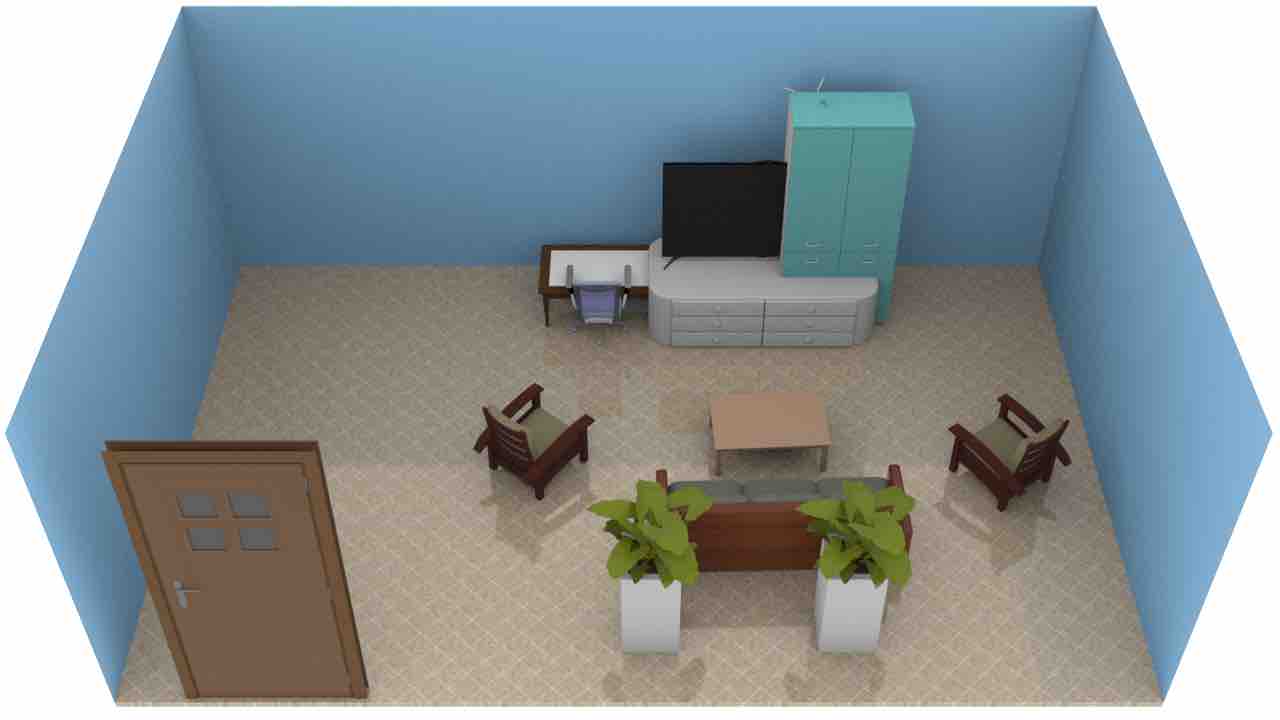}
\label{fig:living2}
\end{subfigure}\hfill
\begin{subfigure}{0.32\textwidth}
\includegraphics[width=\textwidth]{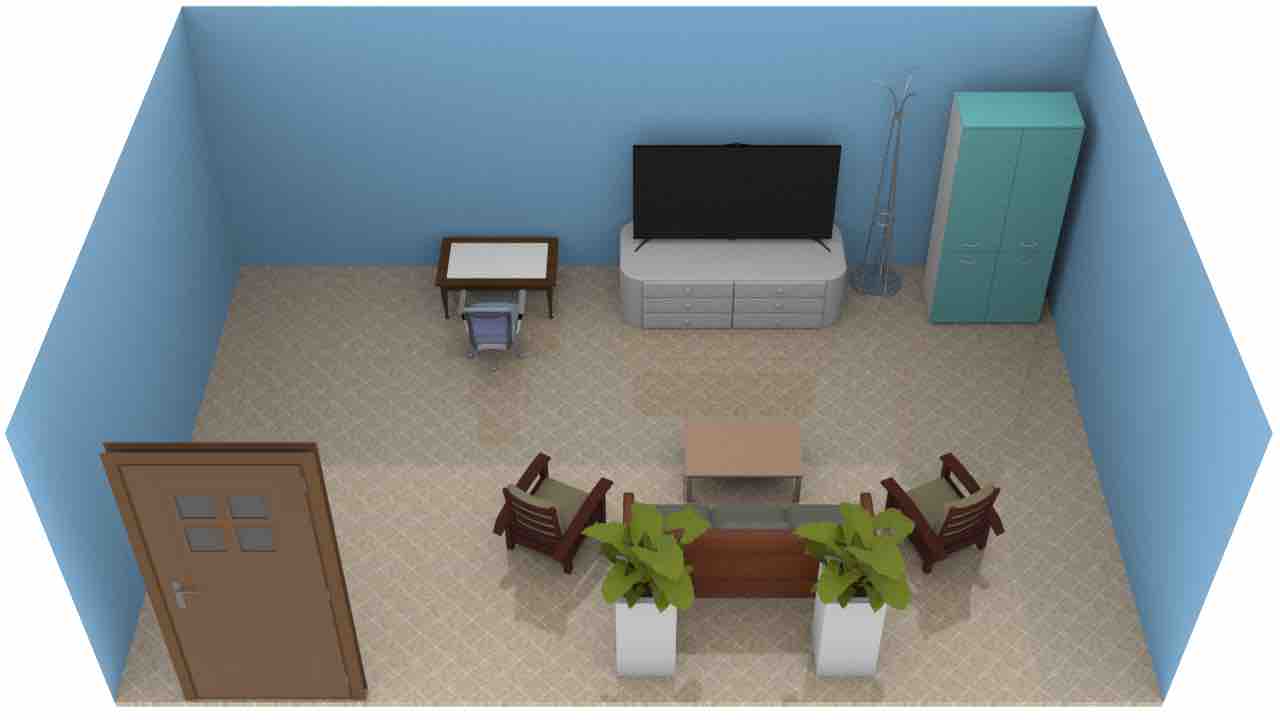}
\label{fig:living3}
\end{subfigure}\\[-1pt]
\parbox[b]{2mm}{\rotatebox[origin=b]{90}{\emph{Desk}}}
\begin{subfigure}{0.32\textwidth}
\includegraphics[width=\textwidth]{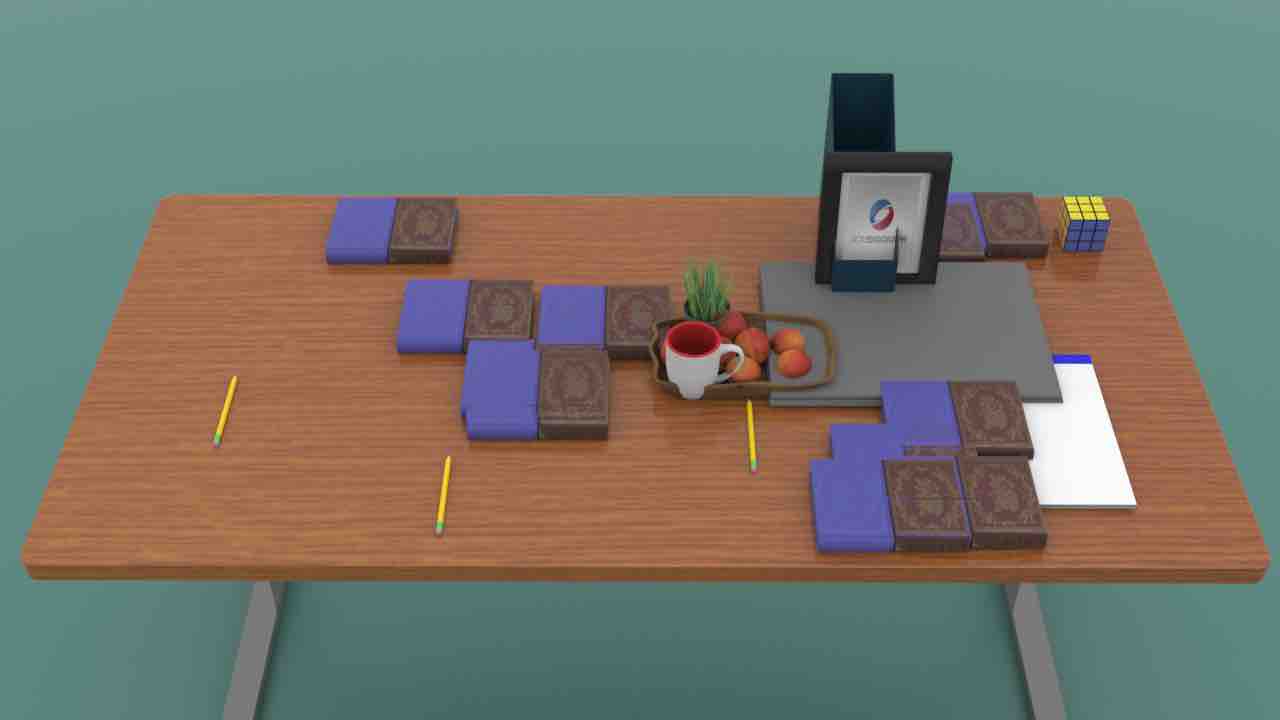}
\caption{Initial Layout}
\label{fig:desk1}
\end{subfigure}\hfill
\begin{subfigure}{0.32\textwidth}
\includegraphics[width=\textwidth]{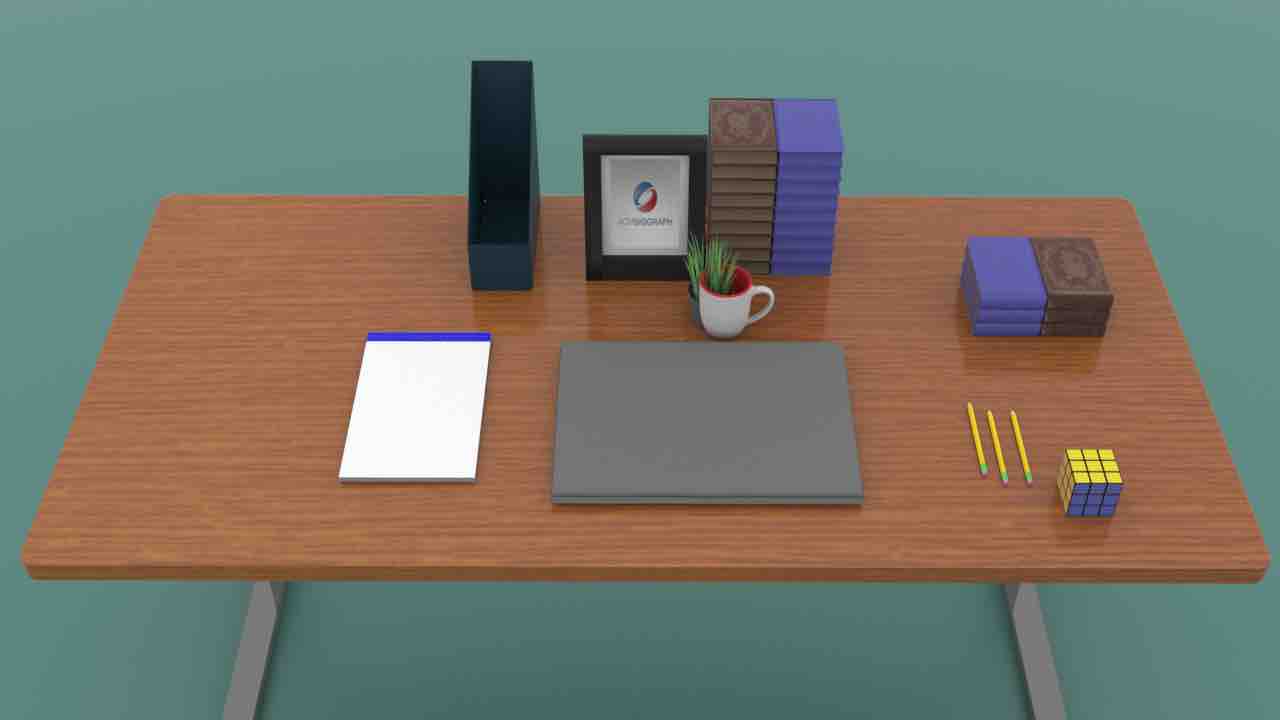}
\caption{Intermediate State}
\label{fig:desk2}
\end{subfigure}\hfill
\begin{subfigure}{0.32\textwidth}
\includegraphics[width=\textwidth]{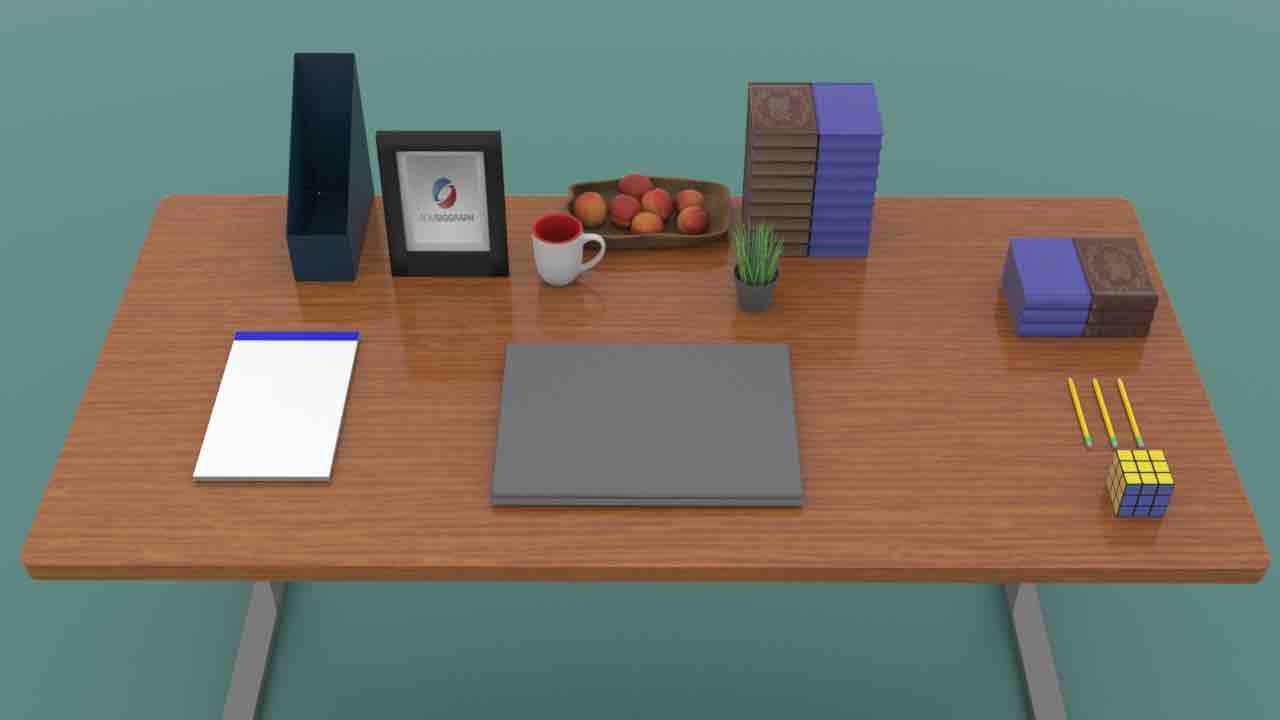}
\caption{Final Layout}
\label{fig:desk3}
\end{subfigure}
\caption{Various synthesized layouts.
\label{fig:comparing_shots}}
\end{figure*}

\subsubsection{Pairwise Orientation Constraint}
\label{constraint_orient_pairwise}

In layout design, the proper orientation of an object relative to
another object in the same furnishing group can promote intimacy or
improve functionality \cite{talbott1999decorating}. For example, a
sofa should face a TV, a coffee table should be parallel to a sofa,
and a seat in a theater should face the stage.

Yu et al.~\cite{yu2011make} propose a pairwise orientation constraint
between interacting layout objects. Similarly, we define an equality
orientation constraint between interacting particles $i$ and $j$. Let
$\theta_i$ and $\theta_i'$ be the current and desired orientation of
particle $i$ relative to $\vec p_j$, and let $\theta_j$ and
$\theta_j'$ be the current and desired orientation of particle $j$
relative to $\vec p_i$. Then the pairwise orientation constraint
functions are defined as
\begin{equation}
C_i(\vec p) = | \theta_i - \theta_i' | ; \quad
C_j(\vec p) = | \theta_j - \theta_j' | ,
\end{equation}
where we calculate the smallest angular difference $\Delta \theta_i$
between $\theta_i$ and $\theta_i'$, and similarly for particle $j$.
This rotational correction is then applied to the particles with
corresponding stiffness $k$. The corrected particle orientations are
$\theta_i + k \Delta\theta_i$ and $\theta_j + k \Delta\theta_j$. The
particle positions $\vec p_i$ and $\vec p_j$ remain unchanged.

\subsubsection{Orientation to Wall Constraint}
\label{constraint_orient_wall}

Some furnishings work best when placed parallel to a wall (e.g., a
table or TV shelf \cite{jones2014beginnings}). Following Yu et
al.~\cite{yu2011make}, we formulate an equality constraint between
particle $i$ and the nearest wall using the constraint function
\begin{equation}
C(\vec p)= |\theta_i - \theta_\mathrm{wall}| ,
\end{equation}
where $\theta_i$ is the orientation of the represented object with
respect to the closest wall point and $\theta_\mathrm{wall}$ is its
desired orientation. To satisfy this constraint, we proceed
analogously to the previous section.

\subsubsection{Vertical Stacking Constraint}

In layout design, accessories serve either a functional or decorative
purpose \cite{jones2014beginnings}. Vertical stacking is a common way
to arrange accessories. For example, books may be stacked in order to
conserve space.

Objects to be stacked can be prespecified. If object $j$ is to be
stacked on object $i$, the vertical distance between the particles
representing these objects should be equal to half the sum of the
objects' heights $h=(h_i + h_j)/2$. We formulate the constraint
function
\begin{equation}
C(\vec p)= z_j - (z_i + h) ,
\end{equation}
involving the $z$ components of $\vec p_i$ and $\vec p_j$, with $z$
axis normal to the ground plane. Hence, particle $j$ should be placed
above particle $i$ (Fig.~\ref{fig:vis_and_stack}b). Additionally, we
constrain the ground plane coordinates of particle $j$ to match those
of particle $i$:
\begin{equation}
C(\vec p)= \norm{x_j - x_i}; \quad
C(\vec p)= \norm{y_j - y_i} .
\end{equation}

\section{Experiments and Results}
\label{sec:expr}

We implemented our layout synthesis system in Python and Cython, and
ran our experiments on a 2.5 GHz Intel i7 Macintosh system.

\begin{figure}
\scriptsize
\textbf{Theater-1 Constraints:}
\begin{itemize}[label={}]
\item Wall distance --- stage
\item Distance --- between each tier of seats and the stage
\item Traffic Lanes --- between the seats and two vectors from the stage
\item Orientation --- between the stage and each seat, which should face the stage
\item Focal point --- for each seat in a seating tier, oriented toward the stage
\end{itemize}
\textbf{Theater-2 Constraints:}
\begin{itemize}[label={}]
\item Heat point --- stage
\item Distance --- between each tier of seats and the stage
\item Traffic Lanes --- between all the seats and two vectors from the stage
\item Orientation --- between the stage and each seat, which should face the stage
\item Grouping --- between all the tiers
\item Grouping --- between chairs in each seating tier
\end{itemize}
\textbf{Picnic Constraints:}
\begin{itemize}[label={}]
\item Focal point --- each table is a focal point for a group
 of 4 chairs
\item Distance --- between each chair in an associated group
\item Distance --- the BBQ grills are linked together
\item Distance --- between each pair of trash cans
\item Heat point --- between each group of trash cans and a picnic layout location
\item Heat point --- on each table to a different layout area
\item Heat point --- on the Carousel to the top-middle corner of the layout
\item Orientation --- between chairs and respective table
\end{itemize}
\textbf{Living-Room Constraints:}
\begin{itemize}[label={}]
\item Focal point --- couch as focal point to table
\item Focal point --- TV as focal point to couch, sofa chairs
\item Focal point --- Table as focal point to office chair
\item Wall distance and orientation --- TV, book case, coat rack,
door, plants
\item Visual balance
\item Orientation --- between objects and their respective focal points
\end{itemize}
\textbf{Desk Constraints:}
\begin{itemize}[label={}]
\item Stacking --- between books, divided into two groups
\item Heat point --- on laptop, to be located near the
front middle of the desk.
\item Heat point --- on notepad to front right of desk
\item Heat point --- on Rubik's cube to front left of desk
\item Distance --- between potted plant and book stack
\item Distance --- between book stack and desk binder
\item Distance --- between binder and photo frame
\item Distance --- between photo frame and mug
\item Distance --- between pencils
\item Focal point --- laptop as focal point to fruit plate
\item Focal point --- Rubik's cube as focal point
to pencil group
\item Focal point  --- Rubik's cube as a focal
point to stack of books
\item Wall distance --- on one stack of books
\end{itemize}
\textbf{Tightly-Packed Bedroom Constraints:}
\begin{itemize}[label={}]
\item Focal point --- each table is a focal point for a group
 of 4 chairs
\item Distance --- between floor lamp, table and chair
\item Distance --- between chair and table
\item Distance --- between bookcase and coat rack
\item Orientation --- between chair and table
\item Wall distance --- for beds, bookcase and table
\end{itemize}
\textbf{Tightly-Packed Picnic Constraints:}
\begin{itemize}[label={}]
\item Distance --- the BBQ grills are linked together
\item Distance --- between each pair of trash cans
\item Heat point --- between each group of trash cans and a picnic layout location
\item Heat point --- on the Carousel to the top-middle corner of the layout
\end{itemize}
\caption{The constraints used in our experiments.}
\label{fig:constraints}
\end{figure}

Fig.~\ref{fig:comparing_shots} shows examples of our experimental
scenarios. In each experiment, the initial object locations and
orientations were set randomly, as shown in
Fig.~\ref{fig:comparing_shots}a. Accessibility and collision
constraints apply and are generated in all experiments. We ran our
layout synthesis algorithm with the constraints described below (refer
to Fig.~\ref{fig:constraints}). Since collision, accessibility, and
wall constraints are treated as hard constraints, for weights in the
energy function (\ref{eq:energy}) we chose 150.0 for the collision and
accessibility constraints, 20.0 for the wall constraints, and 1.0 for
the remaining constraints. We determined these to be suitable weights
experimentally. The iterative procedure terminates when there has been
no improvement to the minimum layout energy for the previous 50
iterations.

\begin{figure}
  \includegraphics[trim={0 0 0 1.8cm},clip,width=\columnwidth]{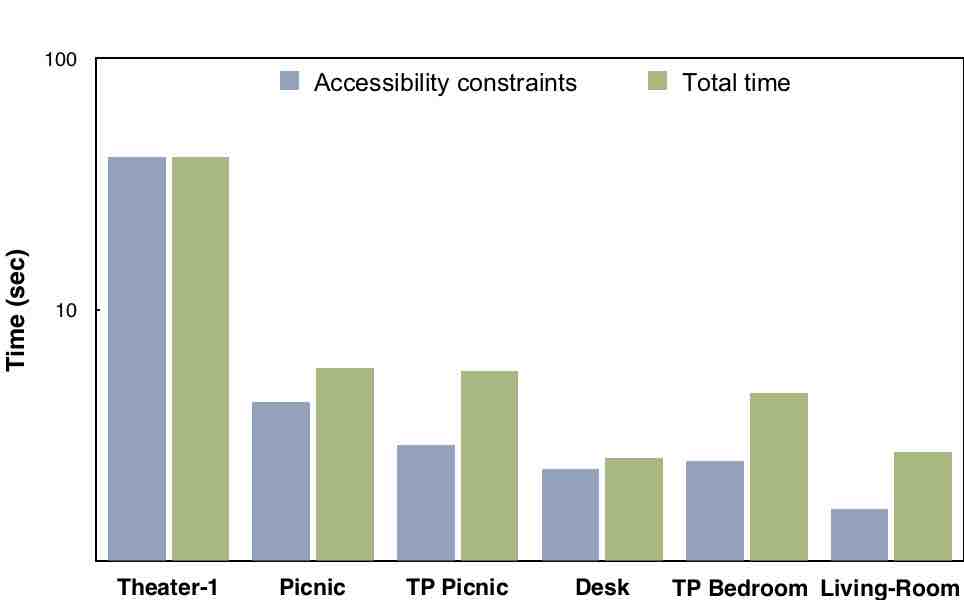}
  \caption{Run-times of our method. TP denotes ``tightly-packed''. The
  major computational cost stems from resolving the accessibility and
  collision constraints, especially when increasing the number of
  layout objects.
  \label{fig:experiments_timings}}
\end{figure}

Table~\ref{table:timings} reports the run-times of our experiments. As
shown in Fig.~\ref{fig:experiments_timings}, most of the computation
time is expended in solving the accessibility and collision
constraints,

\begin{table}
\centering
\fontsize{.85em}{.7em}\selectfont
\begin{tabular}{lrrr}
\toprule
&    \textbf{\# Objects} & \textbf{Our Method (sec)} & \textbf{SA-McMC (sec)} \\ \midrule
Theater-1       &  201   & $39.50$  & $5852$ \\
Theater-2 (arc-0) &  181 & $1.27$  & $\infty$ \\
Theater-2 (arc-1) &  181 & $1.31$  & $\infty$ \\
Theater-2 (arc-2) &  181 & $1.36$  & $\infty$ \\
Theater-2 (seg-0) &  169   & $0.48$  & $\infty$ \\
Theater-2 (seg-1) &  169   & $0.52$  & $\infty$ \\
Theater-2 (seg-2) &  246   & $0.73$  & $\infty$ \\
Picnic        &  77    & $4.77$  & $253$        \\
Living-Room   &  10    & $0.62$  & $27$        \\
Desk          &  21    & $0.69$  & $37$      \\
TP Bedroom & 12 & $0.67$ & $22$ \\
TP Picnic & 53   & $2.42$ &  $109$ \\
\bottomrule
\end{tabular}
\caption{Comparing the run-times of our method versus the baseline
SA-McMC method. Times shown are the mean of 10 runs with the same
starting conditions for both methods. TP denotes ``tightly-packed''.
For the Theater-2 scene, \{arc,seg\}-\# denotes the number of
pathways. Collision detection is disabled in Theater-2. For Theater-2,
we failed to achieve reasonable synthesis results in finite time with
our SA-McMC implementation; better-designed SA-McMC shift moves may
help.
\label{table:timings}}
\end{table}

\subsection{Layout Synthesis}
\label{sec:expr_synthesis}

We next describe the experimental scenarios of
Fig.~\ref{fig:comparing_shots}.

\subsubsection{Theater Variations}
\label{exp:theater}

\begin{figure}
\centering
\includegraphics[width=0.6\columnwidth]{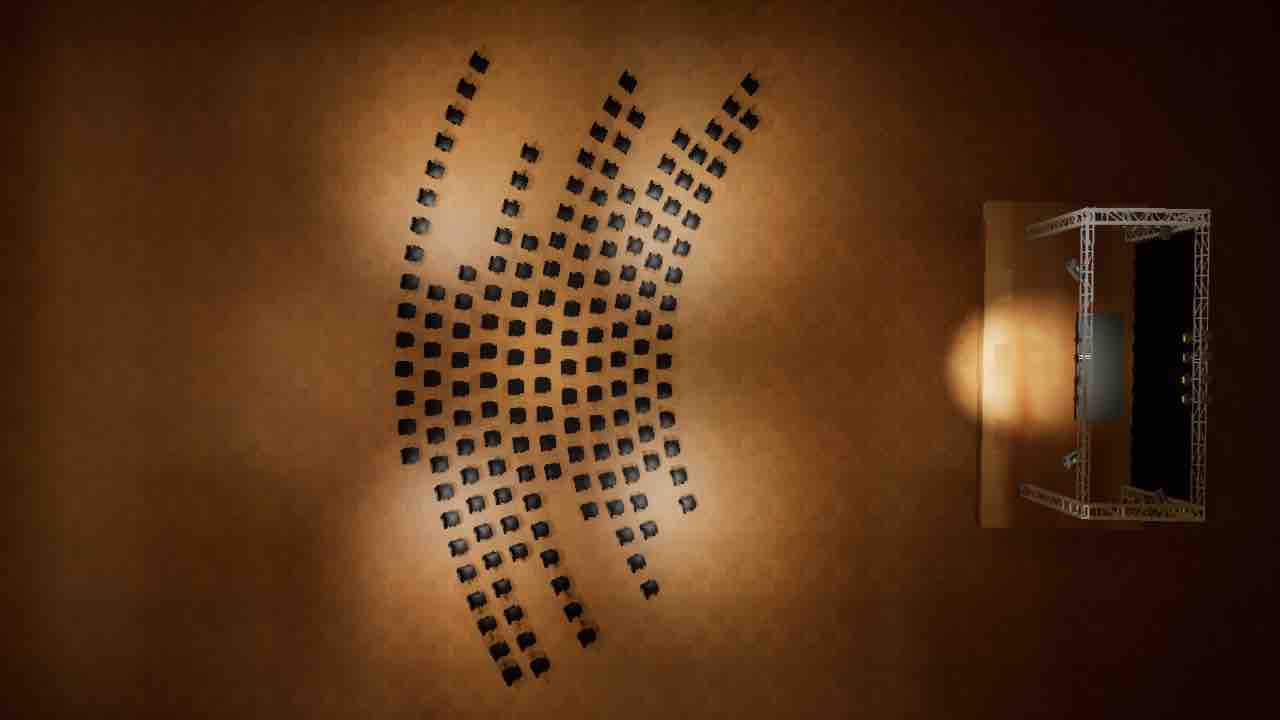}
\caption{Theater with tiers not restricted to front of stage.
\label{fig:fail1}
}
\end{figure}

We demonstrated the efficacy of our method by running our algorithm in
a theater scene with various seating arrangements and two different
constraint strategies:
\begin{enumerate}
\item \label{itm:tv:1} Each chair is at a predefined distance to the
stage due to a focal point constraint, and is constrained not to
collide with other chairs. The chairs are not associated with tiers,
and there are no pairwise distance constraints between chairs in the
same tier (Fig.~\ref{fig:comparing_shots}).

\item \label{itm:tv:2} Each chair is part of a seating tier that is
either a segment or an arc group. Each particle is constrained to have
the same distance from neighboring particles in the tier. Tiers are
constrained to be centered before the stage
(Fig.~\ref{fig:theater_variations}), or layouts like that in
Fig.~\ref{fig:fail1} may result.
\end{enumerate}
The stage is initially located at the front midpoint of the theater.
There are up to $2$ traffic lanes for stage pathways.

\begin{figure}
  \includegraphics[width=0.9\columnwidth]{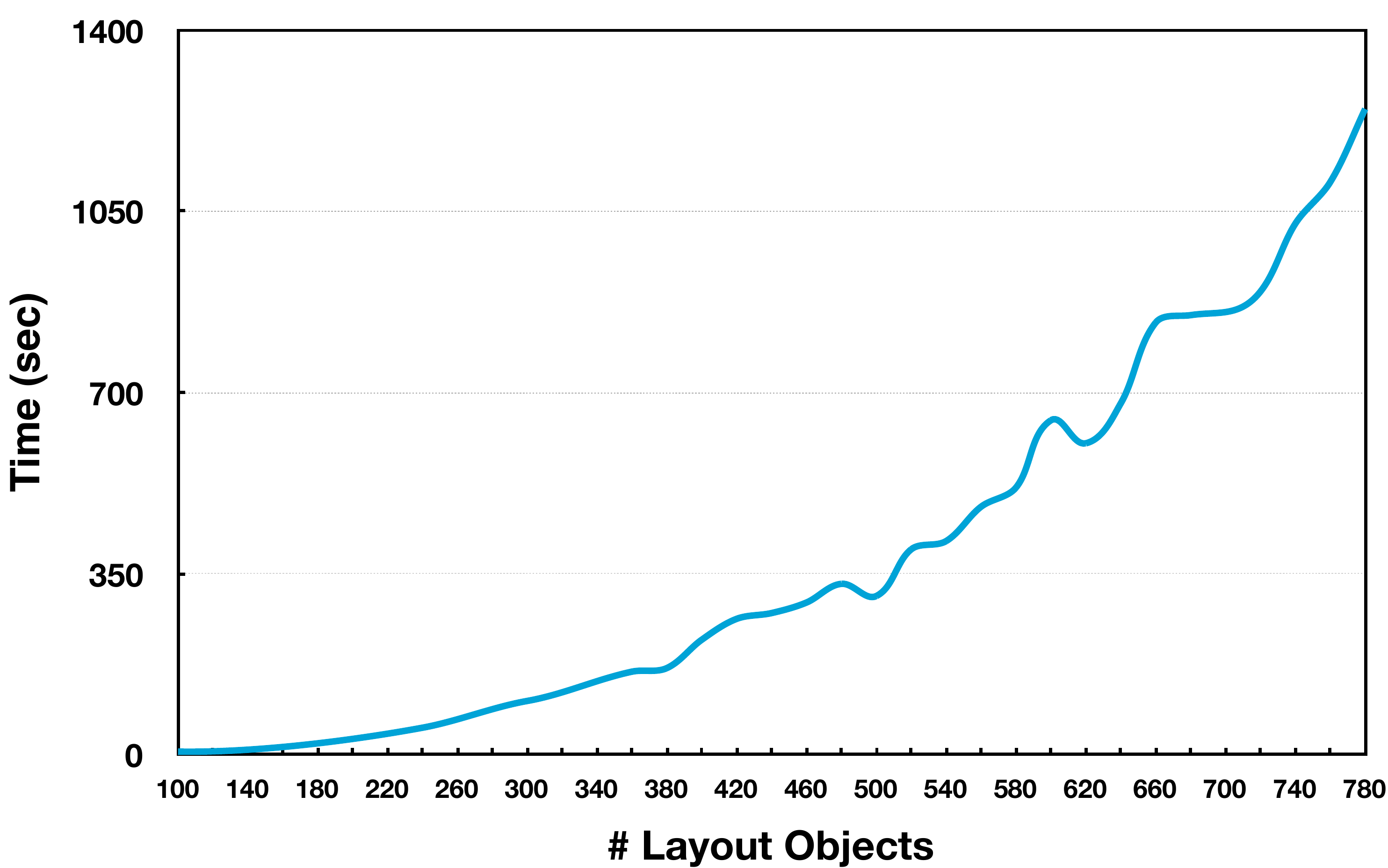}
\caption{Run-times for increasing numbers of theater seats in the
Theater-1 scenario.
\label{fig:scale_timings}}
\end{figure}

With Approach~\ref{itm:tv:1}, we allocated 200 chairs and further
tested the scalability of our algorithm by running additional
experiments with varying numbers of chairs.
Fig.~\ref{fig:scale_timings} plots the run-times. Distances are
enforced through accessibility and collision constraints; even without
distance constraints, the chairs maintain nearly regular intra-tier
distributions.

The scenes synthesized with Approach~\ref{itm:tv:2} involve between
168 to 246 layout objects. Within tiers, the distribution of chairs is
constrained to be around the center of the tier, using a heat point
constraint. The traffic lanes conflict with the pairwise distance
constraints of the chairs in each tier.

\subsubsection{Picnic}
\label{exp:picnic}

The picnic scene consists of 14 tables, 48 chairs, 8 trash cans, 6 BBQ
grills, and a carousel. The main constraints are focal point
constraints between each group of chairs and their table, distances
between chairs around tables, distances between tables, and heat point
constraints to position BBQ grills, trash cans, and picnic tables.

\subsubsection{Living-Room}
\label{exp:living_room}

\begin{figure}
\centering
\includegraphics[width=\columnwidth]{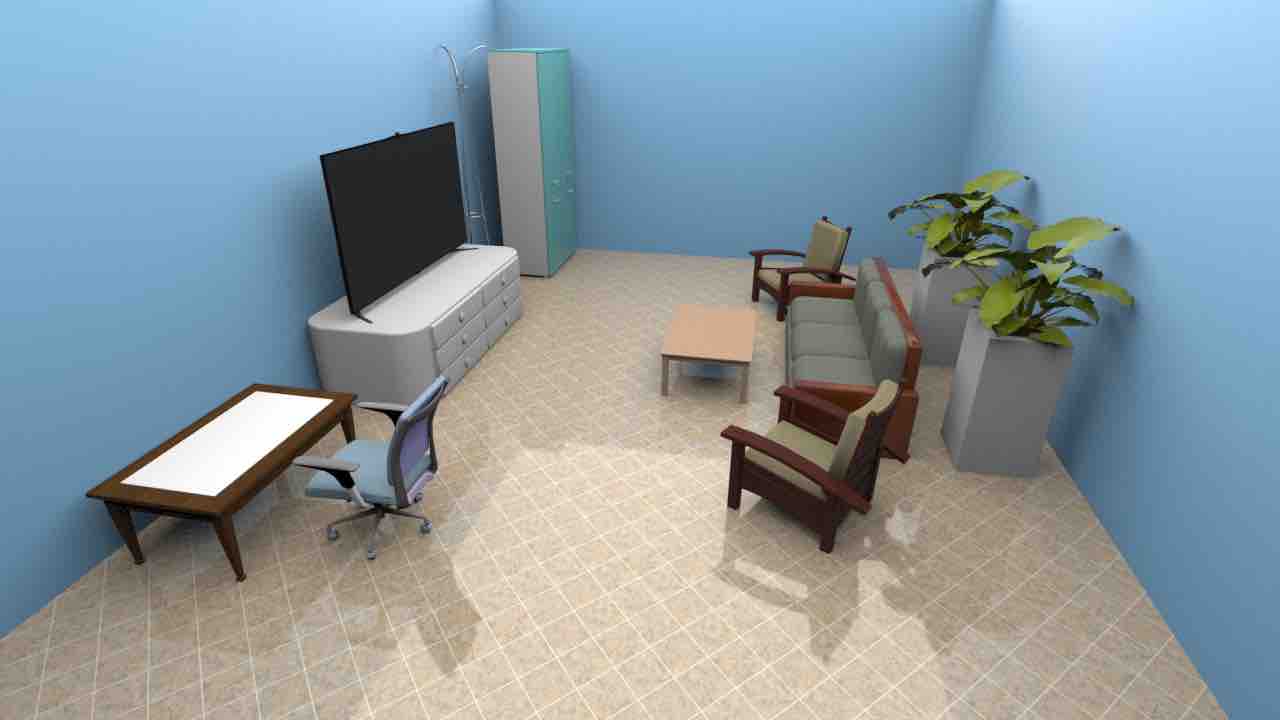}\\[2pt]
\hfill
\includegraphics[width=0.5\columnwidth]{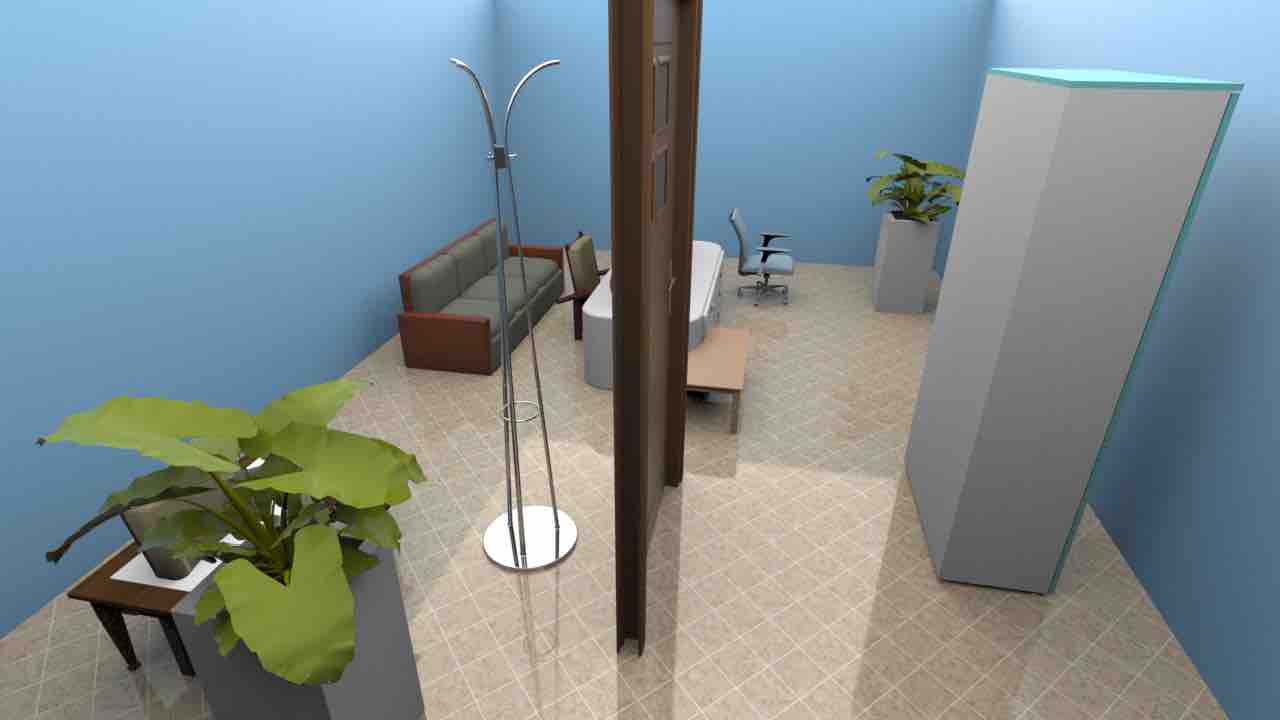}
\caption{Optimized living room layout satisfying criteria such as
distance, viewing angle, focal point grouping, and visual balance,
starting from the initial random layout shown beneath.}
\label{fig:living_room_cam}
\end{figure}

The living-room layout (Fig.~\ref{fig:living_room_cam}) contains 2
chairs, 2 indoor plants, a sofa and armchairs, a coat rack, a door, a
desk, an office chair, and a TV. The main constraints are focal point
constraints between the TV, sofa, and armchairs, as well as wall
constraints on the big furniture objects and plants.

\subsubsection{Desk}
\label{exp:desk}

We also demonstrated the performance of our algorithm for a desk with
small objects, including 12 books, 3 pencils, a food plate, a binder,
a photo frame, a potted plant, a laptop computer, and a mug. The main
constraints are focal point constraints between certain objects, heat
point constraints on different desk parts, and a stacking constraint
for the books.

\subsection{Layout Synthesis in Tightly-Packed Scenarios}

Our method copes well with highly constrained and tightly-packed
settings.

\subsubsection{Tightly-Packed Bedroom}
\label{exp:dense_bedroom}

\begin{figure*}
\centering
\includegraphics[width=0.33\textwidth]{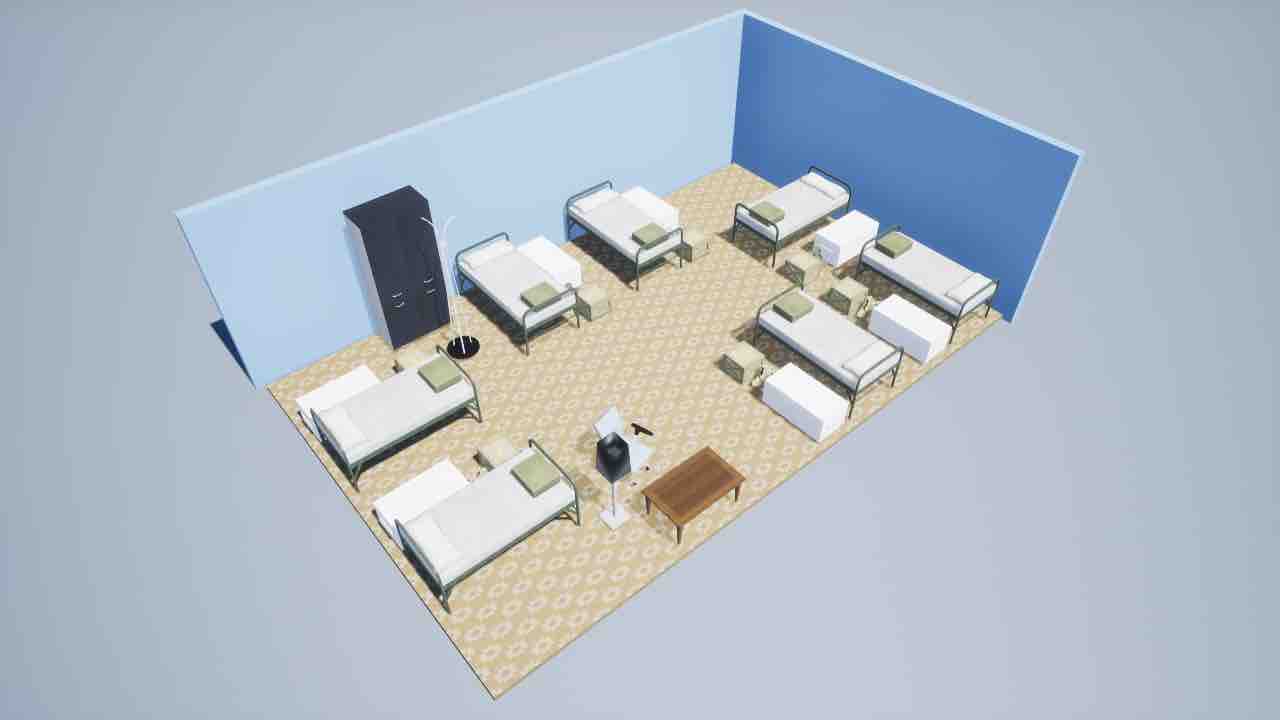}
\includegraphics[width=0.33\textwidth]{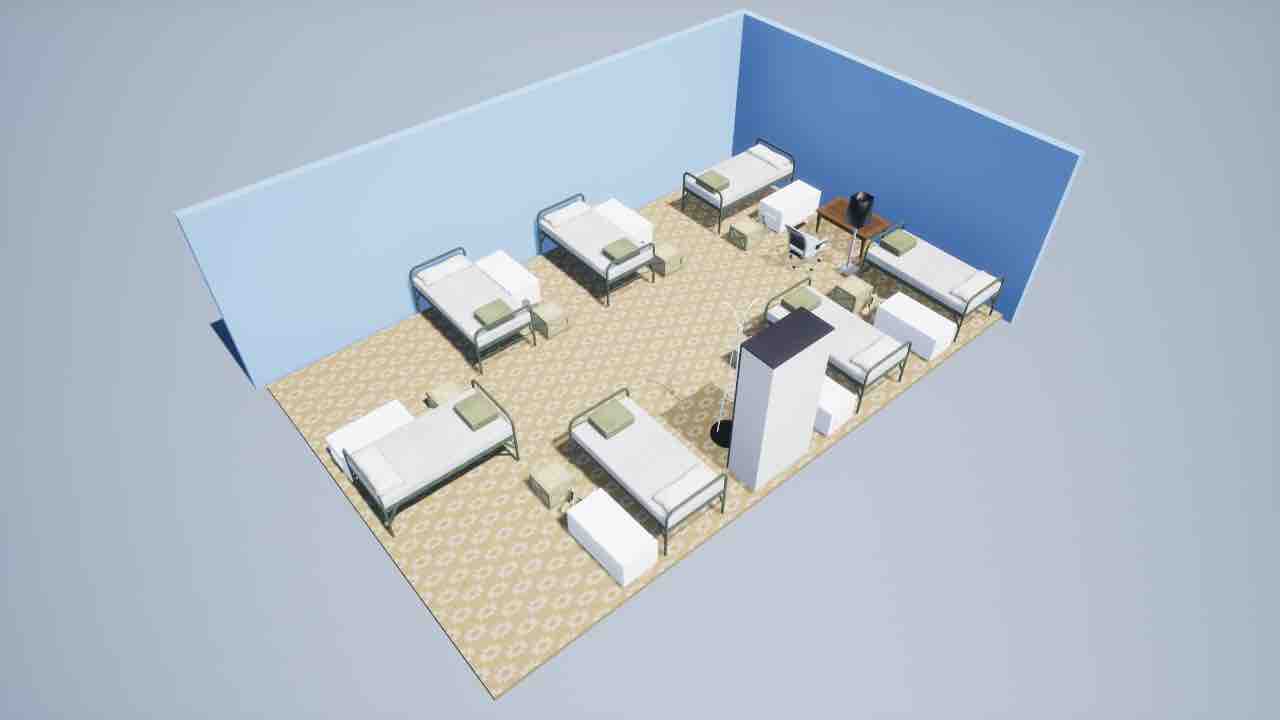}
\includegraphics[width=0.33\textwidth]{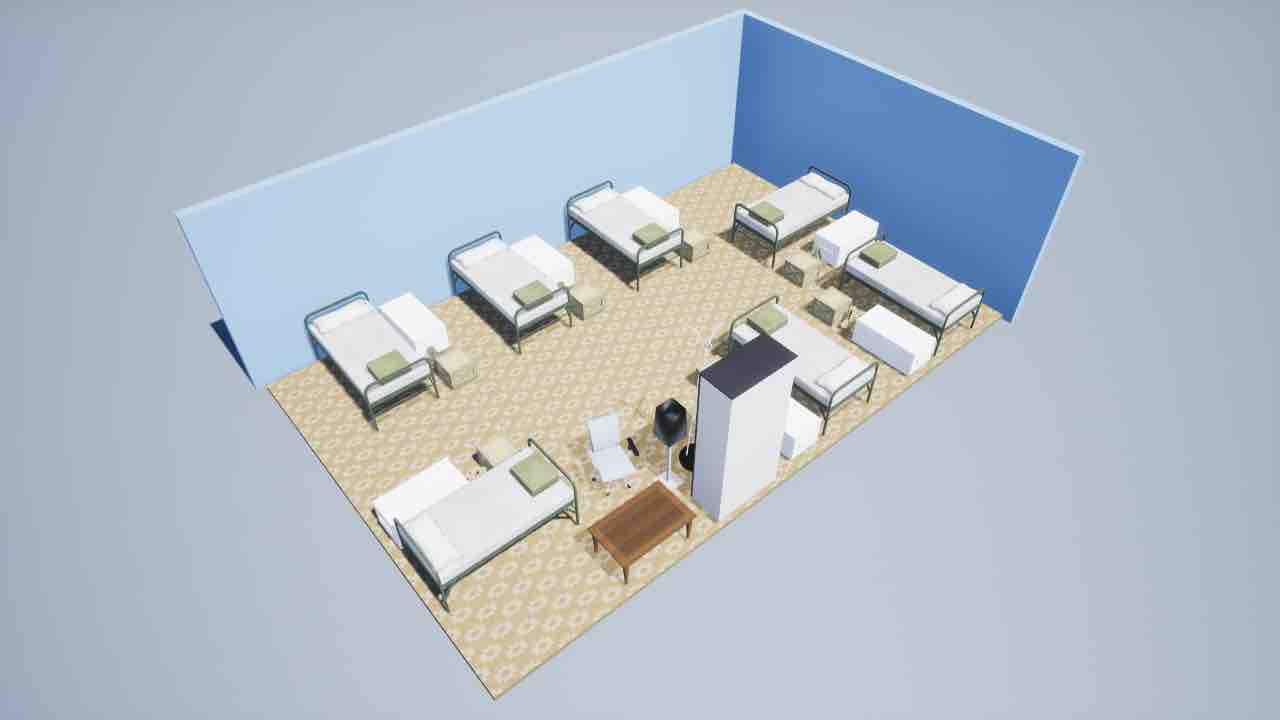}
\\[2pt]
\hfill
\includegraphics[width=0.2\textwidth]{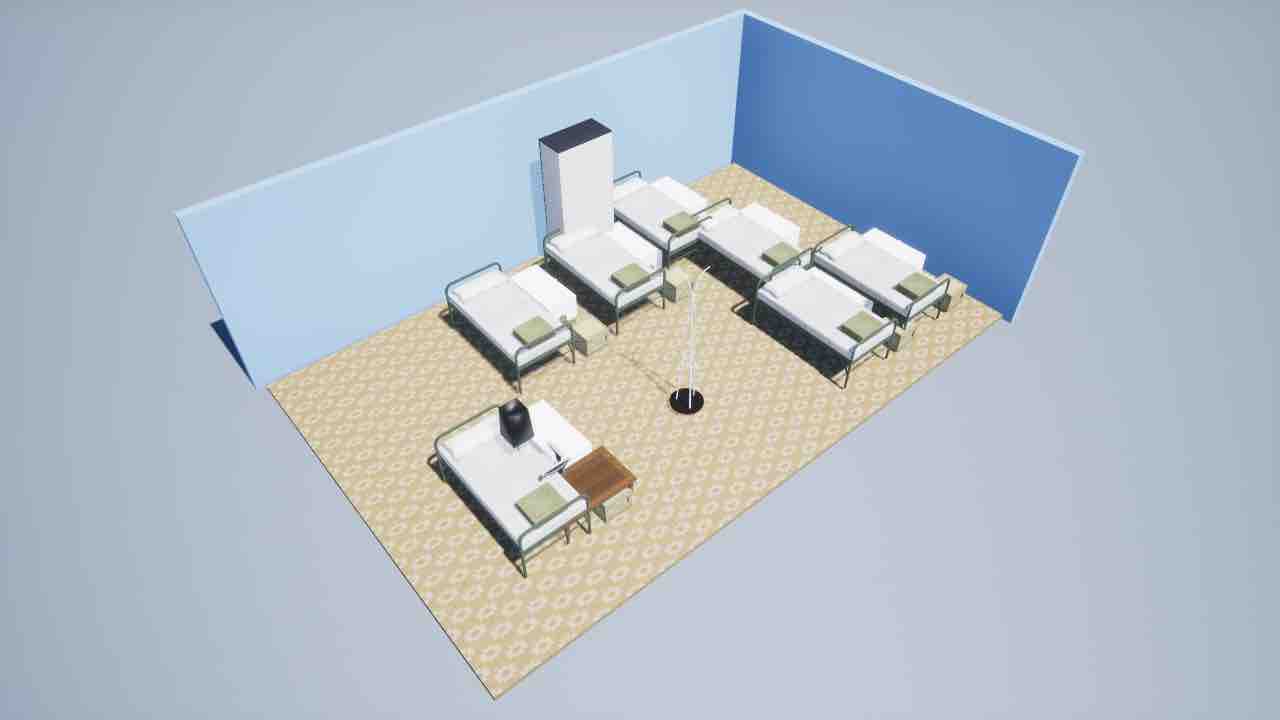}
\hfill
\includegraphics[width=0.2\textwidth]{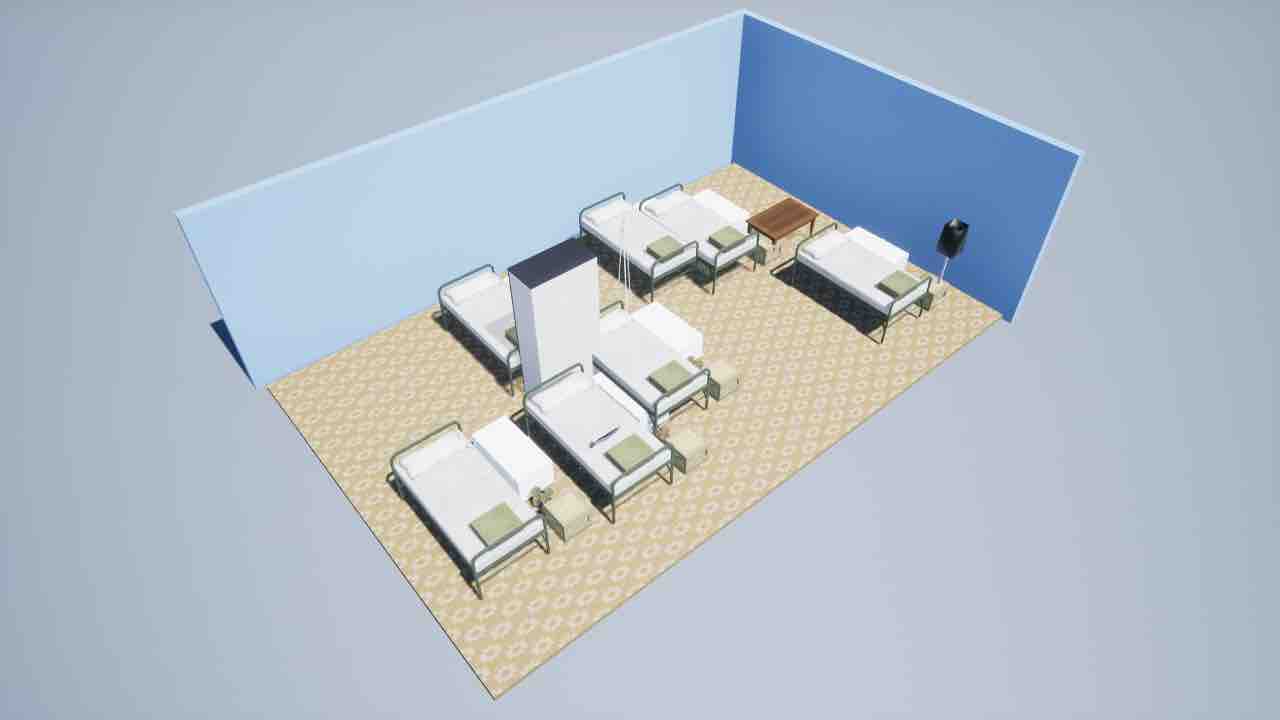}
\hfill
\includegraphics[width=0.2\textwidth]{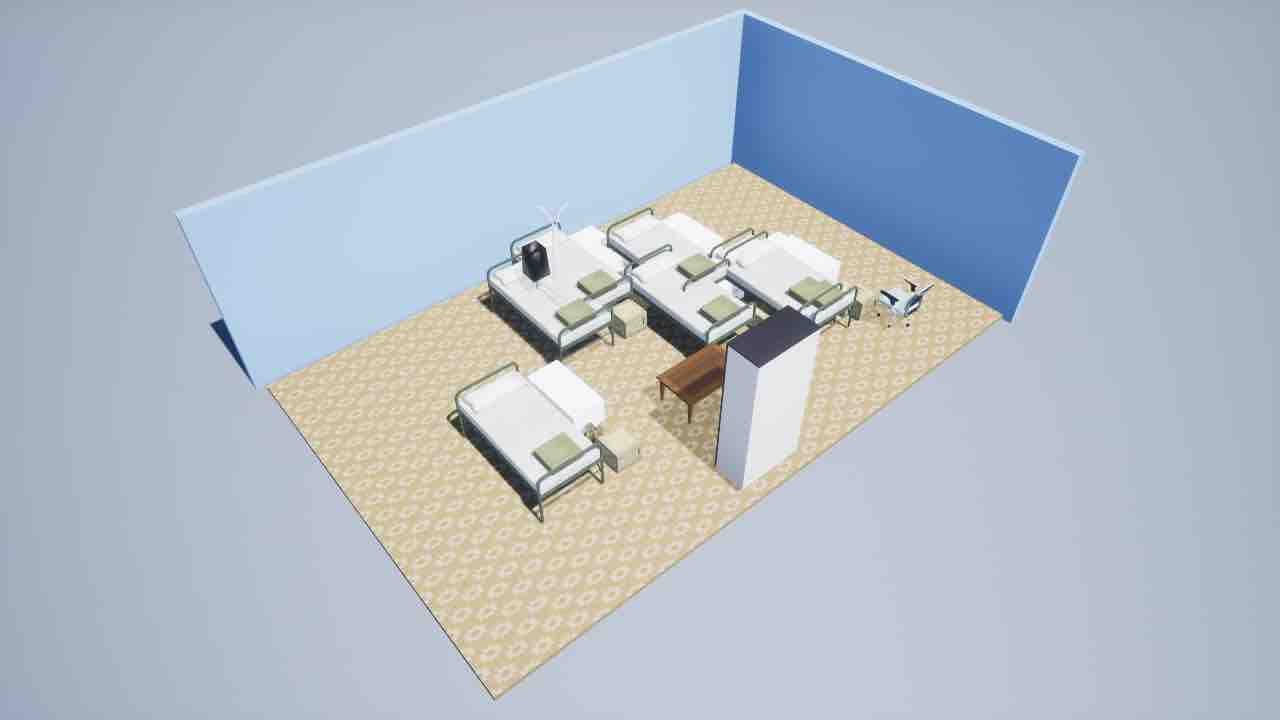}
\caption{Different layout suggestions (top) synthesized by
initializing from different random initial conditions (bottom).
\label{fig:tp_bedroom}}
\end{figure*}

The tightly packed bedroom contains multiple beds and pieces of
furniture. The beds are rigidly grouped together with army style
accessories. The beds, bookcase and table are constrained to be next
to the wall. The coat rack is constrained to be at a certain distance
from the bookcase. We demonstrate that using different initial
conditions results in different suggested layouts. Even though the
space is tight, our method successfully synthesizes different layout
suggestions (Fig.~\ref{fig:tp_bedroom}).

\subsubsection{Tightly-Packed Picnic}
\label{exp:dense_picnic}

This tightly-packed setting demonstrates our method's ability to
synthesize diverse layouts with different numbers and types of
furniture objects. We synthesize a tightly-packed picnic scenario in
two stages. In the first stage, we randomly vary the number of layout
objects of each type, similar to Yeh et al.\cite{yeh2012synthesizing}. The
available layout objects are a superset of the previous picnic
scenario, with an additional rectangular picnic table. For a more uniform
layout, we rigidly attached 4 chairs to each round
picnic table. In the second stage of the synthesis, we run our method
for 270 iterations. Fig.~\ref{fig:tp_picnic} shows these synthesized
layouts.

\begin{figure*}
\centering
\begin{subfigure}{0.495\textwidth}
\includegraphics[width=\textwidth]{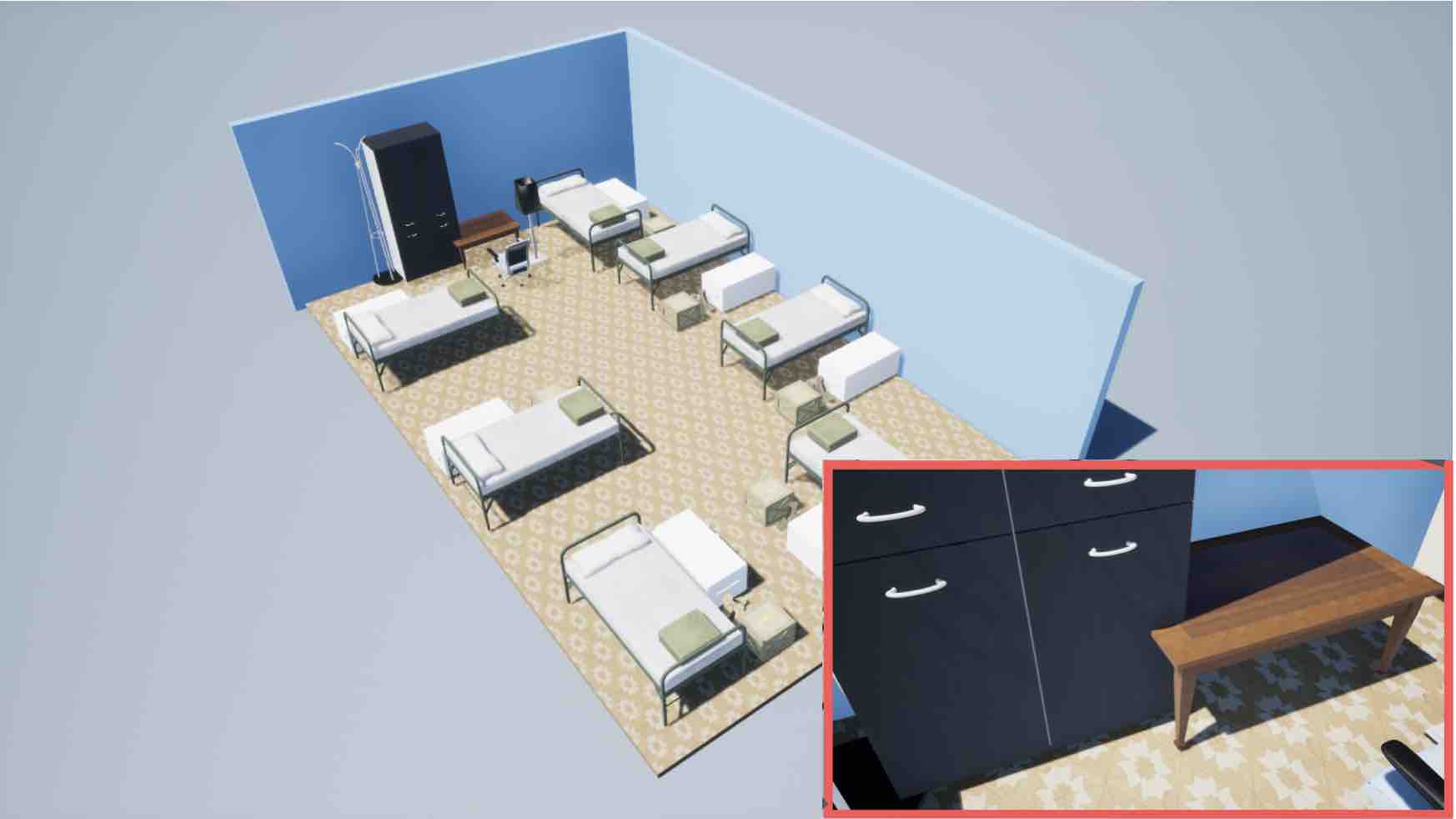}
\caption{ }
\label{fig:tp_bedroom_bad}
\end{subfigure}\hfill
\begin{subfigure}{0.495\textwidth}
\includegraphics[width=\textwidth]{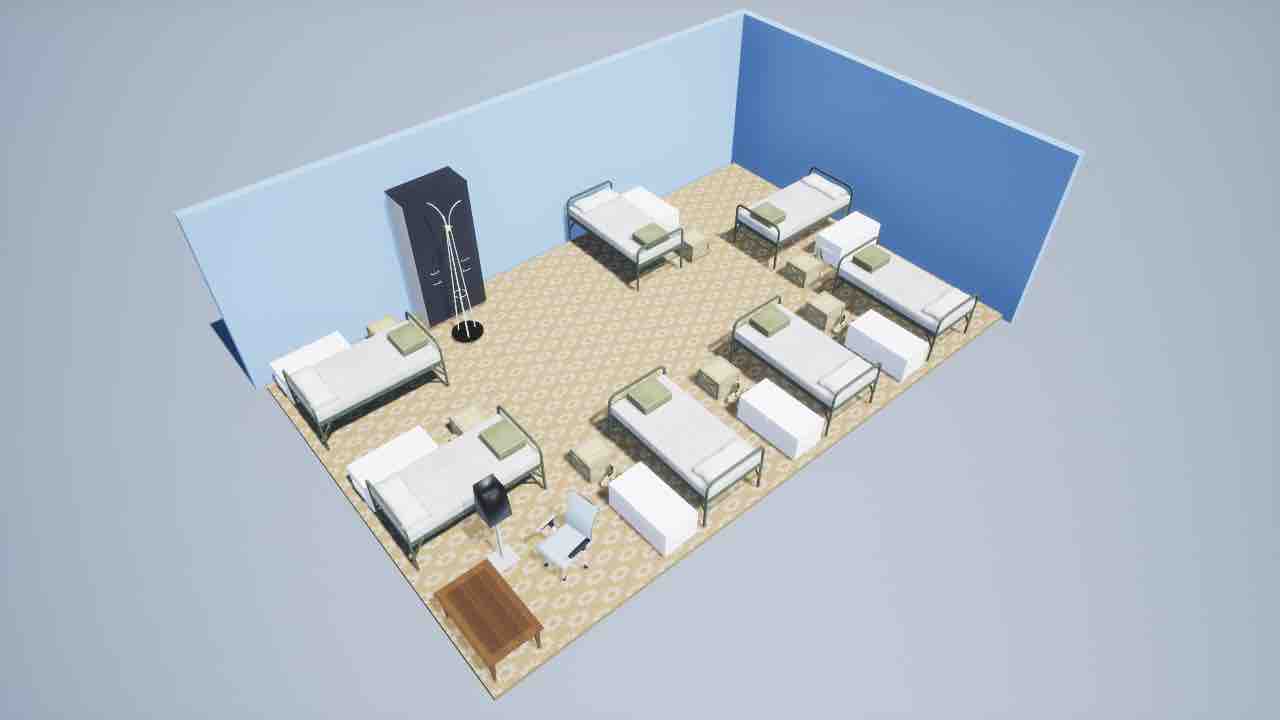}
\caption{ }
\label{fig:tp_bedroom_good}
\end{subfigure}
\caption{(a) A baseline SA-McMC approach struggles with a
tightly-packed bedroom. Using stochastically sampled shift moves
results in objects becoming ``locked'' in configurations that are
probabilistically hard to escape. (b) Our method does not. Since the
table, rack and closet are colliding and constrained to be next to the
wall, the objects manage to escape this configuration following a
positional correction that enforces collision and distance to wall
constraints. See Section \ref{constraint:col}.}
\label{fig:mcmc_collide}
\end{figure*}

\subsection{Comparison to Simulated Annealing}

We compared the performance of our method to a baseline layout
synthesis approach that applies simulated annealing with a
Metropolis-Hastings McMC state-search step, which we denote SA-McMC.
Our implementation is based on code used in \cite{yu2011make}. In this
implementation, the proposal function shifts an attribute of one
layout object in each state-search step. We employed the same energy
function (\ref{eq:energy}) to track the quality of the synthesized
layouts and ran the comparison several times, with different
conditions, such as varying constraint weights $\gamma_i$ and
temperature schedules for the SA-McMC algorithm. For all our
experiments, we used linear, evenly-spaced schedules for about 20,000
iterations, with an additional stopping condition in case the energy
function value did not improve by more than 0.1\% during the previous
1,500 iterations.

Experimentally, we noticed that a baseline SA-McMC approach has
trouble accommodating tightly-packed and constrained layouts, as in
the tightly-packed bedroom and picnic scenarios described in the
previous section. Superficially, synthesized layouts appeared
satisfactory; however, upon closer inspection, they suffered from
unresolved collisions (Fig.~\ref{fig:tp_bedroom_bad}). A possible
explanation is that, unlike our approach, SA-McMC does not exploit
local constraint gradient information when shifting between layout
configurations. In theory, SA-McMC can escape such collisions through
the choice of different parameter settings, using more complex
hand-crafted SA-McMC shifts moves, and/or tuning the weights of
different constraints in the energy. Unfortunately, none of the
settings with which we experimented yielded collision-free layout
suggestions for the tightly-packed scenarios.

\begin{figure}
\centering
\includegraphics[width=\columnwidth]{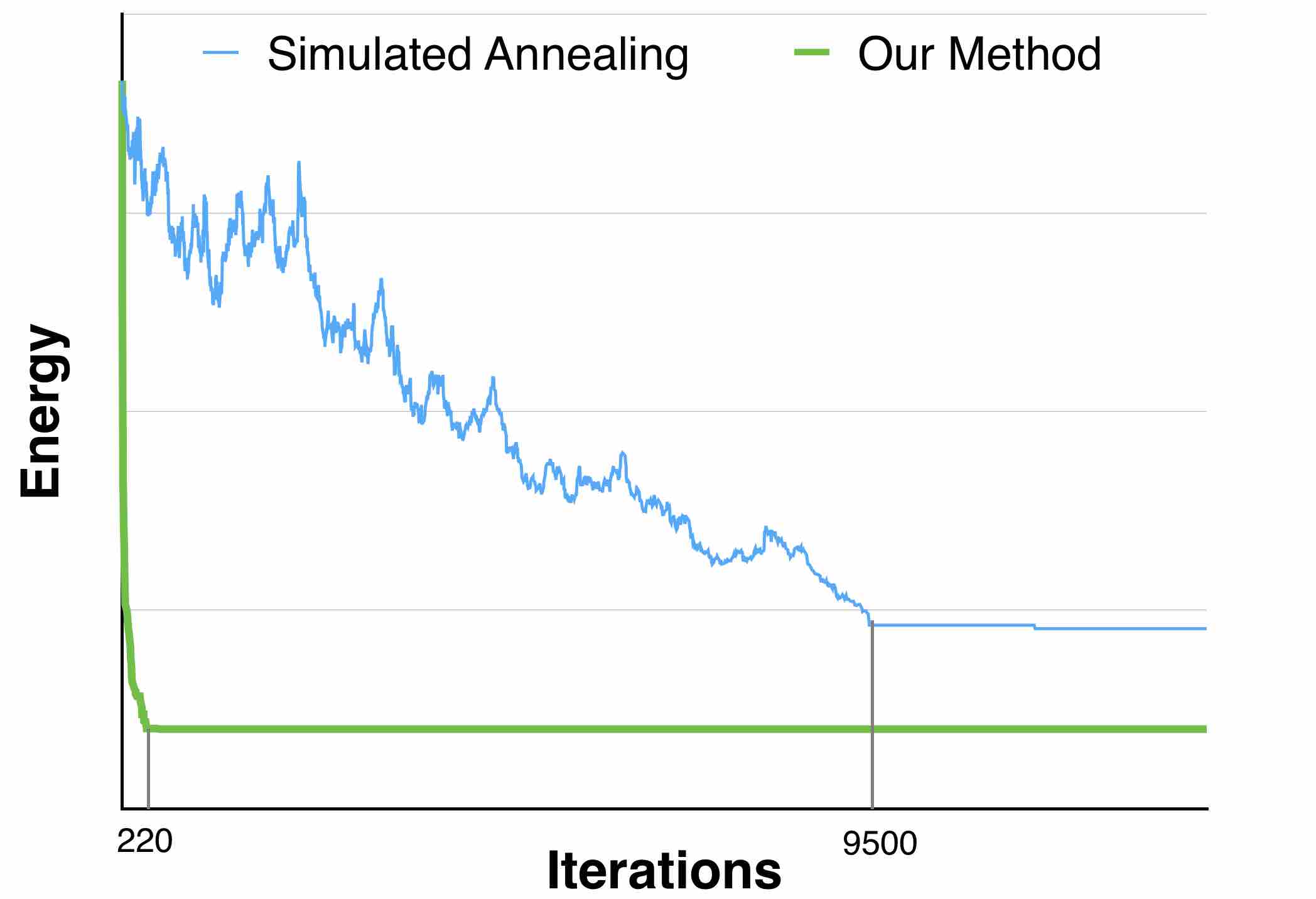}
\caption{Energy plot (linear scale) of our method versus
SA-McMC~\cite{yu2011make} for the tightly packed picnic layout. The
total run-time for SA-McMC was approximately 163 seconds, versus 4.7
seconds for our method. Our method converged to a satisfactory layout
at around iteration 220. SA-McMC converged to a less satisfactory,
higher-energy layout after 9,500 iterations.}
\label{fig:picnic_compare}
\end{figure}

Fig.~\ref{fig:picnic_compare} plots the energy as a function of
iteration number for SA-McMC and our method for the tightly-packed
picnic scenario shown in Fig.~\ref{fig:teaser}. Generally, we observed
that SA-McMC is slower by at least an order of magnitude compared to
our method (Table~\ref{table:timings}). The computational cost of
SA-McMC increases dramatically with the number of objects to be
synthesized. Hence, our method is the one that is more suitable for
interactive applications.

\section{Discussion}
\label{sec:disc-concl}

We have demonstrated that continuous, deterministic, point-based
simulation methods can produce satisfactory layout synthesis results
at low computational cost, at least an order of magnitude faster than
previous methods, which makes it suitable for augmented reality
applications (Appendix~\ref{app:AR}). Layout synthesis objectives are
represented by a set of positional and orientational constraints. Our
approach satisfies the conflicting set of constraints iteratively,
resulting in fast layout synthesis with quality similar to or better
than that of previous stochastic, McMC-based alternatives.

Previous work in layout synthesis does not incorporate constraint
gradient information, but instead uses manually crafted McMC moves
that augment a layout configuration. While this type of method
provides an easy mechanism to explore different layout variations, we
achieve the same effect via different initializations. As observed in
our comparison, the run-times of the baseline SA-McMC method increases
significantly with the number of layout objects. In these approaches,
a user can define a more relaxed set of constraints, which at least in
principle could lead to a desired layout, but at much greater
computational cost. In practice, we observed that these methods
perform poorly in tightly-packed layouts, resulting in unrealistic
collisions between layout items. By contrast, our method is
dramatically faster due to its continuous nature. Notably, it requires
only a few seconds of run-time for dozens of objects and it naturally
scales to hundreds of objects with only moderately increasing
computational cost.

Since layout synthesis poses a non-convex problem, it is
difficult---and fortunately unnecessary---to find the global optimum.
We experimented with nonlinear global optimization solvers
(NLopt~\cite{nlopt2011}), but the results were either poor in quality
(i.e., unrealistic layouts with many colliding layout items), or the
run-times were intractable. In contrast to these solvers, our method
does not directly try to minimize the global energy, but rather
iteratively satisfies individual constraints. This induces positional
corrections to layout items that propagate throughout the layout,
transforming an initially poor, high-energy layout to a satisfactory
low-energy layout. The global energy quantifies the quality of layouts
and enables our comparisons to prior work.

The intended workflow of layout synthesis methods is for the automated
approach to synthesize a variety of viable layouts, from which the
user can select one or more that they prefer. Our method supports this
workflow. It can generate a variety of layouts by repeatedly running
from different random initial conditions (Fig.~\ref{fig:tp_bedroom}).
Ultimately, the quality of configurations is a subjective matter.
Hence, at least from the user's perspective, a global optimum is not
definitive.

\subsection{Limitations}

We observed that solving constraints sequentially, where the new
positions are immediately visible to other constraints, makes quick
progress at first, but the convergence rate slowly decreases as the
iterations progress. For example, in the living-room experiment, the
first few iterations yield a layout that is visually similar to the
final one, whereas later iterations produce smaller refinements of the
layout and resolve cuboid accessibility area intersections.

Like SA-McMC, the layout's energy can increase from one iteration to
the next, which allows to escape suboptimal local minima. Unlike
SA-McMC, our method may not converge in the traditional optimization
sense; however, our termination criterion is based on the satisfaction
of most of the constraints, which is ultimately what matters. In
practice, we never observed outcomes that failed to satisfy layout
objectives. Most starting seeds lead to a satisfactory layout, all
lead to a collision free layout. Although conceptually simple, our
method produces impressive results.

\subsection{Future Work}

In the present study, we did not encode all the constraints that may
be relevant in layout design; however, our method can easily be
extended to a broader set of layout constraints. Incorporating GPU
parallelization can further speed up the procedure, as could a
hierarchical approach, where the layout synthesis problem is broken
into stages. It will also be interesting to adjust the stiffness
factors in a nonuniform manner in an effort to converge to better
global solutions.

Due to the sequential, local constraint satisfaction approach of our
position-based method, we may observe oscillations and collisions
between objects. For example, when there is a collision between the
accessibility areas of two objects, the constraint may be partially
resolved by projecting one of these objects into a collision with a
third object. In future work, we plan to design automatic schemes for
detecting and resolving these conditions.

The layout objects and their relationship to their corresponding
groups can be stochastically sampled from predefined distributions;
e.g., from real-world scene datasets. This is similar to but faster
than applying factor graphs \cite{yeh2012synthesizing}.

\appendices

\section{Augmented Reality Layout Synthesis}
\label{app:AR}

\begin{figure}
\parbox[b]{2mm}{\rotatebox[origin=b]{90}{\emph{Indoor}}}
\begin{subfigure}{0.31\linewidth}
\includegraphics[width=\linewidth]{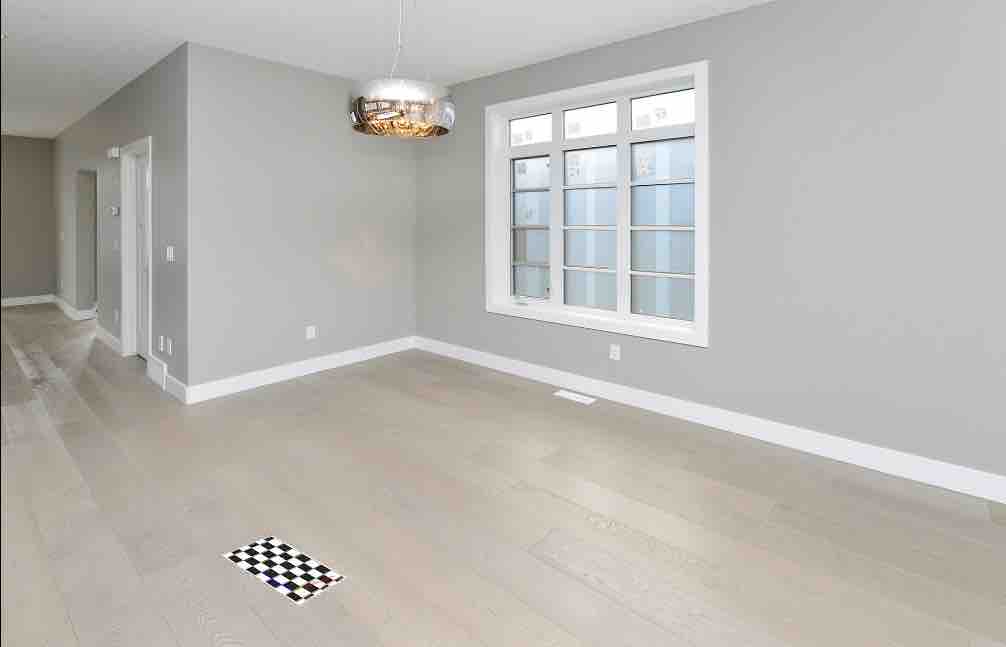}
\label{fig:origina1}
\end{subfigure}
\hfill
\begin{subfigure}{0.31\linewidth}
\includegraphics[width=\linewidth]{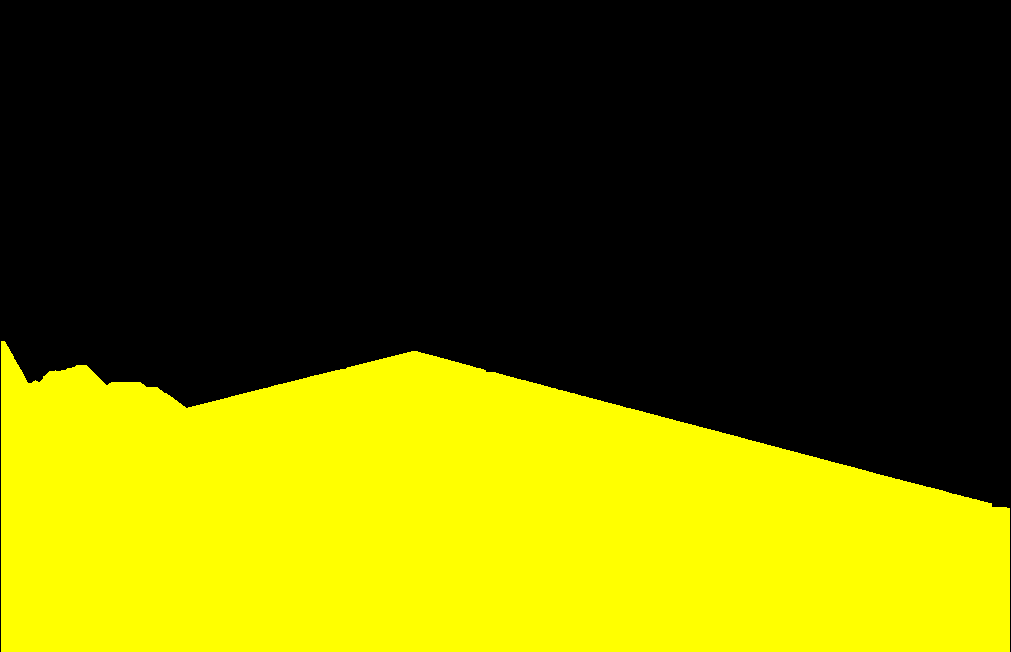}
\label{fig:seg1}
\end{subfigure}
\hfill
\begin{subfigure}{0.31\linewidth}
\includegraphics[width=\linewidth]{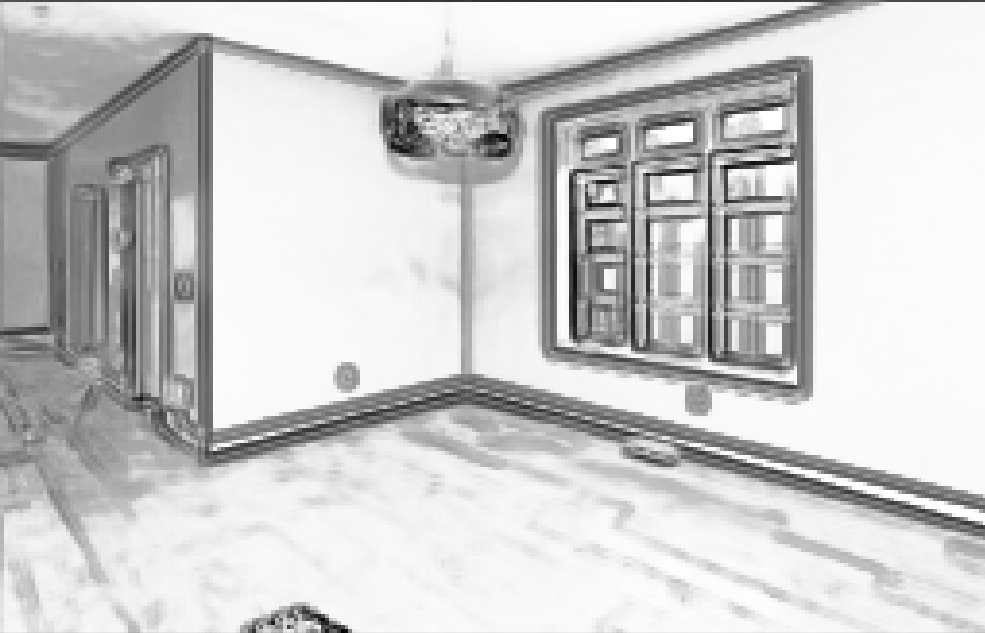}
\label{fig:hed1}
\end{subfigure}\\
\parbox[b]{2mm}{\rotatebox[origin=b]{90}{\emph{Outdoor}}}
\begin{subfigure}{0.31\linewidth}
\includegraphics[width=\linewidth]{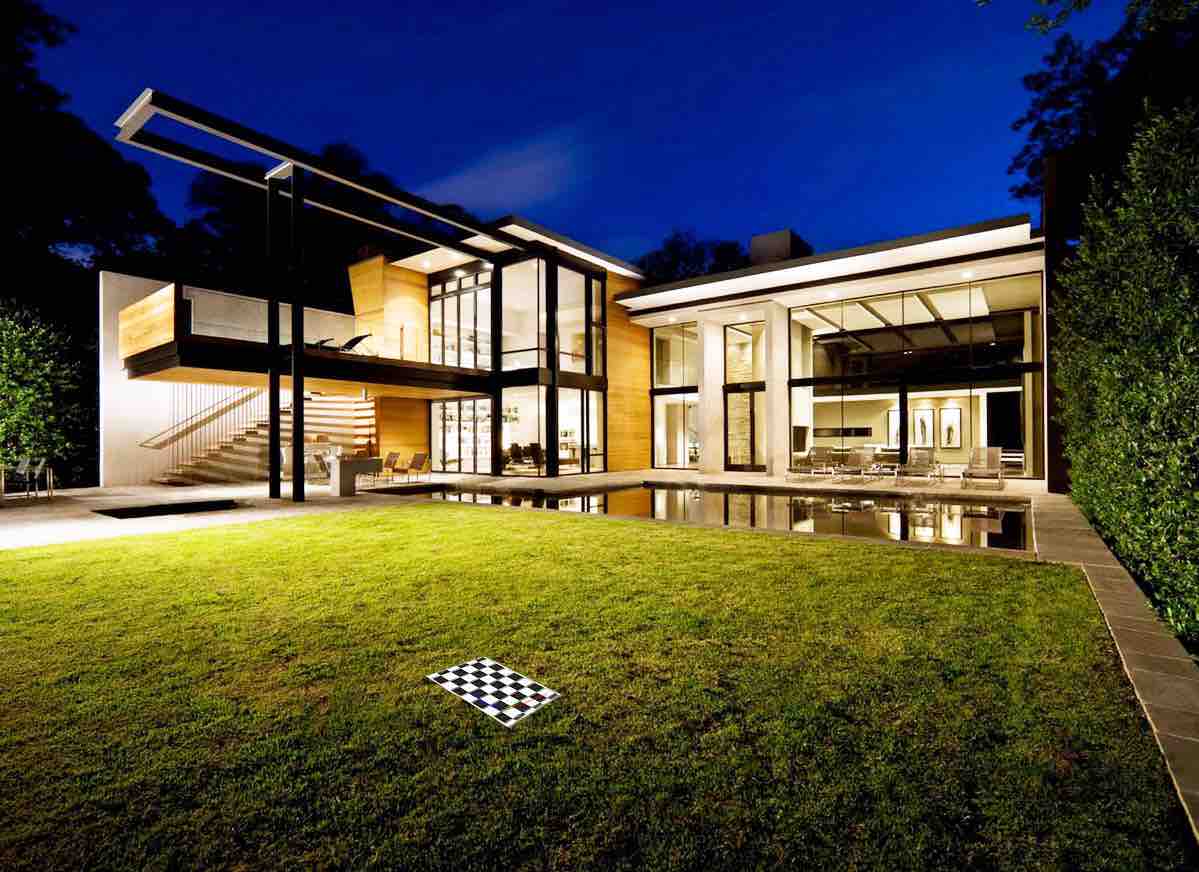}
\label{fig:origina2}
\caption{}
\end{subfigure}
\hfill
\begin{subfigure}{0.31\linewidth}
\includegraphics[width=\linewidth]{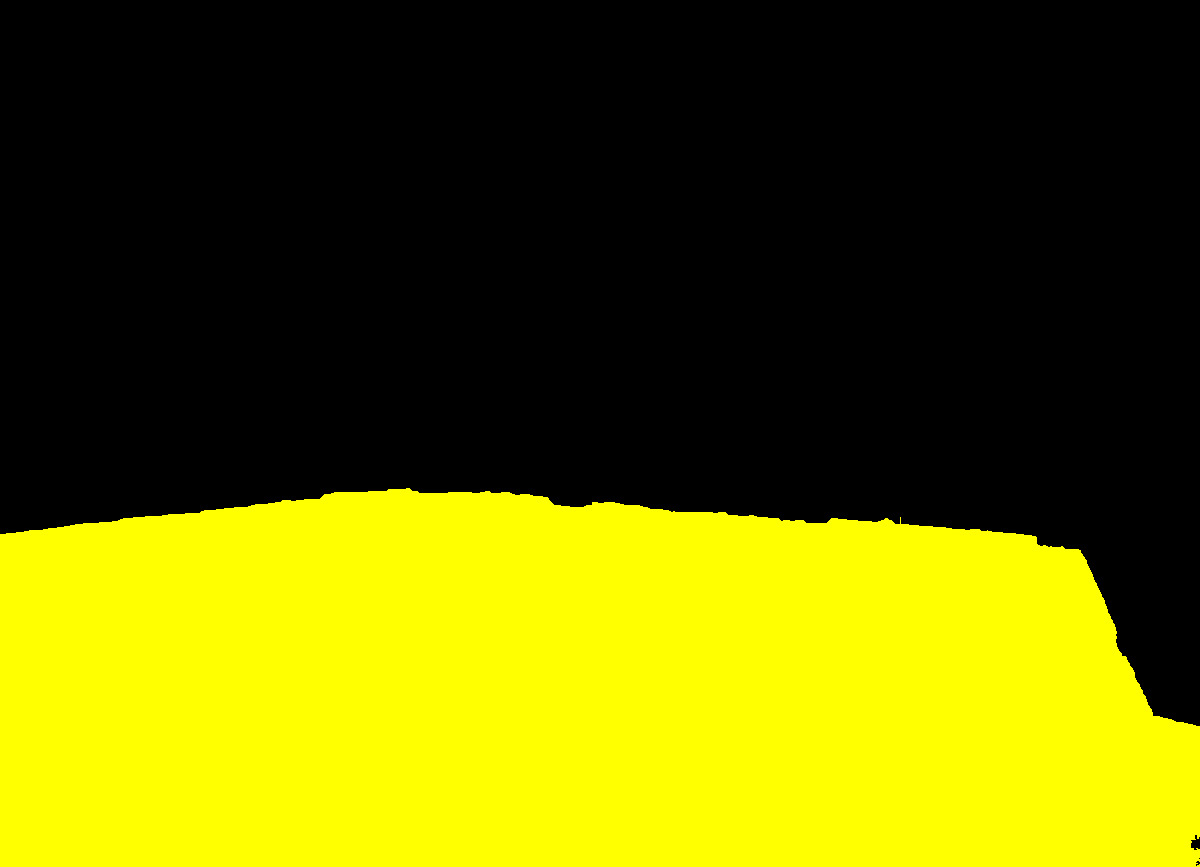}
\label{fig:seg2}
\caption{}
\end{subfigure}
\hfill
\begin{subfigure}{0.31\linewidth}
\includegraphics[width=\linewidth]{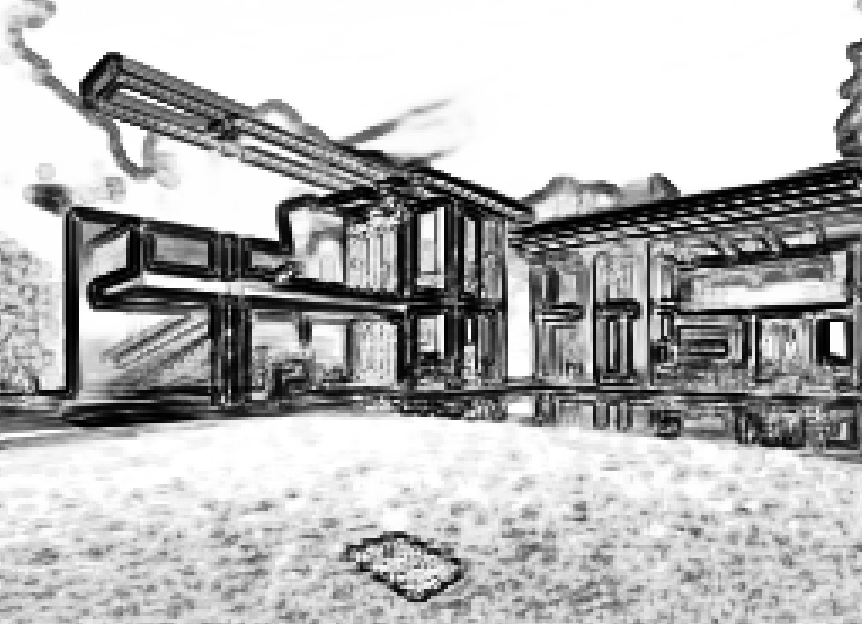}
\label{fig:hed2}
\caption{}
\end{subfigure}

\caption{Segmentation and edge detection. (a) Original images. (b)
Segmented floor/ground. (c) Edge maps.}
\label{fig:seghed}
\end{figure}

\begin{figure}
\centering
\begin{subfigure}{0.49\linewidth}
\centering
\includegraphics[height=2.95cm]{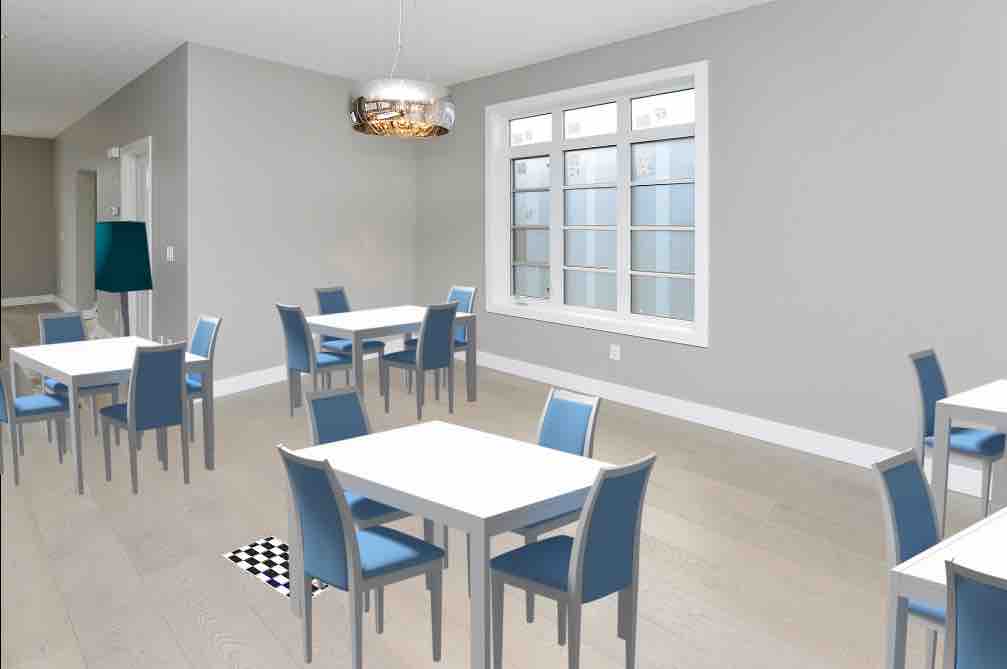}
\caption{}
\end{subfigure}
\hfill
\begin{subfigure}{0.49\linewidth}
\centering
\includegraphics[height=2.95cm]{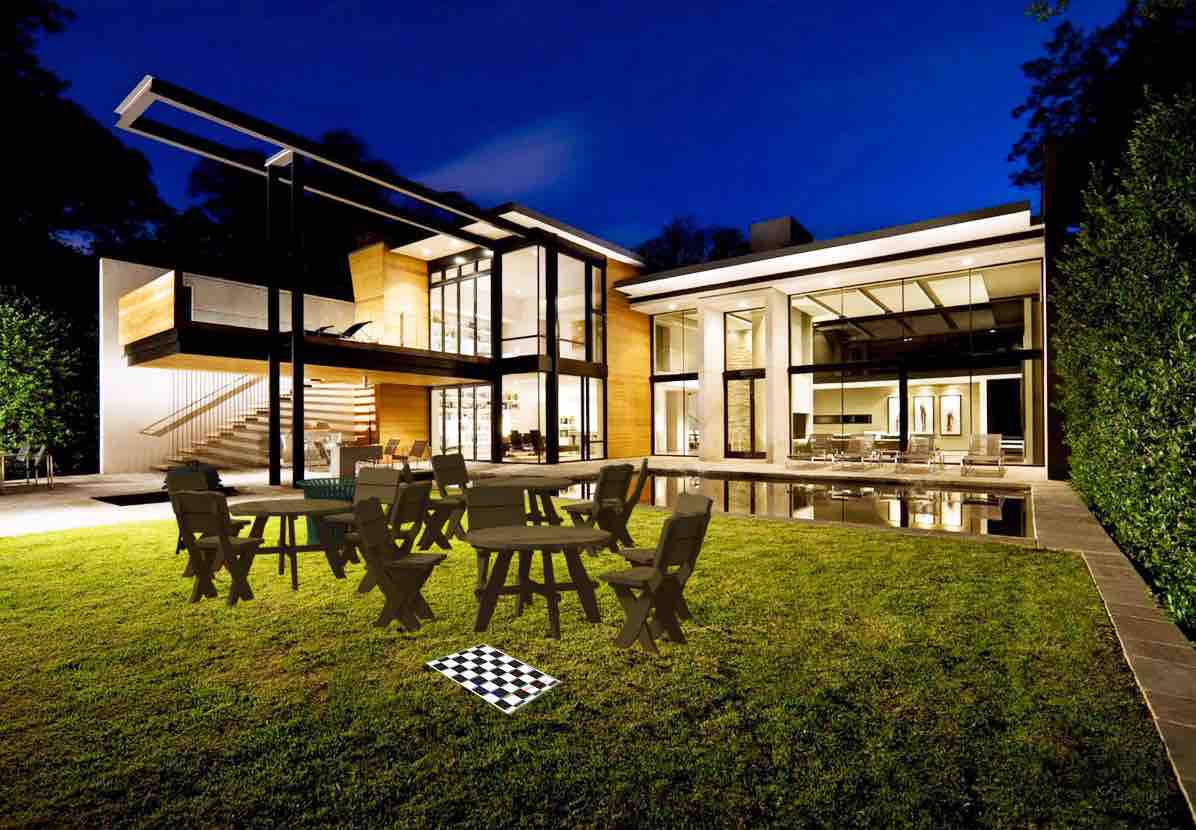}
\caption{}
\end{subfigure}
\caption{Synthesized results from user-provided images. (a)
Synthesized indoor layout. (b) Synthesized outdoor layout.}
\label{fig:stageResult}
\end{figure}

As another use-case example of our method, we demonstrate the fast
synthesis of furniture layouts from and into 2D images of vacant
spaces. After the user uploads an image of an indoor or outdoor space,
selects furnishings, and specifies layout objectives, our system then
automatically analyzes the space using scene understanding algorithms
from computer vision, and finally renders into the original image
optimal layouts of the selected furnishings satisfying the given
objectives. The system works as follows:
\begin{enumerate}

\item
Semantic scene segmentation: Our system employs SegNet
\cite{badrinarayanan2015segnet}, a state-of-the-art, pixel-wise
semantic segmentation network, trained on the SUN-RGBD dataset
\cite{xiao2010sun} with common indoor scene objects, to output 37
categories of per-pixel semantic image labels. GrabCut
\cite{rother2004grabcut} is then applied to segment the pixels labeled
'floor'. Fig.~\ref{fig:seghed} shows examples of the segmentation.

\item
3D scene estimation: We ask users to place a checkerboard calibration
marker into the imaged scene, from which our system estimates the
camera parameters, the orientation and scale of the floor/ground, and
(using Holistically-Nested Edge detection \cite{xie2015holistically})
the height of the scene perpendicular to the estimated ground plane.

\item
Layout synthesis and visualization: The system runs our layout
synthesis method to generate optimal layouts, which it then renders
into the image via a virtual camera with the aforementioned estimated
camera parameters. Fig.~\ref{fig:stageResult} shows examples of the
results.
\end{enumerate}
Reference \cite{weiss2017automated} provides additional details.

\bibliographystyle{IEEEtran}
\bibliography{tvcg18}

\begin{thebibliography}{10}
\providecommand{\url}[1]{#1}
\csname url@samestyle\endcsname
\providecommand{\newblock}{\relax}
\providecommand{\bibinfo}[2]{#2}
\providecommand{\BIBentrySTDinterwordspacing}{\spaceskip=0pt\relax}
\providecommand{\BIBentryALTinterwordstretchfactor}{4}
\providecommand{\BIBentryALTinterwordspacing}{\spaceskip=\fontdimen2\font plus
\BIBentryALTinterwordstretchfactor\fontdimen3\font minus
  \fontdimen4\font\relax}
\providecommand{\BIBforeignlanguage}[2]{{%
\expandafter\ifx\csname l@#1\endcsname\relax
\typeout{** WARNING: IEEEtran.bst: No hyphenation pattern has been}%
\typeout{** loaded for the language `#1'. Using the pattern for}%
\typeout{** the default language instead.}%
\else
\language=\csname l@#1\endcsname
\fi
#2}}
\providecommand{\BIBdecl}{\relax}
\BIBdecl

\bibitem{smelik2014survey}
R.~M. Smelik, T.~Tutenel, R.~Bidarra, and B.~Benes, ``A survey on procedural
  modelling for virtual worlds,'' in \emph{Comp. Graphics Forum}, vol.~33,
  no.~6, 2014, pp. 31--50.

\bibitem{chib1995understanding}
S.~Chib and E.~Greenberg, ``Understanding the {M}etropolis-{H}astings
  algorithm,'' \emph{The American Statistician}, vol.~49, no.~4, pp. 327--335,
  1995.

\bibitem{yu2011make}
L.-F. Yu, S.~K. Yeung, C.-K. Tang, D.~Terzopoulos, T.~F. Chan, and S.~Osher,
  ``Make it home: {Automatic} optimization of furniture arrangement,''
  \emph{ACM Trans. Graphics}, vol.~30, no.~4, p.~86, 2011.

\bibitem{merrell2011interactive}
P.~Merrell, E.~Schkufza, Z.~Li, M.~Agrawala, and V.~Koltun, ``Interactive
  furniture layout using interior design guidelines,'' in \emph{ACM Trans.
  Graphics}, vol.~30, no.~4, 2011, p.~87.

\bibitem{yeh2012synthesizing}
Y.-T. Yeh, L.~Yang, M.~Watson, N.~D. Goodman, and P.~Hanrahan, ``Synthesizing
  open worlds with constraints using locally annealed reversible jump {McMC},''
  \emph{ACM Trans. Graphics}, vol.~31, no.~4, p.~56, 2012.

\bibitem{fu_siga17}
Q.~Fu, X.~Chen, X.~Wang, S.~Wen, B.~Zhou, and H.~Fu, ``Adaptive synthesis of
  indoor scenes via activity-associated object relation graphs,'' \emph{ACM
  Trans. Graphics}, vol.~36, no.~6, p. 201, 2017.

\bibitem{feng2016crowd}
T.~Feng, L.-F. Yu, S.-K. Yeung, K.~Yin, and K.~Zhou, ``Crowd-driven mid-scale
  layout design,'' \emph{ACM Trans. Graphics}, vol.~35, no.~4, 2016.

\bibitem{fisher2012example}
M.~Fisher, D.~Ritchie, M.~Savva, T.~Funkhouser, and P.~Hanrahan,
  ``Example-based synthesis of {3D} object arrangements,'' \emph{ACM Trans.
  Graphics}, vol.~31, no.~6, p. 135, 2012.

\bibitem{peng2014computing}
C.-H. Peng, Y.-L. Yang, and P.~Wonka, ``Computing layouts with deformable
  templates,'' \emph{ACM Trans. Graphics}, vol.~33, no.~4, p.~99, 2014.

\bibitem{wu2018}
W.~Wu, L.~Fan, L.~Liu, and P.~Wonka, ``Miqp-based layout design for building
  interiors,'' in \emph{Comp. Graphics Forum}, 2018.

\bibitem{majerowicz2014filling}
L.~Majerowicz, A.~Shamir, A.~Sheffer, and H.~H. Hoos, ``Filling your shelves:
  {Synthesizing} diverse style-preserving artifact arrangements,'' \emph{IEEE
  Trans. on Visualization and Comp. Graphics}, vol.~20, no.~11, pp. 1507--1518,
  2014.

\bibitem{Bao:2013:GEG:2461912.2461977}
F.~Bao, D.-M. Yan, N.~J. Mitra, and P.~Wonka, ``Generating and exploring good
  building layouts,'' \emph{ACM Trans. Graphics}, vol.~32, no.~4, 2013.

\bibitem{zhu2012motion}
L.~Zhu, W.~Xu, J.~Snyder, Y.~Liu, G.~Wang, and B.~Guo, ``Motion-guided
  mechanical toy modeling.'' \emph{ACM Trans. Graphics}, vol.~31, no.~6, p.
  127, 2012.

\bibitem{cao2012automatic}
Y.~Cao, A.~B. Chan, and R.~W. Lau, ``Automatic stylistic manga layout,''
  \emph{ACM Trans. Graphics}, vol.~31, no.~6, p. 141, 2012.

\bibitem{cao2014look}
Y.~Cao, R.~W. Lau, and A.~B. Chan, ``Look over here: Attention-directing
  composition of manga elements,'' \emph{ACM Trans. Graphics}, vol.~33, no.~4,
  p.~94, 2014.

\bibitem{reinert2013interactive}
B.~Reinert, T.~Ritschel, and H.-P. Seidel, ``Interactive by-example design of
  artistic packing layouts,'' \emph{ACM Trans. Graphics}, vol.~32, no.~6, p.
  218, 2013.

\bibitem{terzopoulos1987elastically}
D.~Terzopoulos, J.~Platt, A.~Barr, and K.~Fleischer, ``Elastically deformable
  models,'' \emph{ACM Trans. Graphics}, vol.~21, no.~4, pp. 205--214, 1987.

\bibitem{manteaux2016adaptive}
P.-L. Manteaux, C.~Wojtan, R.~Narain, S.~Redon, F.~Faure, and M.-P. Cani,
  ``Adaptive physically based models in computer graphics,'' in \emph{Comp.
  Graphics Forum}, 2016.

\bibitem{qin1996d}
H.~Qin and D.~Terzopoulos, ``D-nurbs: a physics-based framework for geometric
  design,'' \emph{IEEE Trans. on Visualization and Comp. Graphics}, vol.~2,
  no.~1, pp. 85--96, 1996.

\bibitem{attar2009physics}
R.~Attar, R.~Aish, J.~Stam, D.~Brinsmead, A.~Tessier, M.~Glueck, and A.~Khan,
  ``Physics-based generative design,'' in \emph{CAAD Futures Conf.}, 2009, pp.
  231--244.

\bibitem{harada1995interactive}
M.~Harada, A.~Witkin, and D.~Baraff, ``Interactive physically-based
  manipulation of discrete/continuous models,'' \emph{ACM Trans. Graphics}, pp.
  199--208, 1995.

\bibitem{arvin2002modeling}
S.~A. Arvin and D.~H. House, ``Modeling architectural design objectives in
  physically based space planning,'' \emph{Automation in Construction},
  vol.~11, no.~2, pp. 213--225, 2002.

\bibitem{muller2007position}
M.~M{\"u}ller, B.~Heidelberger, M.~Hennix, and J.~Ratcliff, ``Position based
  dynamics,'' \emph{Virtual Reality Interactions and Physical Simulations
  (VRIPHYS)}, vol.~18, no.~2, pp. 109--118, 2007.

\bibitem{stam2009nucleus}
J.~Stam, ``Nucleus: {Towards} a unified dynamics solver for computer
  graphics,'' in \emph{Proc. IEEE Int. Conf. on Comp.-Aided Design and Comp.
  Graphics}, 2009, pp. 1--11.

\bibitem{deul2016position}
C.~Deul, P.~Charrier, and J.~Bender, ``Position-based rigid-body dynamics,''
  \emph{Comp. Animation and Virtual Worlds}, vol.~27, no.~2, pp. 103--112,
  2016.

\bibitem{macklin2013position}
M.~Macklin and M.~M{\"u}ller, ``Position based fluids,'' \emph{ACM Trans.
  Graphics}, vol.~32, no.~4, p. 104, 2013.

\bibitem{weiss_mig17}
T.~Weiss, A.~Litteneker, C.~Jiang, and D.~Terzopoulos, ``Position-based
  multi-agent dynamics for real-time crowd simulation,'' in \emph{Motion in
  Games}, ser. MIG '17.\hskip 1em plus 0.5em minus 0.4em\relax ACM, 2017, pp.
  10:1--10:8.

\bibitem{bender2014survey}
J.~Bender, M.~M{\"u}ller, M.~A. Otaduy, M.~Teschner, and M.~Macklin, ``A survey
  on position-based simulation methods in computer graphics,'' in \emph{Comp.
  Graphics Forum}, vol.~33, no.~6, 2014, pp. 228--251.

\bibitem{Algower2003}
E.~L. Algower and K.~Georg, \emph{Introduction to Numerical Continuation
  Methods}.\hskip 1em plus 0.5em minus 0.4em\relax SIAM, 2003, vol.~45.

\bibitem{macklin:2014:unified}
M.~Macklin, M.~M{\"u}ller, N.~Chentanez, and T.~Kim, ``Unified particle physics
  for real-time applications,'' \emph{ACM Trans Graph}, vol.~33, no.~4, p. 104,
  2014.

\bibitem{umetani2014position}
N.~Umetani, R.~Schmidt, and J.~Stam, ``Position-based elastic rods,'' in
  \emph{Proc. of the ACM SIGGRAPH/Eurographics Symp. on Comp. Animation}, 2014,
  pp. 21--30.

\bibitem{dechiara2001time}
J.~DeChiara, J.~Panero, and M.~Zelnik, \emph{Time-Saver Standards for Interior
  Design and Space Planning}.\hskip 1em plus 0.5em minus 0.4em\relax
  McGraw-Hill, 2001.

\bibitem{jones2014beginnings}
L.~M. Jones and P.~S. Allen, \emph{Beginnings of Interior Environments}.\hskip
  1em plus 0.5em minus 0.4em\relax Pearson, 2014.

\bibitem{deasy1990designing}
C.~Deasy and T.~E. Lasswell, \emph{Designing Places for People}.\hskip 1em plus
  0.5em minus 0.4em\relax Whitney, 1990.

\bibitem{lok2004evaluation}
S.~Lok, S.~Feiner, and G.~Ngai, ``Evaluation of visual balance for automated
  layout,'' in \emph{Proc. of the 9th Int. Conf. on Intelligent user
  interfaces}, 2004, pp. 101--108.

\bibitem{talbott1999decorating}
C.~Talbott, M.~Matthews, and C.~Cosentino, \emph{Decorating for Good: {A}
  Step-by-step Guide to Rearranging What You Already Own}.\hskip 1em plus 0.5em
  minus 0.4em\relax C. Potter, 1999.

\bibitem{nlopt2011}
\BIBentryALTinterwordspacing
S.~G. Johnson, \emph{The NLopt nonlinear-optimization package}, 2011. [Online].
  Available: \url{http://ab-initio.mit.edu/nlopt}
\BIBentrySTDinterwordspacing

\bibitem{badrinarayanan2015segnet}
V.~Badrinarayanan, A.~Kendall, and R.~Cipolla, ``Segnet: A deep convolutional
  encoder-decoder architecture for image segmentation,'' \emph{arXiv preprint
  arXiv:1511.00561}, 2015.

\bibitem{xiao2010sun}
J.~Xiao, J.~Hays, K.~A. Ehinger, A.~Oliva, and A.~Torralba, ``Sun database:
  Large-scale scene recognition from abbey to zoo,'' in \emph{Comp. Vis. and
  Pattern Recognition (CVPR)}, 2010, pp. 3485--3492.

\bibitem{rother2004grabcut}
C.~Rother, V.~Kolmogorov, and A.~Blake, ``Grabcut: Interactive foreground
  extraction using iterated graph cuts,'' \emph{ACM Trans. Graphics}, vol.~23,
  no.~3, pp. 309--314, 2004.

\bibitem{xie2015holistically}
S.~Xie and Z.~Tu, ``Holistically-nested edge detection,'' in \emph{Proc. of the
  Int. Conf. on Comp. Vis. (ICCV)}, 2015, pp. 1395--1403.

\bibitem{weiss2017automated}
T.~Weiss, M.~Nakada, and D.~Terzopoulos, ``Automated layout synthesis and
  visualization from images of interior or exterior spaces,'' in \emph{IEEE
  CVPR Workshop on Vis. Meets Cognition}, 2017, pp. 41--47.

\end{thebibliography}

\begin{IEEEbiography}[{\includegraphics[width=1in,height=1.25in,clip,keepaspectratio]{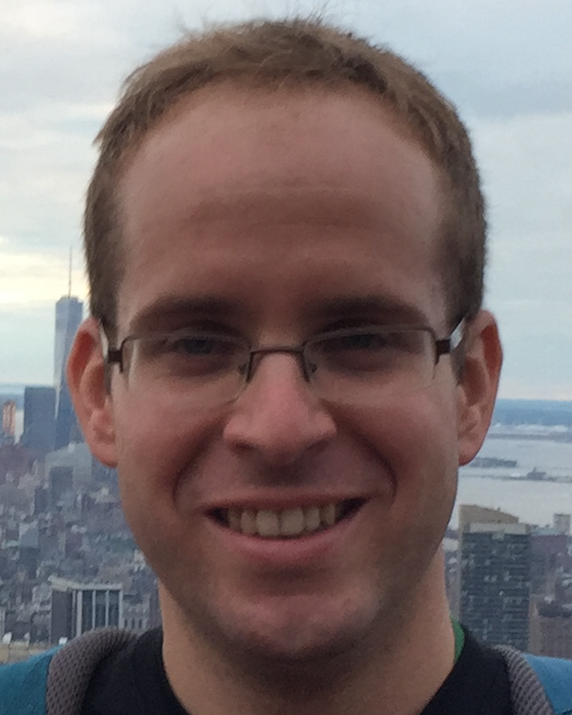}}]{Tomer Weiss}
received his PhD degree in computer science at the University of
California, Los Angeles, in 2018 and his BSc degree in Computer
Science from Tel Aviv University in 2013. Currently, he is an
operations research engineer at Wayfair, Inc., and a research
scientist in the UCLA Computer Graphics \& Vision Laboratory. His
research interests include computer graphics and optimization methods.
\end{IEEEbiography}
\begin{IEEEbiography}[{\includegraphics[width=1in,height=1.25in,clip,keepaspectratio]{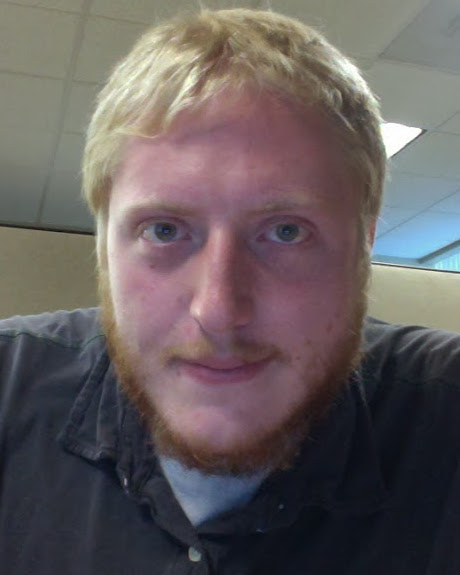}}]{Alan Litteneker}
is a PhD student at the University of California, Los Angeles, and a
member of the UCLA Computer Graphics \& Vision Laboratory. He received
his BSc degree in computer science from Chapman University in 2013.
His research interests include computer graphics and cinematography.
\end{IEEEbiography}
\begin{IEEEbiography}[{\includegraphics[width=1in,height=1.25in,clip,keepaspectratio]{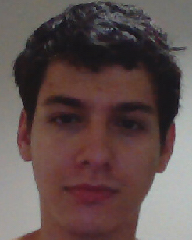}}]{Noah Duncan}
is Co-Founder and Chief Technology Officer at WorkPatterns, Inc. He received
his PhD degree in computer science from the University of California,
Los Angeles, in 2017 and his BSc degree in computer science from
Harvey Mudd College in 2012. His research interests include computer
graphics, with a focus on open-ended design problems.
\end{IEEEbiography}
\vskip -18pt plus -1fil
\begin{IEEEbiography}[{\includegraphics[width=1in,height=1.25in,clip,keepaspectratio]{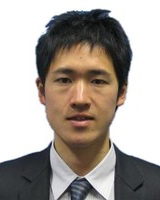}}]{Masaki Nakada}
is a postdoctoral scholar in the UCLA Computer Graphics \& Vision
Laboratory. He obtained his PhD degree in computer science from the
University of California, Los Angeles, in 2017 and his MS and BSc
degrees in Applied Physics from Waseda University in Tokyo, Japan. His
main research interests span machine learning, neuroscience, computer
vision, computer graphics, and biomechanics.
\end{IEEEbiography}
\begin{IEEEbiography}[{\includegraphics[width=1in,height=1.25in,clip,keepaspectratio]{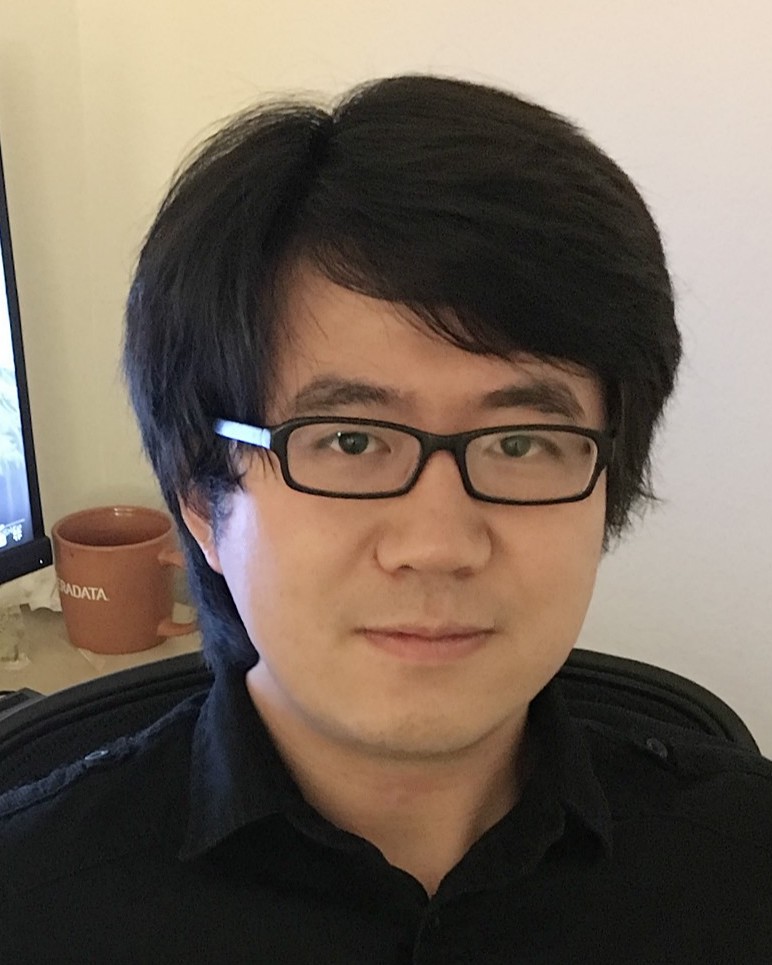}}]{Chenfanfu Jiang}
is an assistant professor of Computer and Information Science at the
University of Pennsylvania. He received his PhD degree in Computer
Science from the University of California, Los Angeles, in 2015 and was
awarded the 2015 UCLA Henry Samueli School of Engineering and Applied
Science Edward K.~Rice Outstanding Doctoral Student Award. He received
his BS degree in Physics in the Class for the Gifted Young (SCGY) of
the University of Science and Technology of China (USTC) in 2010. His
primary research focus is physics-based simulation and visual
computing, and their overlap with computer vision, robotics, cognitive
science, and medicine.
\end{IEEEbiography}
\begin{IEEEbiography}[{\includegraphics[width=1in,height=1.25in,clip,keepaspectratio]{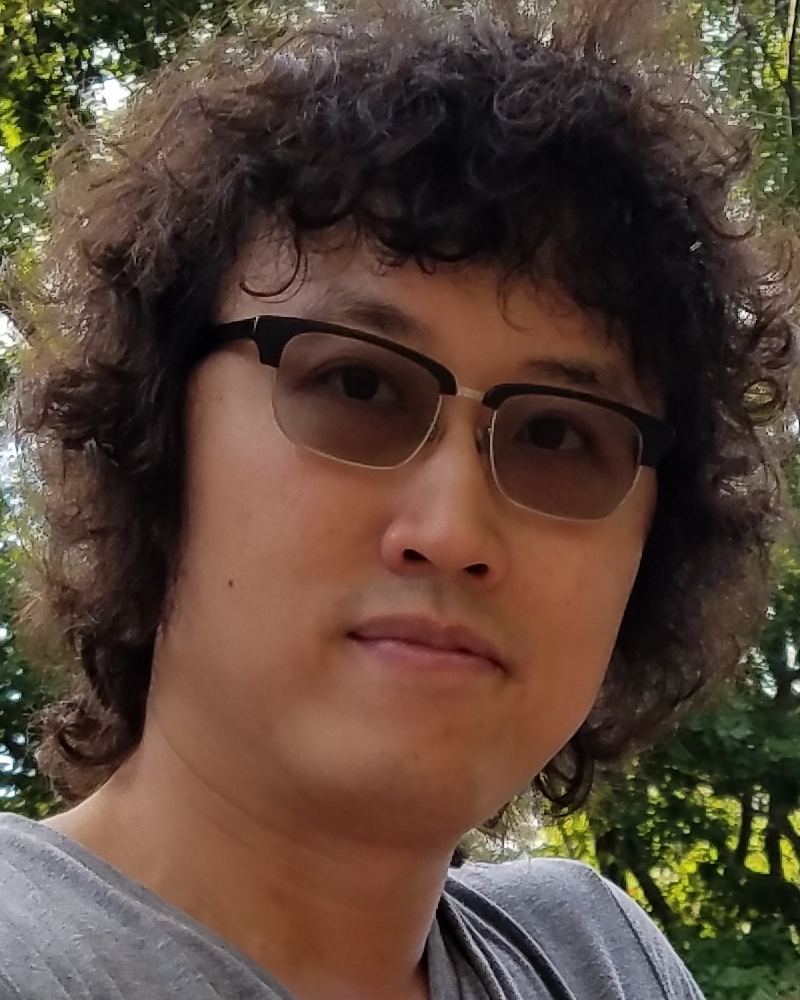}}]{Lap-Fai (Craig) Yu}
is an assistant professor at the University of Massachusetts Boston,
where he directs the Graphics and Virtual Environment Laboratory. He
received his BEng and MPhil degrees in computer science from the Hong
Kong University of Science and Technology (HKUST) in 2007 and 2009,
respectively, and his PhD degree in computer science from the
University of California, Los Angeles, in 2013, where he received the
Cisco Outstanding Graduate Research Award. He is also a recipient of
the Award of Excellence from Microsoft Research Asia. He has been a
visiting scholar at Stanford University and a visiting scientist at
the Massachusetts Institute of Technology. His research interests
include computer graphics, computer vision, and virtual reality.
\end{IEEEbiography}
\begin{IEEEbiography}[{\includegraphics[width=1in,height=1.25in,clip,keepaspectratio]{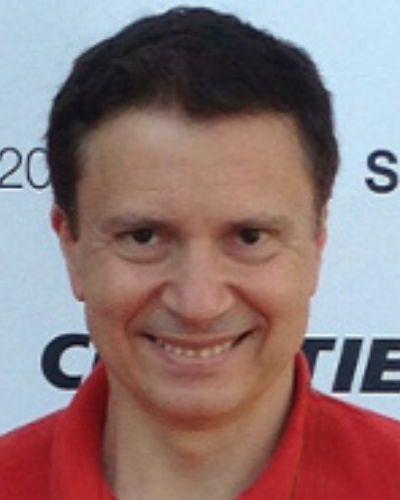}}]{Demetri Terzopoulos}
is a Chancellor's Professor of Computer Science at the University of
California, Los Angeles, where he holds the rank of Distinguished
Professor and directs the UCLA Computer Graphics \& Vision Laboratory.
He is also Co-Founder and Chief Scientist of VoxelCloud, Inc.
He graduated from McGill University and received his PhD degree ('84)
in artificial intelligence from MIT. He is or was a Guggenheim Fellow,
a Fellow of the ACM, a Fellow of the IEEE, a Fellow of the Royal
Society of London, a Fellow of the Royal Society of Canada, a member
of the European Academy of Sciences and the New York Academy of
Sciences, and a life member of Sigma Xi. His many awards include an
Academy Award for Technical Achievement from the Academy of Motion
Picture Arts and Sciences for his pioneering work on physics-based
computer animation, and the inaugural Computer Vision Distinguished
Researcher Award from the IEEE for his pioneering and sustained
research on deformable models and their applications. ISI and other
indexes list him among the most highly-cited authors in engineering
and computer science, with more than 400 published research papers and
several volumes, primarily in computer graphics, computer vision,
medical imaging, computer-aided design, and artificial
intelligence/life. He has given hundreds of invited talks around the
world about his research, including more than 100 distinguished
lectures and keynote/plenary addresses. He joined UCLA in 2005 from
New York University, where he held the Henry and Lucy Moses
Professorship in Science and was Professor of Computer Science and
Mathematics at NYU's Courant Institute of Mathematical Sciences.
Previously, he was Professor of Computer Science and Professor of
Electrical \& Computer Engineering at the University of Toronto.
Before becoming an academic in 1989, he was a Program Leader at
Schlumberger corporate research centers in California and Texas.
\end{IEEEbiography}

\end{document}